\documentclass[aps,prd,notitlepage,amsmath,amssymb,preprintnumbers,nofootinbib,longbibliography,superscriptaddress,10pt]{revtex4-1}
\usepackage[utf8]{inputenc} %utf8
\usepackage{graphicx}
\usepackage{url}
\usepackage[bookmarks, pagebackref=false]{hyperref}
\usepackage[usenames,dvipsnames]{xcolor}
\usepackage{todonotes}
\usepackage{tikz}
\usetikzlibrary{shapes,decorations.pathreplacing,decorations.markings,positioning}
 \allowdisplaybreaks
\makeatletter
\pgfdeclareshape{crossed circle}
{
  \inheritsavedanchors[from=circle]
  \inheritanchorborder[from=circle]
  \inheritanchor[from=circle]{north}
  \inheritanchor[from=circle]{north west}
  \inheritanchor[from=circle]{north east}
  \inheritanchor[from=circle]{center}
  \inheritanchor[from=circle]{west}
  \inheritanchor[from=circle]{east}
  \inheritanchor[from=circle]{mid}
  \inheritanchor[from=circle]{mid west}
  \inheritanchor[from=circle]{mid east}
  \inheritanchor[from=circle]{base}
  \inheritanchor[from=circle]{base west}
  \inheritanchor[from=circle]{base east}
  \inheritanchor[from=circle]{south}
  \inheritanchor[from=circle]{south west}
  \inheritanchor[from=circle]{south east}

  \inheritbackgroundpath[from=circle]

  \foregroundpath{
    \centerpoint
    \pgf@xc =\radius
    \advance\pgf@x by-0.707107\pgf@xc
    \advance\pgf@y by-0.707107\pgf@xc
    \pgf@xa=\pgf@x \pgf@ya=\pgf@y  % Store coordinates
    \centerpoint
    \pgf@xc=\radius
    \advance\pgf@x by0.707107\pgf@xc
    \advance\pgf@y by0.707107\pgf@xc
    \pgf@xb=\pgf@x \pgf@yb=\pgf@y  % Store coordinates
    \pgfpathmoveto{\pgfqpoint{\pgf@xa}{\pgf@ya}}
    \pgfpathlineto{\pgfqpoint{\pgf@xb}{\pgf@yb}}
    \pgfpathmoveto{\pgfqpoint{\pgf@xa}{\pgf@yb}}
    \pgfpathlineto{\pgfqpoint{\pgf@xb}{\pgf@ya}}
    \pgfsetarrowsstart{}
    \pgfsetarrowsend{}
 }
}
\pgfdeclareshape{crossed rectangle}
{
  \inheritsavedanchors[from=rectangle]
  \inheritanchorborder[from=rectangle]
  \inheritanchor[from=rectangle]{north}
  \inheritanchor[from=rectangle]{north west}
  \inheritanchor[from=rectangle]{north east}
  \inheritanchor[from=rectangle]{center}
  \inheritanchor[from=rectangle]{west}
  \inheritanchor[from=rectangle]{east}
  \inheritanchor[from=rectangle]{mid}
  \inheritanchor[from=rectangle]{mid west}
  \inheritanchor[from=rectangle]{mid east}
  \inheritanchor[from=rectangle]{base}
  \inheritanchor[from=rectangle]{base west}
  \inheritanchor[from=rectangle]{base east}
  \inheritanchor[from=rectangle]{south}
  \inheritanchor[from=rectangle]{south west}
  \inheritanchor[from=rectangle]{south east}

  \inheritbackgroundpath[from=rectangle]

  \foregroundpath{
    \southwest \pgf@xa=\pgf@x \pgf@ya=\pgf@y
    \northeast \pgf@xb=\pgf@x \pgf@yb=\pgf@y
    \pgfpathmoveto{\pgfqpoint{\pgf@xa}{\pgf@ya}}
    \pgfpathlineto{\pgfqpoint{\pgf@xb}{\pgf@yb}}
    \pgfpathmoveto{\pgfqpoint{\pgf@xa}{\pgf@yb}}
    \pgfpathlineto{\pgfqpoint{\pgf@xb}{\pgf@ya}}
    \pgfsetarrowsstart{}
    \pgfsetarrowsend{}
 }
}
\pgfdeclarelayer{edgelayer}
\pgfdeclarelayer{nodelayer}
\pgfsetlayers{edgelayer,nodelayer,main}
\tikzset{
	PL/.style={myred,thick},
	PR/.style={myblue,thick}
}
\tikzset{
	none/.style={inner sep=0pt},
	h4/.style = {
		minimum size=12pt,
		fill={rgb,255: red,128; green,128; blue,0}, 
		draw={rgb,255: red,128; green,128; blue,0}, 
		shape=circle, 
		scale=0.3
	},
	ffh/.style={
		fill=mycyan,
		shape=circle,
		scale=0.3
	},
	Op/.style={
		draw,
		fill=white,
		shape=crossed circle,
		scale=0.5
	},
	ChFlip/.style={
		draw,
		fill=white,
		shape=crossed rectangle,
		scale=0.5
	},
	onepi/.style={
		circle,
		minimum height=15pt,
		minimum width=15pt,
		fill=gray
	},
	scalar/.style={
		dashed,
		draw={rgb,255: red,255; green,128; blue,0}
	},
	fermion/.style={
		postaction=decorate,
		decoration={
    			markings,
    			mark=at position #1 with {
				\node[transform shape,
				isosceles triangle,
				inner sep=0mm,
				minimum width=3pt,
				xshift=-0.5pt,
				draw=none,
				fill] {};
		},
		}
	},
	fermion/.default={0.5},
	HtoHbar/.style n args={3}{
		preaction=decorate,decoration={
			markings,
			mark=at position #1 with {
		\draw[stealth-,thin,#2] (-1.5mm,1.5mm) -- (1.5mm,1.5mm);
		\draw[-stealth,thin,#3] (-1.5mm,-1.5mm) -- (1.5mm,-1.5mm);
			}
		}
	},
	HtoHbar/.default={0.4}{PL}{PR},
	HtoHbar'/.style n args={3}{
		preaction=decorate,decoration={
			markings,
			mark=at position #1 with {
		\draw[-stealth,thin,#2] (-1.5mm,1.5mm) -- (1.5mm,1.5mm);
		\draw[stealth-,thin,#3] (-1.5mm,-1.5mm) -- (1.5mm,-1.5mm);
			}
		}
	},
	HtoHbar'/.default={0.4}{PR}{PL}
}
\tikzset{every picture/.style={baseline=+12pt,scale=0.5}}

\tikzset{
	v-vertex/.pic = {
			\begin{pgfonlayer}{nodelayer}
				\node [style=none,minimum size=4pt,label={[xshift=-2pt,yshift=-4pt]$a$}] (a) at (-7.5, 2.5) {};
				\node [style=none,minimum size=4pt,label={[xshift=2pt,yshift=-4pt]$b$}] (b) at (-4.5, 2.5) {};
				\node [style=none,minimum size=4pt,label={[xshift=-2pt,yshift=-13pt]$d$}] (d) at (-7.5, -0.5) {};
				\node [style=none,minimum size=4pt,label={[xshift=2pt,yshift=-13pt]$c$}] (c) at (-4.5, -0.5) {};
			\end{pgfonlayer}
			\begin{pgfonlayer}{edgelayer}
				\draw [style={fermion},PL, looseness=1.3] (a.south) to[out=-45,in=45] (d.north);
				\draw [style={fermion},PR, looseness=1.3] (b.west) to[out=-135,in=-45] (a.east);
				\draw [style={fermion},PL, looseness=1.3] (c.north) to[out=135,in=-135] (b.south);
				\draw [style={fermion},PR, looseness=1.3] (d.east) to[out=45,in=135] (c.west);
			\end{pgfonlayer}
	},
	u-vertex/.pic = {
			\begin{pgfonlayer}{nodelayer}
				\node [style=none,minimum size=4pt,label={[xshift=-2pt,yshift=-4pt]$a$}] (a) at (-7.5, 2.5) {};
				\node [style=none,minimum size=4pt,label={[xshift=2pt,yshift=-4pt]$b$}] (b) at (-4.5, 2.5) {};
				\node [style=none,minimum size=4pt,label={[xshift=-2pt,yshift=-13pt]$d$}] (d) at (-7.5, -0.5) {};
				\node [style=none,minimum size=4pt,label={[xshift=2pt,yshift=-13pt]$c$}] (c) at (-4.5, -0.5) {};
			\end{pgfonlayer}
			\begin{pgfonlayer}{edgelayer}
				\draw [style={fermion},PL, looseness=1.3] (a.south) to[out=-45,in=-135] (b.south);
				\draw [style={fermion},PR, looseness=1.3] (b.west) to[out=-135,in=-45] (a.east);
				\draw [style={fermion},PL, looseness=1.3] (c.north) to[out=135,in=45] (d.north);
				\draw [style={fermion},PR, looseness=1.3] (d.east) to[out=45,in=135] (c.west);
			\end{pgfonlayer}
	},
	O3_ex/.pic={
	\begin{pgfonlayer}{nodelayer}
     	\node [style=ChFlip] (0) at (-5, 1) {};
		\node [style=none,minimum size=4pt,label={[xshift=-1pt,yshift=-4pt]$a$}] (a) at (-7.5, 2.5) {};
		\node [style=none,minimum size=4pt,label={[xshift=1pt,yshift=-4pt]$b$}] (b) at (-5, 2.5) {};
		\node [style=none,minimum size=4pt,label={[xshift=2pt,yshift=-4pt]$c$}] (c) at (-2.5, 2.5) {};
	\end{pgfonlayer}
	\begin{pgfonlayer}{edgelayer}
		\draw [style={fermion},PL] (a.south) to[out=-45,in=-135] (b.south);
		\draw [style={fermion},PR] (b.west) to[out=-135,in=-45] (a.east);
		\draw [style={fermion},PR] (0.east) to[out=15,in=-135] (c.south);
		\draw [style={fermion},PL] (c.west) to[out=-160,in=55] (0.north);
		\end{pgfonlayer}
	},
	O2_ex/.pic={
	\begin{pgfonlayer}{nodelayer}
		\node [style=ChFlip] (0) at (-5, 1) {};
		\node [style=none,minimum size=4pt,label={[xshift=-1pt,yshift=-4pt]$a$}] (a) at (-7.5, 2.5) {};
		\node [style=none,minimum size=4pt,label={[xshift=0pt,yshift=-4pt]$b$}] (b) at (-5, 2.5) {};
		\node [style=none,minimum size=4pt,label={[xshift=2pt,yshift=-4pt]$c$}] (c) at (-2.5, 2.5) {};

	\end{pgfonlayer}
	\begin{pgfonlayer}{edgelayer}
		\draw [style={fermion},PR] (b.west) to[out=-135,in=-45] (a.east);
		\draw [style={fermion},PL] (c.west) to[out=-135,in=-45] (b.east);
		\draw [style={fermion},PR] (0.east) to[out=15,in=-135] (c.south);
		\draw [style={fermion},PL] (a.south) to[out=-45,in=175] (0.west) ;
	\end{pgfonlayer}
}
}

\makeatother
\DeclareMathOperator{\Tr}{Tr}
\def\hc{\ensuremath{\mathrm{h.c.}}}

\definecolor{myred}{HTML}{FF2445}
\definecolor{myred}{HTML}{FF4D67}

\definecolor{myblue}{HTML}{5B7EB7}
\definecolor{myblue}{HTML}{608A91}
\definecolor{myblue}{HTML}{A4A8D1}

\definecolor{mycyan}{HTML}{75A0E6}

\def\rc{\textcolor{myred}{r_c}}
\def\rf{\textcolor{myred}{r_f}}

\begin{document}

\title{\textbf{\huge Asymptotic safety in the Litim-Sannino model at four loops}}

\author{A.V. {\sc Bednyakov}}\email{ bednya@jinr.ru}
\author{A.I. {\sc Mukhaeva}}\email{ mukhaeva@theor.jinr.ru}

\affiliation{%
	Joint Institute for Nuclear Research, Joliot-Curie, 6, Dubna 141980, Russia
}

\begin{abstract}
We consider a four-dimensional $SU(N_c)$ gauge theory coupled to $N_f$ species of color fermions and $N_f^2$ colorless scalars.
The quantum field theory possesses a weakly interacting ultraviolet fixed point that we determine from beta functions computed up to four-loop order in the gauge coupling, and up to three-loop order in the Yukawa and quartic scalar couplings. The fixed point has one relevant direction giving rise to asymptotic safety. We compute fixed point values of dimensionless couplings together with the corresponding scaling exponents up to the first three nontrivial orders in Veneziano parameter $\epsilon$, both for infinite and finite number of colors $N_c$. 
We also consider anomalous dimensions for fields, scalar mass squared, and a class of dimension-three operators. Contrary to previous studies, we take into account possible mixing of the latter and compute eigenvalues of the corresponding matrix. Further, we investigate the size of the conformal window in the Veneziano limit and its dependence on $N_c$.
\end{abstract}
\maketitle

\maketitle
\section{Introduction}

The study of asymptotic behavior of the dimensionless couplings in quantum field theory (QFT) provides important information both for the Standard Model (SM) and Beyond the Standard Model (BSM) scenarios. One of these behaviors is known as asymptotic freedom  \cite{Gross:1973id, Politzer:1973fx}, which is a defining feature of quantum chromodynamics. This behavior entails a decrease in the value of a coupling with the energy scale. Thus, in the deep ultraviolet (UV), this coupling tends to approach the Gaussian noninteractive fixed point (FP). 

Asymptotic safety (AS) is an extension of the concept of asymptotic freedom, as outlined in the work of S. Weinberg \cite{Weinberg:1980gg}. In AS, the coupling in the deep UV also reaches a fixed point, but unlike in asymptotic freedom, the fixed point value is not zero, which means that the theory remains interactive. Such theories are referred to as asymptotically safe.

The concept of asymptotic safety was initially introduced by S. Weinberg in the late 1970s as a means of achieving nonperturbative renormalizability for the four-dimensional theory of gravity \cite{Weinberg:1980gg}. However, in recent years, AS has been widely applied in the context of gauge theories to address issues with $U(1)$ gauge couplings (Landau pole) by stabilizing them at an interactive fixed point at a certain scale. 
Indications of AS has been found in simple \cite{Pica:2010xq,Litim:2014uca}, semisimple \cite{Bond:2017lnq}, and supersymmetric gauge theories coupled to matter \cite{Bajc:2016efj,Bond:2017suy,Bajc:2023uls}.  
From the phenomenological point of view, properties of such UV interactive fixed points for matter fields can be transmitted down to the low-energy regime and lead us to some phenomenological predictions, see, e.g., recent reviews \cite{Eichhorn:2022gku,Bednyakov:2023fmc}. 

An example of a UV-complete particle theory with a weakly interacting fixed point is a model  of $N_f$ fermions coupled to $SU(N_c)$ gauge fields and elementary scalars through gauge and Yukawa interactions \cite{Antipin:2013pya,Litim:2014uca}. In the large-$N$ Veneziano limit, the fixed point can be systematically studied in perturbation theory using a small control parameter $\epsilon$, allowing for the extraction of specific details of theory. Previous studies \cite{Litim:2014uca,Bond:2017tbw,Bond:2021tgu,Litim:2015iea}  have identified critical couplings and universal exponents up to second order in $\epsilon$, including finite $N$ corrections. 
Phenomenological applications of the model and its extensions can be found in  Refs.~\cite{Litim:2015iea,Sannino:2014lxa,Nielsen:2015una,Rischke:2015mea,Codello:2016muj,Bond:2017wut,Dondi:2017civ,Kowalska:2017fzw,Abel:2017ujy,Christiansen:2017qca,Sannino:2018suq,Barducci:2018ysr,Hiller:2019mou,Hiller:2019tvg,Hiller:2020fbu,Bissmann:2020lge,Bause:2021prv,Hiller:2022hgt,Hiller:2022rla}.  %Some applications within the framework of quantum gravity can be found, e.g., in the review \cite{Eichhorn:2022gku}. 
It is also worth mentioning that the model was the first nonsupersymmetric theory investigated in the large global charge limit~\cite{Orlando:2019hte,Antipin:2022naw}. The holographic description of the model was considered in Ref.~\cite{Rey:2019wve}. 

%Examples in the context of Gauge-Yukawa models can be found in Refs.~\cite{Litim:2015iea,Sannino:2014lxa,Nielsen:2015una,Rischke:2015mea,Codello:2016muj,Bond:2017wut,Dondi:2017civ,Kowalska:2017fzw,Abel:2017ujy,Christiansen:2017qca,Sannino:2018suq,Barducci:2018ysr,Hiller:2019mou,Hiller:2019tvg,Hiller:2020fbu,Bissmann:2020lge,Bause:2021prv,Hiller:2022hgt,Hiller:2022rla,Bednyakov:2023fmc}.  Some applications within the framework of quantum gravity can be found, e.g., in the review \cite{Eichhorn:2022gku}. 

In this paper, we confirm the study \cite{Litim:2023tym} of the UV critical theory and provide the fixed point couplings and conformal data up to third order in $\epsilon$ (both for infinite and finite-$N_c$ scenarios) by considering four-loop gauge, three-loop Yukawa, and quartic $\beta$ functions. We also determine three-loop anomalous dimensions for dimension-three operators and discuss peculiarities in their computations as compared to Ref.~\cite{Litim:2023tym}. 

The paper is organized as follows. Section \ref{sec:Model} provides the main information about the considered model and operators. In Sec.\ref{sec:Calc_methods} we give the details of computation of the renormalization-group (RG) functions. In Sec.\ref{sec:Res} we demonstrate our results for fixed points, anomalous dimensions and scaling exponents for finite $N_c$. The bounds on the conformal window are given in Sec.\ref{sec:conf_window}. We conclude in Sec.\ref{sec:Concl}. The full expressions for beta functions and various anomalous dimensions can be found in Appendixes \ref{sec:beta_func}, \ref{sec:ADM_fields_mass}, and \ref{sec:ADM_3}. App.\ref{sec:other_fits} contains additional information related to finite-$N_c$ results.

\section{Model description}\label{sec:Model}
\begin{table}[]
    \centering
    \begin{tabular}{|c|c|c|c|}
    \hline
       Field  & $SU(N_c)$& $U_L(N_f)$& $U_R(N_f)$ \\
       \hline
       $\psi_L$  & $N_c$& $N_f$& $1$ \\
       $\psi_R$  & $N_c$& $1$& $N_f$ \\
       \hline
       $H$ & $1$& $N_f$& $\bar{N_f}$ \\
       \hline
    \end{tabular}
    \caption{Representations of matter fields under gauge $SU(N_c)$ and flavor 
    $U_L(N_f)$ and $U_R(N_f)$ groups.}
    \label{tab:LS_groups}
\end{table}

We consider four-dimensional theory with $SU(N_c)$ gauge fields coupled to $N_f$ massless Dirac fermions and a scalar singlet  field $H$. 
The last is uncharged under the gauge group and carries two flavor indices, such that it can be written as a $N_f \times N_f$ complex matrix. The corresponding Lagrangian is
\begin{align}
    \mathcal{L}  &= -\frac{1}{4}F^{a\mu\nu}F_{\mu\nu}^a+\mathcal{L}_{gf} + \mathcal{L}_{gh}{}\nonumber\\ 
			& \hspace{0.5cm}
    +\Tr(\bar{\psi}i\hat D \psi)+\Tr(\partial^\mu H^\dagger \partial_\mu H ) - y\Tr[\bar{\psi}( H \mathcal{P}_R +\ H^\dagger \mathcal{P}_L)\psi]{}\nonumber\\ 
			& \hspace{0.5cm} - m^2\Tr(H^\dagger H) - u\Tr((H^\dagger H)^2) - v(\Tr(H^\dagger H))^2,
    \label{eq:Lag}
\end{align}
where $F^a_{\mu\nu}$ is the field strength of the gauge bosons $G^a_\mu$ with $a=1,\ldots,N_c^2-1$. The trace in Eq.~\eqref{eq:Lag} runs over both color and flavor indices and $\psi = \psi_L+\psi_R$ are fermions with $\mathcal{P}_{L/R}  = \frac{1}{2}(1\pm \gamma_5)$.
In what follows we use a linear $R_\xi$-gauge with $\mathcal{L}_{gf} = - \frac{1}{2\xi} (\partial_\mu G^a_\mu)^2$ together with the corresponding ghost Lagrangian $\mathcal{L}_{gh}$. 

The theory \eqref{eq:Lag} is invariant under global $G = U_L(N_f) \times U_R(N_f)$ ``flavor'' symmetry corresponding to independent unitary rotations of left- and right-handed chiral fermions. The matrix scalar field $H$ is colorless but transform under $G$ (see Table~\ref{tab:LS_groups}). 
In this paper we also consider a class of operators that breaks the flavor symmetry down to diagonal $U(N_f)$. One can introduce independent couplings for these operators resulting in the following additional contribution to the Lagrangian \eqref{eq:Lag}
\begin{align}
	\delta \mathcal{L} & = 
	- m_\psi \Tr (\bar \psi \psi) 
	- \frac{h_2}{2}
	\left[
	\Tr (H H^\dagger H) + \hc   
	\right]
	- \frac{h_3}{2}
	\left[
	\Tr (H H^\dagger) \Tr(H) + \hc   
	\right] \nonumber \\
	& \equiv - m_\psi O_1 - h_2 O_2 - h_3 O_3
	= - \vec{\kappa} \cdot \vec{O}.
	\label{eq:dLag}
\end{align}
The choice of the $G$-breaking terms is dictated by the fact that the operator $O_1 = \bar \psi \psi$, also considered in Ref.~\cite{Litim:2023tym}, mixes under renormalization with the two operators coupled to $h_{2,3}$. 

The model has four dimensionless couplings:  gauge coupling $g$, the Yukawa $y$, and two quartic scalar couplings $u$ and $v$. One usually introduces a set of rescaled couplings \cite{tHooft:1973alw}
\begin{align}
    \alpha_g =\frac{g^2 N_c}{(4\pi^2)}, \quad \alpha_y =\frac{y^2 N_c}{(4\pi^2)},\nonumber\\
    \alpha_u =\frac{u N_f}{(4\pi^2)}, \quad \alpha_v =\frac{v N_f^2}{(4\pi^2)}.
    \label{eq:dim4_couplings_VL}
\end{align}

The latter has been done since we consider the Veneziano limit \cite{Veneziano:1976wm} with $N_f, N_c \to \infty$. One also introduces a parameter
\begin{align}
    \epsilon & \equiv \frac{N_f}{N_c}-\frac{11}{2}
    \label{eq:eps_def}
\end{align}
that becomes continuous and may take any value between $(-\frac{11}{2},\infty)$.
The benefit of the Veneziano limit is that it allows systematic expansions in a small parameter. 
In our work we suppose that
\begin{align}
    0<\epsilon\leq 1,
\end{align}
and treat it as a small control parameter for perturbativity. 

In order to study dimension-three operators in the Veneziano limit, we rescale the corresponding couplings and the operators as
\begin{align}
	m_\psi' & = m_\psi \sqrt{N_c},  & h'_2 & = h_2 N_f, & h'_3 & = h_3 N_f^2, 
	\label{eq:dim3_couplings_VL}\\
	O'_1  & = O_1/\sqrt{N_c}, & O'_2 & = O_2/N_f, & O'_3 &= O_3/N_f^2.   
	\label{eq:dim3_ops_rescaled}
\end{align}

This allows one to absorb all corrections with positive powers of $N_c$ and $N_f$ appearing in the beta functions for $\vec{\kappa}$ into the rescaled couplings given in Eqs.~\eqref{eq:dim4_couplings_VL} and \eqref{eq:dim3_couplings_VL}. It is worth noticing that the parameters $m_\phi$, $h_1$ and $h_2$  are rescaled in the same way as the dimensionless couplings $y$, $u$, and $v$, respectively. 

\section{Calculation methods}\label{sec:Calc_methods}
The beta functions and anomalous dimensions in the $\overline{\mathrm{MS}}$ scheme can be computed by standard methods, so we omit full description of the techniques refereeing to appropriate literature. In this section, we only discuss peculiarities of the current calculation.  Let us mention here that we do not rescale the couplings and operators according to Eqs.~\eqref{eq:dim4_couplings_VL} and \eqref{eq:dim3_couplings_VL} in our explicit computation. The transition to the Veneziano-limit normalization is carried out at the final stage. Nevertheless, we present all our results in terms of $\alpha_{g,y,u,v}$ and for the rescaled operators \eqref{eq:dim3_ops_rescaled}.  

\begin{figure*}[t]
\begin{align*}
\begin{tikzpicture}
	\begin{pgfonlayer}{nodelayer}
		\node [style=ffh] (0) at (-6, 1) {};
		\node [style=none,label={[xshift=1pt,yshift=-2pt]$\psi_{\alpha,L}^{i}$}] (1) at (-7.5, 2.5) {};
		\node [style=none,label={[xshift=-5pt,yshift=-14pt]$\psi_{\beta,R}^{j}$}] (2) at (-7.5, -0.5) {};
		\node [style=none,label={[xshift=5pt,yshift=-1pt]$\phi^a$}] (3) at (-4.5, 1) {};
	\end{pgfonlayer}
	\begin{pgfonlayer}{edgelayer}
		%\draw [style=scalarRL] (0) to (3.center);
		\draw [style=scalar,HtoHbar] (0) to (3.center);
		\draw [style=fermion,PR] (2.center) to (0);
		\draw [style=fermion,PL] (0) to (1.center);
	\end{pgfonlayer}
\end{tikzpicture}
& = 
- i y  T^a_{ij}  \delta_{\alpha \beta},
\hspace{4.5cm}
\begin{tikzpicture}
	\begin{pgfonlayer}{nodelayer}
		\node [style=ffh] (0) at (-6, 1) {};
		\node [style=none,label={[xshift=1pt,yshift=-2pt]$\psi_{\alpha,R}^{i}$}] (1) at (-7.5, 2.5) {};
		\node [style=none,label={[xshift=-5pt,yshift=-14pt]$\psi_{\beta,L}^{j}$}] (2) at (-7.5, -0.5) {};
		\node [style=none,label={[xshift=2pt,yshift=-1pt]$\phi^a$}] (3) at (-4.5, 1) {};
	\end{pgfonlayer}
	\begin{pgfonlayer}{edgelayer}
		%\draw [style=scalarLR] (0) to (3.center);
		\draw [style=scalar,HtoHbar={0.4}{PR}{PL}] (0) to (3.center);
		\draw [style=fermion,PL] (2.center) to (0);
		\draw [style=fermion,PR] (0) to (1.center);
	\end{pgfonlayer}
\end{tikzpicture}
= - i y \cdot \bar T^a_{ij} \cdot \delta_{\alpha\beta}
	%\label{eq:ffh_vertex}
\\
\begin{tikzpicture}
	\begin{pgfonlayer}{nodelayer}
		\node [style=h4] (0) at (-6, 1) {};
		\node [style=none,label={[xshift=-10pt,yshift=-2pt]$\phi^a$}] (5) at (-7.5, 2.5) {};
		\node [style=none,label={[xshift=10pt,yshift=-2pt]$\phi^b$}] (2) at (-4.5, 2.5) {};
		\node [style=none,label={[xshift=-10pt,yshift=-12pt]$\phi^d$}] (3) at (-7.5, -0.5) {};
		\node [style=none,label={[xshift=10pt,yshift=-12pt]$\phi^c$}] (4) at (-4.5, -0.5) {};
	\end{pgfonlayer}
	\begin{pgfonlayer}{edgelayer}
		\draw [style=scalar] (5.center) to (0);
		\draw [style=scalar] (2.center) to (0);
		\draw [style=scalar] (0) to (4.center);
		\draw [style=scalar] (0) to (3.center);
	\end{pgfonlayer}
\end{tikzpicture}
& = 
- i 2 \{ u \cdot [ \underbrace{\Tr(T^a \bar T^b) \Tr(T^c \bar T^d)}_{\tikz{\pic[scale=0.3]{u-vertex}}} + \text{11 perms.}]
+v \cdot [ \underbrace{\Tr (T^a \bar T^b T^c \bar T^d)}_{\tikz{\pic[scale=0.3]{v-vertex}}} + \text{11 perms.}]\}
	%\label{eq:h4_vertex},
\end{align*}
\caption{Feynman rules for Yukawa and quartic vertices.
The indices $i,j=1,\ldots,N_f$ count fermion generations (flavor), while $\alpha,\beta=1,\ldots,N_c$ correspond to the $SU(N_c)$ gauge group. Flavor flows for left- and right-handed fermions are indicated.
	}
	\label{fig:O4_feynman_rules}
\end{figure*}
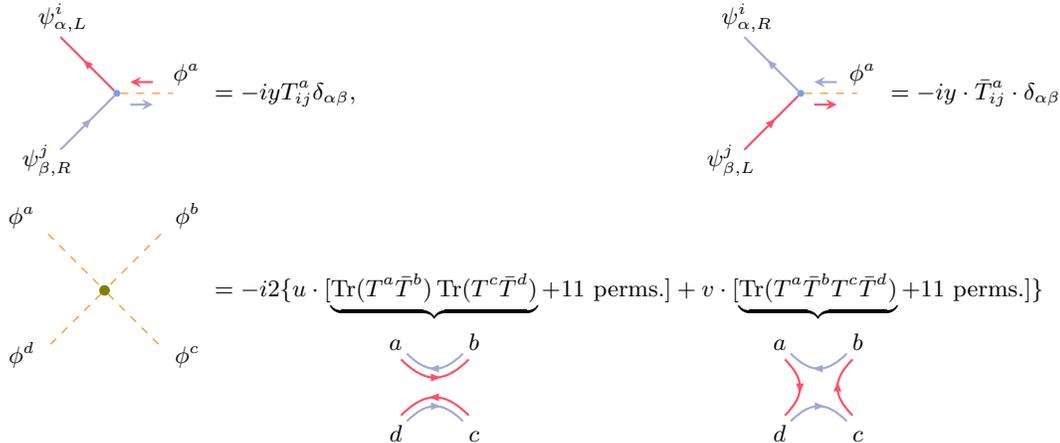

In order to compute the required RG functions, we rewrite the Lagrangian \eqref{eq:Lag} in terms of real scalars $\phi^a$ utilizing a decomposition ($a=1,\ldots,2 N_f^2$) \cite{Bednyakov:2021ojn}
\begin{align}
	H = \phi^a T^a, \qquad H^\dagger = \phi^a \bar T^a,
	\qquad \bar T^a \equiv T^{a\dagger}
	\label{eq:H_decomposition}
\end{align}
with $T^a$ being complex $N_f \times N_f$ matrices\footnote{One can also use a decomposition in terms of the identity matrix and $SU(N_f)$ generators as, e.g., in Refs.~\cite{Orlando:2019hte,Antipin:2022naw}.}  normalized as \begin{align}
	\Tr(T^a \bar T^b) + \Tr(T^b \bar T^a)  = \delta^{ab}
\end{align}
and satisfying the following identities
\begin{align}
	T^a_{ij} T^a_{kl} = \bar T^a_{ij} \bar T^a_{kl} = 0, \qquad 
	T^a_{ij} \bar T^a_{kl} = \delta_{il} \delta_{jk}.
	\label{eq:Tmatrix_rules}
\end{align}
Such a decomposition gives rise to the Feynman rules (see Fig.~\ref{fig:O4_feynman_rules}) for the vertices involving $\phi^a$.
In Fig.~\ref{fig:O4_feynman_rules} we also indicate the flavor ``flow'' for the double-trace and single-trace scalar couplings that can be associated with the left ($L$) and right ($R$) chiral fermions. In the absence of $U_L(N_f) \times U_R(N_f)$ breaking, the ``left-'' and ``right-handed'' flows are ``conserved'' separately. 

We implemented the rules in the \texttt{DIANA} \cite{Tentyukov:1999is} package based on \texttt{QGRAF} \cite{Nogueira:1991ex} and utilize  \texttt{FORM} \cite{Vermaseren:1992vn} to deal with index contractions and to compute Feynman integrals via the \texttt{MATAD} \cite{Steinhauser:2000ry} code. In order to derive RG equations for $u$, $v$ and $y$, we generate Green functions corresponding to radiative corrections for the tree-level Yukawa and quartic vertices up to three loops and extract local divergent terms by applying suitable projectors. Our explicit calculations heavily rely on well-known infrared rearrangement (IRR) trick \cite{Vladimirov:1979zm,Misiak:1994zw}, which allows one to deal only with fully massive vacuum integrals.  
	The fermion-fermion-scalar interaction involves the $\gamma_5$-matrix, which requires special treatment in dimensional regularization (see, e.g., Ref.~\cite{Jegerlehner:2000dz}). In this paper, we restrict ourselves to the seminaive approach \cite{Chetyrkin:2012rz,Bednyakov:2013eba} and by explicit computation we prove that potential ambiguities do not appear in the final result for the RG functions, thus, providing an independent cross-check of the results obtained in Ref.~\cite{Litim:2023tym}. 
	As for the four-loop gauge-coupling beta function the $\gamma_5$-ambiguity can be fixed by means of Weyl consistency conditions \cite{Jack:2013sha,Antipin:2013sga,Poole:2019txl,Poole:2019kcm}. Moreover, we do not carry out explicit computations here, but use the \texttt{RGBeta} code \cite{Thomsen:2021ncy} extended to 432-order in Refs.~\cite{Bednyakov:2021qxa,Davies:2021mnc}. 

	Let us now switch to the discussion of a family of dimension-three operators that includes $O_1 = \bar \psi \psi$.  In dimensionally regularized theory\footnote{Note the difference between the Veneziano parameter $\epsilon$ and that of dimensional regularization $\varepsilon$.} ($d=4 - 2\varepsilon$) within the $\overline{\mathrm{MS}}$-scheme, we have a relation between bare and renormalized quantities
\begin{align}
	\vec{\kappa}_R \cdot [\vec{O}]_R & =
	\mu^{2\varepsilon} \cdot (\vec{\kappa})_0 \cdot (\vec{O})_0  = 
	\mu^{2\varepsilon} \cdot (\vec{\kappa})_R \cdot \mathcal{Z}^T_\kappa \cdot \mathcal{Z}^{-1}_O \cdot [\vec{O}]_R,
\end{align}
where all couplings in $\vec{\kappa}_R$, and all operators in $\vec{O}_R$ have in our case the mass dimension one and three, respectively. The corresponding renormalization matrices $\mathcal{Z}_{\kappa}$ and $\mathcal{Z}_O$ are defined as 
\begin{align}
	\vec{\kappa}_0 &= \mathcal{Z}_\kappa(\mu,\alpha(\mu),\varepsilon) \cdot \vec{\kappa}_R, &  
	\mathcal{Z}_\kappa(\mu,\alpha(\mu),\varepsilon) &= 
	\begin{pmatrix} 1 & & \\
		& \mu^{\varepsilon} & \\
		& & \mu^{\varepsilon}
	\end{pmatrix}
	\cdot
	Z_{\kappa}(\alpha(\mu), \varepsilon),\\
		[\vec{O}]_R & = \mathcal{Z}_O(\mu,\alpha(\mu), \varepsilon) \cdot (\vec{O})_0, 
			    & 
		\mathcal{Z}_O(\mu, \alpha(\mu), \varepsilon) &= 
		Z_O(\alpha(\mu), \varepsilon)
		\cdot 
		\begin{pmatrix} \mu^{2\varepsilon} & & \\
		& \mu^{3 \varepsilon} & \\
		& & \mu^{3 \varepsilon}
	\end{pmatrix}.
	\label{eq:O123_renorm_no_EOMs}
\end{align}	
The renormalization matrices involve poles in $\varepsilon$ and depend on the renormalization scale $\mu$ together with the running dimensionless couplings from \eqref{eq:Lag} denoted collectively by $\alpha(\mu)$. They satisfy
\begin{align}
	\mu^{2\varepsilon} \mathcal{Z}_{\kappa}^T  = \mathcal{Z}_O\Rightarrow Z^T_\kappa = Z_O 
\end{align}
 In what follows, we will routinely use $\mathcal{Z}$ and $Z$ to denote renormalization constants with and without explicit dependence on the renormalization scale $\mu$.
The diagonal matrices involving powers of $\mu^\varepsilon$ account for mass dimensions of bare couplings and operators. We include them in the definition of $\mathcal{Z}$s for convenience, since we can write the beta functios of $\vec{\kappa}_R$ in a compact form 
\begin{align}
	\frac{d}{d \ln \mu} \vec{\kappa} \equiv \dot{\vec{\kappa}}
	= \vec{\beta}_\kappa = \gamma_\kappa(\alpha) \cdot \vec{\kappa}
	\quad \gamma_\kappa (\alpha)  = \dot{\mathcal{Z}}^{-1}_{\kappa} \cdot \mathcal{Z}_\kappa  = - \mathcal{Z}^{-1}_\kappa \cdot \dot{\mathcal{Z}}_\kappa, 
\end{align}		
and relate the matrix anomalous dimension $\gamma_\kappa$ to the anomalous dimensions of the dimension-three operators
\begin{align}
	\gamma_O(\alpha)  \equiv  - \dot{\mathcal{Z}}_O \cdot \mathcal{Z}^{-1}_O =
	- 2 \varepsilon - \dot{\mathcal{Z}}^T_\kappa \cdot (\mathcal{Z}^{T})^{-1}_\kappa = - 2\varepsilon + \gamma^T_\kappa(\alpha).
	\label{eq:gammaO}
\end{align}
Both $\gamma_O(\alpha)$ and $\gamma_\kappa(\alpha)$ should be finite in the limit $\varepsilon \to 0$, which serves as a welcome check of the computation. As a consequence, in $d=4$ we have 
\begin{align}
	\gamma_O = \gamma_\kappa^T
\end{align}
This relation can be used in two ways. Given the beta functions of (a closed set of) the dimension-1 couplings, one can extract the matrix anomalous dimension for the corresponding operators. In case we know $\gamma_O$, it is possible to reconstruct the $\overline{\mathrm{MS}}$ beta functions for the operator couplings, e.g., via 
\begin{align}
	\beta_{m_\psi} & = m_\psi (\gamma_O)_{11} + h_{2} (\gamma_O)_{21} + h_3 (\gamma_O)_{31},  \\  
	\beta_{h_2} & = m_\psi (\gamma_O)_{12} + h_{2} (\gamma_O)_{22} + h_3 (\gamma_O)_{32},  \\  
	\beta_{h_3} & = m_\psi (\gamma_O)_{13} + h_{2} (\gamma_O)_{23} + h_3 (\gamma_O)_{33}. 
\end{align}

In this paper we explicitly carry out the renormalization of dimension-three operators and compute $\gamma_O$ up to three loops. In addition, we also cross-check our results at lower loops by adding \eqref{eq:dLag} to the Litim-Sannino model \eqref{eq:Lag} implemented in public computer codes \texttt{ARGES} \cite{Litim:2020jvl} and \texttt{RGBeta}\footnote{Correct treatment of dimension-1 couplings is implemented in RGBeta since version 1.1.5.} \cite{Thomsen:2021ncy} and computing $\vec{\beta}_{\kappa}$. 

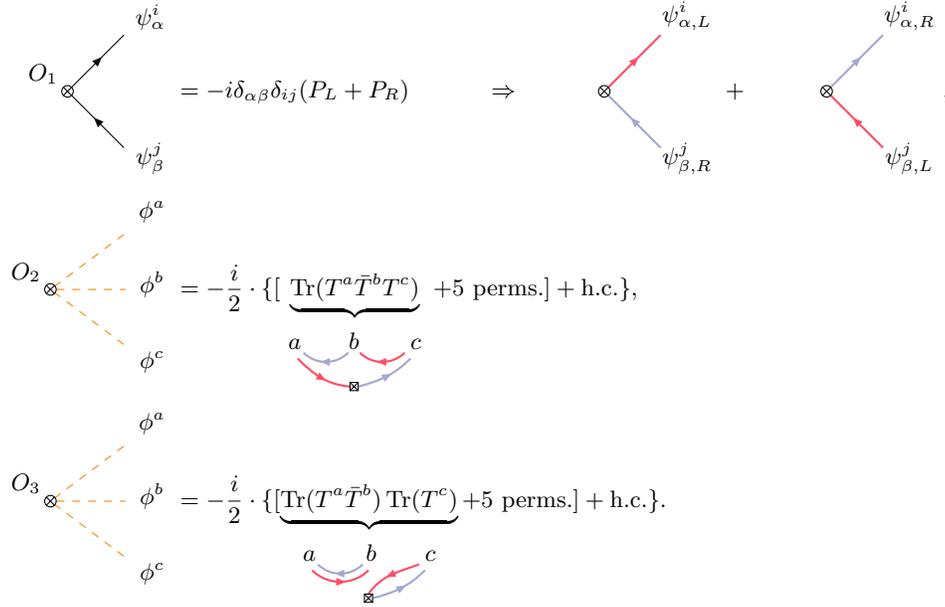
\begin{figure*}[t]
\begin{align*}
   \begin{tikzpicture}
	\begin{pgfonlayer}{nodelayer}
		\node [style=Op] (0) at (-6, 1) {};
                \node [style=none,label={[xshift=5pt,yshift=-1pt]$O_1$}] (Op) at (-7, 1) {};
		\node [style=none,label={[xshift=+10pt,yshift=-2pt]$\psi_{\alpha}^{i}$}] (1) at (-4.5, 2.5) {};
		\node [style=none,label={[xshift=+10pt,yshift=-14pt]$\psi_{\beta}^{j}$}] (2) at (-4.5, -0.5) {};
	\end{pgfonlayer}
	\begin{pgfonlayer}{edgelayer}
		\draw [style=fermion] (2.center) to (0);
		\draw [style=fermion] (0) to (1.center);
	\end{pgfonlayer}
\end{tikzpicture}
& = -i \delta_{\alpha\beta} \delta_{ij} (P_L + P_R) 
\hspace{1cm} \Rightarrow \hspace{1cm} 
   \begin{tikzpicture}
	\begin{pgfonlayer}{nodelayer}
		\node [style=Op] (0) at (-6, 1) {};
		\node [style=none,label={[xshift=10pt,yshift=-2pt]$\psi_{\alpha,L}^{i}$}] (1) at (-4.5, 2.5) {};
		\node [style=none,label={[xshift=10pt,yshift=-14pt]$\psi_{\beta,R}^{j}$}] (2) at (-4.5, -0.5) {};
	\end{pgfonlayer}
	\begin{pgfonlayer}{edgelayer}
		\draw [style=fermion,PR] (2.center) to (0);
		\draw [style=fermion,PL] (0) to (1.center);
	\end{pgfonlayer}
\end{tikzpicture}
+
   \begin{tikzpicture}
	\begin{pgfonlayer}{nodelayer}
		\node[style=none] (x) at (-8,1) {};
		\node [style=Op] (0) at (-6, 1) {};
		\node [style=none,label={[xshift=10pt,yshift=-2pt]$\psi_{\alpha,R}^{i}$}] (1) at (-4.5, 2.5) {};
		\node [style=none,label={[xshift=10pt,yshift=-14pt]$\psi_{\beta,L}^{j}$}] (2) at (-4.5, -0.5) {};
	\end{pgfonlayer}
	\begin{pgfonlayer}{edgelayer}
		\draw [style=fermion,PL] (2.center) to (0);
		\draw [style=fermion,PR] (0) to (1.center);
	\end{pgfonlayer}
\end{tikzpicture},
%\label{eq:O1_fr}
\\
   \begin{tikzpicture}
        \begin{pgfonlayer}{nodelayer}
                \node [style=Op] (0) at (-6, 1) {};
                \node [style=none,label={[xshift=5pt,yshift=-1pt]$O_2$}] (3) at (-7, 1) {};
                \node [style=none,label={[xshift=10pt,yshift=0pt]$\phi^a$}] (1) at (-4, 2.5) {};
                \node [style=none,label={[xshift=10pt,yshift=-7pt]$\phi^b$} ] (4) at (-4, 1) {};
                \node [style=none,label={[xshift=10pt,yshift=-14pt]$\phi^c$}] (2) at (-4, -0.5) {};
       \end{pgfonlayer}
        \begin{pgfonlayer}{edgelayer}
                %\draw [style=scalar] (0) to (3.center);
                \draw [style=scalar] (2) to (0.center); 
                \draw [style=scalar] (1) to (0.center);
                \draw [style=scalar] (0) to (4.center);
        \end{pgfonlayer}
        \end{tikzpicture}
& = - \frac{i}{2} \cdot \{ [ \underbrace{\Tr( T^a \bar T^b T^c)}_{\tikz{\pic[scale=0.3]{O2_ex}}}  + \text{5 perms.}] + \hc\},
%\label{eq:O2_fr}
	\\
   \begin{tikzpicture}
        \begin{pgfonlayer}{nodelayer}
                \node [style=Op] (0) at (-6, 1) {};
                \node [style=none,label={[xshift=5pt,yshift=-1pt]$O_3$}] (3) at (-7, 1) {};
                \node [style=none,label={[xshift=10pt,yshift=0pt]$\phi^a$}] (1) at (-4, 2.5) {};
                \node [style=none,label={[xshift=10pt,yshift=-7pt]$\phi^b$} ] (4) at (-4, 1) {};
                \node [style=none,label={[xshift=10pt,yshift=-14pt]$\phi^c$}] (2) at (-4, -0.5) {};
        \end{pgfonlayer}
        \begin{pgfonlayer}{edgelayer}
                %\draw [style=scalar] (0) to (3.center);
                \draw [style=scalar] (2) to (0.center); 
                \draw [style=scalar] (1) to (0.center);
                \draw [style=scalar] (0) to (4.center);
        \end{pgfonlayer}
        \end{tikzpicture}
& = - \frac{i}{2} \cdot \{ [ \underbrace{\Tr( T^a \bar T^b ) \Tr (T^c)}_{\tikz{\pic[scale=0.3]{O3_ex}}}  + \text{5 perms.}] + \hc\}.
%\label{eq:O3_fr}
\end{align*}
	\caption{Feynman rules for dimension-three operators.
All the operators break the flavor symmetry $G$ by ``flipping'' the ``chirality'' of the flavor flow, which in the case of scalar operators we indicate by a box with a cross inside.  
	}
	\label{fig:O3_feynman_rules}
\end{figure*}
To find three-loop corrections to $Z_O$, we consider insertions of the operators $O_{1-3}$ into one-particle irreducible (1PI) Green's functions. The corresponding Feynman rules are given in Fig.~\ref{fig:O3_feynman_rules}.
The renormalized 1PI Green functions are finite and are related to the bare ones via 
\begin{align}
	\langle [O^i]_R \cdot \tilde O^j \rangle_{\text{1PI}} & = (Z_O)_{ik} \cdot \langle O^k_0  \cdot \tilde O^j_0  \rangle_{\text{1PI}} \cdot Z_{j,ext}. %= (Z_O)_{ij} \cdot Z^j_{fields} + \ldots = \text{finite},
	\label{eq:op_insertions_ren}
\end{align}
Here $\tilde O_j$ is a product of external fields at different space-time points (either bare or renormalized) corresponding to the local operator $O_j$ with a property that $\langle O^i \cdot \tilde O^j \rangle_{tree} \propto \delta^{ij}$. The factors $Z_{j,ext}$ account for the external fields renormalization entering $\tilde O_j$, e.g., $Z_{1,ext} = Z_\psi$, $Z_{2,ext} = Z_{3,ext} = Z_H^{3/2}$ for $\psi_0 = \sqrt{Z_\psi} \cdot \psi$, and 
 $H_0 = \sqrt{Z_H} \cdot H$. We determine the matrix elements of $(Z_O)_{ij}$ in the $\overline{\mathrm{MS}}$ scheme order-by-order from the requirement that there are no $\varepsilon$ poles in the rhs  \eqref{eq:op_insertions_ren}.

\begin{figure*}[t]
	\begin{tabular}{c@{\hskip 2cm}c}
	\begin{tikzpicture}[scale=1.5]
        \begin{pgfonlayer}{nodelayer}
		\node [ellipse,minimum height=80pt,
			       minimum width=60pt,
			       fill=gray,
			       fill opacity=0.2] at (-5,1) {};
                \node [style=Op] (0) at (-6, 1) {};
                \node [style=ffh] (1) at (-4.5, 2.5) {};
                \node [style=ffh] (2) at (-4.5, -0.5) {};
                \node [style=none,label={[xshift=5pt,yshift=-1pt]$O_1$}] (3) at (-7, 1) {};
                \node [style=ffh] (4) at (-4.5, 1) {};

                \node [style=ffh] (5) at (-3, 2.5) {};
                \node [style=ffh] (6) at (-3, 1) {};
                \node [style=ffh] (7) at (-3, -0.5) {};

                \node [style=none] (9) at (-2, -1.5) {};
                \node [style=none] (10) at (-2, 3.5) {};
        \end{pgfonlayer}
        \begin{pgfonlayer}{edgelayer}
                %\draw [style=scalar] (0) to (3.center);
                \draw [style=fermion,PR] (2) to (0.center);
                \draw [style=fermion,PL] (0) to (1.center);
                \draw [style=fermion,PR] (1) to (4.center);
                \draw [style=fermion,PL] (4) to (2.center);

		\draw [style=scalar, HtoHbar'={0.5}{PL}{PR}] (1) to (5.center);
                \draw [style=scalar, HtoHbar'={0.5}{PR}{PL}] (4) to (6.center);
                \draw [style=scalar, HtoHbar'={0.5}{PL}{PR}] (2) to (7.center);

                \draw [style=fermion,PL] (7) to (6.center);
                \draw [style=fermion,PR] (6) to (5.center);

                \draw [style=fermion,PR] (9) to (7.center);
                \draw [style=fermion,PL] (5) to (10.center);
        \end{pgfonlayer}
        \end{tikzpicture}
	&
	\begin{tikzpicture}[scale=1.5]
        \begin{pgfonlayer}{nodelayer}
                \node [style=Op] (0) at (-6, 1) {};
                \node [style=ffh] (1) at (-4, 2.5) {};
                \node [style=ffh] (2) at (-4, -0.5) {};
                \node [style=none,label={[xshift=5pt,yshift=-1pt]$O_2$}] (3) at (-7, 1) {};
                \node [style=ffh] (4) at (-4, 1) {};

                \node [style=none] (9) at (-3, -1.5) {};
                \node [style=none] (10) at (-3, 3.5) {};
        \end{pgfonlayer}
        \begin{pgfonlayer}{edgelayer}
                %\draw [style=scalar] (0) to (3.center);
                \draw [style=scalar,HtoHbar'={0.4}{PR}{PL}] (2) to (0.center); 
                \draw [style=scalar,HtoHbar'={0.4}{PR}{PL}] (1) to (0.center);
                \draw [style=scalar,HtoHbar'={0.65}{PR}{PL}] (0) to (4.center); 
                \draw [style=fermion,PR] (4) to (1.center);
                \draw [style=fermion,PL] (2) to (4.center);

                \draw [style=fermion,PR] (9) to (2.center);
                \draw [style=fermion,PL] (1) to (10.center);
        \end{pgfonlayer}
        \end{tikzpicture}
	\\
		(a) & (b) 
	\end{tabular}
	\caption{A three-loop diagram (a) with a $O_1=\bar \psi \psi$ insertion that requires a two-loop counterterm due to $O_2$ (b). The grey blob gives rise to a one-loop contribution to $(Z_O)_{21}$. }
	\label{fig:O1_O2_mixing}
\end{figure*}
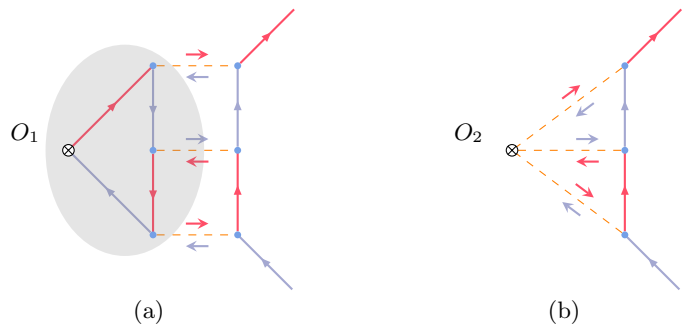

It is important to stress that if we ignore $O_{2,3}$ (or the $h_{2,3}$ couplings in the Lagrangian \eqref{eq:dLag}), we can compute $(Z_O)_{11}$ and $(\gamma_{O})_{11}$ (the latter is denoted by $\gamma_{m_{\psi}}$ in Ref.~\cite{Litim:2023tym}) by considering  $\langle O_1(x) \psi(y) \bar \psi(z) \rangle_{\text{1PI}}$ up to the two-loop order without hitting any difficulties. At three loops there are diagrams (see,e.g., Fig.~\ref{fig:O1_O2_mixing}a), which require an insertion of the $O_2$ operator as a counterterm (Fig.~\ref{fig:O1_O2_mixing}b). Because of this, we are forced to include $O_2$ in the game. The latter mixes with $O_1$ already at the one-loop order\footnote{ Speaking in other terms, the sole introduction of $m_\psi$ in \eqref{eq:dLag} will radiatively generate a $h_2$ coupling at one-loop level (corresponding to a grey blob in Fig.~\ref{fig:O1_O2_mixing}a).}. Moreover, starting from one loop, the $O_3$ operator is needed to account for all the divergences appearing in four-point functions $\langle O_2 \phi^a \phi^b\phi^c\rangle_{\text{1PI}}$.   

We also compute the divergences  of the two-point functions $\langle O_{i}(x) \phi^a(y) \rangle_{\text{1PI}}$ for $i=1,2,3$ and extract the mixing of $O_{i}$ with $O_4=1/2(\partial^2 \Tr(H) + \hc)$, thus, modifying \eqref{eq:O123_renorm_no_EOMs} as
\begin{align}
	[ \vec{O} ]_R & = \mathcal{Z}_O \cdot (\vec{O})_0 + \vec{\mathcal{Z}} (O_4)_0, \\ 
	[O_4]_R &   = \mathcal{Z}_H^{-1/2} (O_4)_0
\end{align}
This gives rise to a $4\times4$ mixing renormalization matrix $\mathcal{\bar Z}$ entering
\begin{align}
	[O_i]_R & = (\mathcal{\bar Z}_O)_{ij} (O_j)_0, \qquad 
	\mathcal{\bar Z}_O = \bar Z_O \cdot 
	\begin{pmatrix} 
		\mu^{2\epsilon} & & &\\
		  & \mu^{3\epsilon} & & \\
		  &   & \mu^{3\epsilon} &  \\
		  &   &   & \mu^{\epsilon}
	\end{pmatrix},
	\qquad
	\bar \gamma_O = - \dot{\mathcal{\bar Z}}_O \cdot \mathcal{\bar Z}_O^{-1}
	\label{eq:anom4x4}
\end{align}
where $i=1,\ldots,4$ and $\bar Z_O$ is schematically depicted in Fig.~\ref{fig:ZO4x4}. If we take into account the equations of motion (EOM) for the bare operators % in the form
 \begin{align}
	 (O_4)_0 + \vec{\Lambda}_0 \cdot (\vec{O})_0 = 0, 
	 \qquad \vec{\Lambda}_0 = \left\{ y_0/2,  2 u_0 , 2 v_0 \right\}
	  \equiv \left\{ \mu^\varepsilon Z_y  (y/2) , \mu^{2 \varepsilon} Z_u (2 u) , \mu^{2 \varepsilon} Z_v (2 v) \right\}
	 \label{eq:EOM}
 \end{align}
 where $Z_{y,u,v}$ are renormalization constants for the corresponding couplings, and express $O_4$ as a linear combination of $\vec{O}$, we obtain another $3 \times 3$ matrix 
\begin{align}
	\mathcal{Z}_O \to \mathcal{\tilde Z}_O \equiv \mathcal{Z}_O - \vec{\mathcal{Z}} \otimes \vec{\Lambda}_0,
	\label{eq:Z3x3_EOMs}
\end{align}
 schematically presented in Fig.~\ref{fig:ZO3x3}. The corresponding anomalous dimension is given by 
 \begin{align}
	 \tilde \gamma_O \equiv - \dot{\tilde{\mathcal{Z}}}_O
	 \cdot \tilde{\mathcal{Z}}^{-1}_O
	 \label{eq:gammaO_EOMs}
 \end{align}
 and is different from $\gamma_O$ \eqref{eq:gammaO}. At a fixed point, $\tilde \gamma^*_O$ ($\gamma^*_O$) has a eigenvalue $\gamma^*_H$ ($-\gamma^*_H$). The other two eigenvalues of $\gamma^*_O$ coincide with those of $\tilde \gamma^*_O$. 
This can be interpreted as the fact that only two of the considered dimension-three eigenoperators are independent, while the remaining one is a descendent of $\Tr(H) + \hc$
\begin{figure}[t]
\begin{align*}
\bar Z_O & = \begin{pmatrix}
	Z_O & \vec{Z} \\
	\vec{0} & Z^{-1/2}_H
\end{pmatrix}
\equiv \mathcal{K R}'
	\begin{pmatrix}
\begin{tikzpicture}
        \begin{pgfonlayer}{nodelayer}
		\node [onepi] (1pi) at (-4.2,1) {};
                \node [style=none,label={[xshift=5pt,yshift=-1pt]$O_1$}] (Op) at (-7, 1) {};
                \node [style=Op] (0) at (-6, 1) {};
                \node [style=none] (9) at (-3, -0.5) {};
                \node [style=none] (10) at (-3, 2.5) {};
        \end{pgfonlayer}
        \begin{pgfonlayer}{edgelayer}
                %\draw [style=scalar] (0) to (3.center);
		\draw [style={fermion,PR}] (1pi.center) to[out=-135,in=-45] (0.center);
		\draw [style={fermion,PL}] (0.center) to[out=45,in=135] (1pi.center);
                \draw [style=fermion,PR] (9) to (1pi.center);
                \draw [style=fermion,PL] (1pi) to (10.center);
        \end{pgfonlayer}
        \end{tikzpicture}
	&
	\begin{tikzpicture}
        \begin{pgfonlayer}{nodelayer}
		\node [onepi] (1pi) at (-4.2,1) {};
                \node [style=none,label={[xshift=5pt,yshift=-1pt]$O_1$}] (Op) at (-7, 1) {};
                \node [style=Op] (0) at (-6, 1) {};
                \node [style=none] (c) at (-3, -0.5) {};
                \node [style=none] (a) at (-3, 2.5) {};
                \node [style=none] (b) at (-3, 1) {};
        \end{pgfonlayer}
        \begin{pgfonlayer}{edgelayer}
                %\draw [style=scalar] (0) to (3.center);
		\draw [style={fermion,PR}] (1pi.center) to[out=-135,in=-45] (0.center);
		\draw [style={fermion,PL}] (0.center) to[out=45,in=135] (1pi.center);
                \draw [style=scalar] (c) to (1pi.center);
                \draw [style=scalar] (a) to (1pi.center);
                \draw [style=scalar] (b) to (1pi.center);
        \end{pgfonlayer}
        \end{tikzpicture}
	&
	\begin{tikzpicture}
        \begin{pgfonlayer}{nodelayer}
		\node [onepi] (1pi) at (-4.2,1) {};
                \node [style=Op] (0) at (-6, 1) {};
                \node [style=none,label={[xshift=5pt,yshift=-1pt]$O_1$}] (Op) at (-7, 1) {};
                \node [style=none] (9) at (-3, -0.5) {};
                \node [style=none] (10) at (-3, 2.5) {};
                \node [style=none] (11) at (-3, 1) {};
        \end{pgfonlayer}
        \begin{pgfonlayer}{edgelayer}
                %\draw [style=scalar] (0) to (3.center);
		\draw [style={fermion,PR}] (1pi.center) to[out=-135,in=-45] (0.center);
		\draw [style={fermion,PL}] (0.center) to[out=45,in=135] (1pi.center);
                \draw [style=scalar] (9) to (1pi.center);
                \draw [style=scalar] (10) to (1pi.center);
                \draw [style=scalar] (11) to (1pi.center);
        \end{pgfonlayer}
        \end{tikzpicture}
	&
\begin{tikzpicture}
        \begin{pgfonlayer}{nodelayer}
		\node [onepi] (1pi) at (-4.2,1) {};
                \node [style=none,label={[xshift=5pt,yshift=-1pt]$O_1$}] (Op) at (-7, 1) {};
                \node [style=Op] (0) at (-6, 1) {};
                \node [style=none] (11) at (-3, 1) {};
        \end{pgfonlayer}
        \begin{pgfonlayer}{edgelayer}
                %\draw [style=scalar] (0) to (3.center);
		\draw [style={fermion,PR}] (1pi.center) to[out=-135,in=-45] (0.center);
		\draw [style={fermion,PL}] (0.center) to[out=45,in=135] (1pi.center);
                \draw [style=scalar] (11) to (1pi.center);
        \end{pgfonlayer}
\end{tikzpicture}
\\[1cm]
\begin{tikzpicture}
        \begin{pgfonlayer}{nodelayer}
		\node [onepi] (1pi) at (-4.3,1) {};
                \node [style=none,label={[xshift=5pt,yshift=-1pt]$O_2$}] (Op) at (-7, 1) {};
                \node [style=Op] (0) at (-6, 1) {};
                \node [style=none] (9) at (-3, -0.5) {};
                \node [style=none] (10) at (-3, 2.5) {};
        \end{pgfonlayer}
        \begin{pgfonlayer}{edgelayer}
                %\draw [style=scalar] (0) to (3.center);
		\draw [style=scalar] (0.center) to[out=-45,in=-135] (1pi.center);
		\draw [style=scalar] (0.center) to[out=45,in=135] (1pi.center);
		\draw [style=scalar] (0.center) to (1pi.center);
                \draw [style=fermion,PR] (9) to (1pi.center);
                \draw [style=fermion,PL] (1pi) to (10.center);
        \end{pgfonlayer}
        \end{tikzpicture}
	&
	\begin{tikzpicture}
        \begin{pgfonlayer}{nodelayer}
		\node [onepi] (1pi) at (-4.2,1) {};
                \node [style=none,label={[xshift=5pt,yshift=-1pt]$O_2$}] (Op) at (-7, 1) {};
                \node [style=Op] (0) at (-6, 1) {};
                \node [style=none] (9) at (-3, -0.5) {};
                \node [style=none] (10) at (-3, 2.5) {};
                \node [style=none] (11) at (-3, 1) {};
        \end{pgfonlayer}
        \begin{pgfonlayer}{edgelayer}
                %\draw [style=scalar] (0) to (3.center);
		\draw [style=scalar] (0.center) to[out=-45,in=-135] (1pi.center);
		\draw [style=scalar] (0.center) to[out=45,in=135] (1pi.center);
		\draw [style=scalar] (0.center) to (1pi.center);
                \draw [style=scalar] (9) to (1pi.center);
                \draw [style=scalar] (10) to (1pi.center);
                \draw [style=scalar] (11) to (1pi.center);
        \end{pgfonlayer}
        \end{tikzpicture}
	&
	\begin{tikzpicture}
        \begin{pgfonlayer}{nodelayer}
		\node [onepi] (1pi) at (-4.2,1) {};
                \node [style=Op] (0) at (-6, 1) {};
                \node [style=none,label={[xshift=5pt,yshift=-1pt]$O_2$}] (Op) at (-7, 1) {};
                \node [style=none] (9) at (-3, -0.5) {};
                \node [style=none] (10) at (-3, 2.5) {};
                \node [style=none] (11) at (-3, 1) {};
        \end{pgfonlayer}
        \begin{pgfonlayer}{edgelayer}
                %\draw [style=scalar] (0) to (3.center);
		\draw [style=scalar] (0.center) to[out=-45,in=-135] (1pi.center);
		\draw [style=scalar] (0.center) to[out=45,in=135] (1pi.center);
		\draw [style=scalar] (0.center) to (1pi.center);
                \draw [style=scalar] (9) to (1pi.center);
                \draw [style=scalar] (10) to (1pi.center);
                \draw [style=scalar] (11) to (1pi.center);
        \end{pgfonlayer}
        \end{tikzpicture}
	&
\begin{tikzpicture}
        \begin{pgfonlayer}{nodelayer}
		\node [onepi] (1pi) at (-4.2,1) {};
                \node [style=none,label={[xshift=5pt,yshift=-1pt]$O_2$}] (Op) at (-7, 1) {};
                \node [style=Op] (0) at (-6, 1) {};
                \node [style=none] (11) at (-3, 1) {};
        \end{pgfonlayer}
        \begin{pgfonlayer}{edgelayer}
                %\draw [style=scalar] (0) to (3.center);
		\draw [style=scalar] (0.center) to[out=-45,in=-135] (1pi.center);
		\draw [style=scalar] (0.center) to[out=45,in=135] (1pi.center);
		\draw [style=scalar] (0.center) to (1pi.center);
                \draw [style=scalar] (11) to (1pi.center);
        \end{pgfonlayer}
\end{tikzpicture}
\\[1cm]
\begin{tikzpicture}
        \begin{pgfonlayer}{nodelayer}
		\node [onepi] (1pi) at (-4.3,1) {};
                \node [style=none,label={[xshift=5pt,yshift=-1pt]$O_3$}] (Op) at (-7, 1) {};
                \node [style=Op] (0) at (-6, 1) {};
                \node [style=none] (9) at (-3, -0.5) {};
                \node [style=none] (10) at (-3, 2.5) {};
        \end{pgfonlayer}
        \begin{pgfonlayer}{edgelayer}
                %\draw [style=scalar] (0) to (3.center);
		\draw [style=scalar] (0.center) to[out=-45,in=-135] (1pi.center);
		\draw [style=scalar] (0.center) to[out=45,in=135] (1pi.center);
		\draw [style=scalar] (0.center) to (1pi.center);
                \draw [style=fermion,PR] (9) to (1pi.center);
                \draw [style=fermion,PL] (1pi) to (10.center);
        \end{pgfonlayer}
        \end{tikzpicture}
	&
	\begin{tikzpicture}
        \begin{pgfonlayer}{nodelayer}
		\node [onepi] (1pi) at (-4.2,1) {};
                \node [style=none,label={[xshift=5pt,yshift=-1pt]$O_3$}] (Op) at (-7, 1) {};
                \node [style=Op] (0) at (-6, 1) {};
                \node [style=none] (9) at (-3, -0.5) {};
                \node [style=none] (10) at (-3, 2.5) {};
                \node [style=none] (11) at (-3, 1) {};
        \end{pgfonlayer}
        \begin{pgfonlayer}{edgelayer}
                %\draw [style=scalar] (0) to (3.center);
		\draw [style=scalar] (0.center) to[out=-45,in=-135] (1pi.center);
		\draw [style=scalar] (0.center) to[out=45,in=135] (1pi.center);
		\draw [style=scalar] (0.center) to (1pi.center);
                \draw [style=scalar] (9) to (1pi.center);
                \draw [style=scalar] (10) to (1pi.center);
                \draw [style=scalar] (11) to (1pi.center);
        \end{pgfonlayer}
        \end{tikzpicture}
	&
	\begin{tikzpicture}
        \begin{pgfonlayer}{nodelayer}
		\node [onepi] (1pi) at (-4.2,1) {};
                \node [style=Op] (0) at (-6, 1) {};
                \node [style=none,label={[xshift=5pt,yshift=-1pt]$O_3$}] (Op) at (-7, 1) {};
                \node [style=none] (9) at (-3, -0.5) {};
                \node [style=none] (10) at (-3, 2.5) {};
                \node [style=none] (11) at (-3, 1) {};
        \end{pgfonlayer}
        \begin{pgfonlayer}{edgelayer}
                %\draw [style=scalar] (0) to (3.center);
		\draw [style=scalar] (0.center) to[out=-45,in=-135] (1pi.center);
		\draw [style=scalar] (0.center) to[out=45,in=135] (1pi.center);
		\draw [style=scalar] (0.center) to (1pi.center);
                \draw [style=scalar] (9) to (1pi.center);
                \draw [style=scalar] (10) to (1pi.center);
                \draw [style=scalar] (11) to (1pi.center);
        \end{pgfonlayer}
        \end{tikzpicture}
	&
\begin{tikzpicture}
        \begin{pgfonlayer}{nodelayer}
		\node [onepi] (1pi) at (-4.2,1) {};
                \node [style=none,label={[xshift=5pt,yshift=-1pt]$O_3$}] (Op) at (-7, 1) {};
                \node [style=Op] (0) at (-6, 1) {};
                \node [style=none] (11) at (-3, 1) {};
        \end{pgfonlayer}
        \begin{pgfonlayer}{edgelayer}
                %\draw [style=scalar] (0) to (3.center);
		\draw [style=scalar] (0.center) to[out=-45,in=-135] (1pi.center);
		\draw [style=scalar] (0.center) to[out=45,in=135] (1pi.center);
		\draw [style=scalar] (0.center) to (1pi.center);
                \draw [style=scalar] (11) to (1pi.center);
        \end{pgfonlayer}
\end{tikzpicture}
\\[1cm]
		0 & 0 & 0 &  
\begin{tikzpicture}
        \begin{pgfonlayer}{nodelayer}
		\node [onepi] (1pi) at (-4.2,1) {};
		\node [style=none,label={[xshift=5pt,yshift=-1pt]$\phantom{O_4}$}] (Op) at (-7, 1) {};
                \node [style=none] (0) at (-5.5, 1) {};
                \node [style=none] (11) at (-3, 1) {};
        \end{pgfonlayer}
        \begin{pgfonlayer}{edgelayer}
                %\draw [style=scalar] (0) to (3.center);
		\draw [style=scalar] (0.center) to (1pi.center);
                \draw [style=scalar] (11) to (1pi.center);
        \end{pgfonlayer}
\end{tikzpicture}
\end{pmatrix}
\end{align*}
\caption{One-particle-irreducible Green functions used to extract matrix elements of $Z_O$. The factor $Z_H^{-1/2}$ enters the matrix $\bar Z_O$ due to $[O_4]_R = Z_H^{-1/2} (O_4)_0$.}
\label{fig:ZO4x4}
\end{figure}

Before discussing results, let us note how our approach differs from that of Ref.~\cite{Litim:2023tym}. The authors of Ref.~\cite{Litim:2023tym} utilized the same IRR technique and the seminaive treatment of $\gamma_5$ in their three-loop calculations. However, instead of \texttt{DIANA} and \texttt{MATAD}, they used a private framework to deal with Feynman diagrams generated by \texttt{QGRAF}. In our study we \emph{do not} rely on the dummy-field method (see, e.g., Ref.~\cite{Schienbein:2018fsw} and references therein) to derive the beta functions of dimensional couplings, but extract them from the Green's functions with insertions of the corresponding operators. 
Moreover, the authors of Ref.~\cite{Litim:2023tym} considered only the $O_1= \bar \psi \psi$ operator and did not account for possible mixing with $O_{2,3}$. Because of this, the anomalous dimension\footnote{Corresponding to  $(\gamma_O)_{11} = (\gamma_{\kappa})_{11}$ in our notation.} denoted by $\gamma_{m_\psi}$ is \emph{not} an eigenvalue of the corresponding matrix and, if computed at a fixed point, does \emph{not} represent a correction to canonical scaling. 

%While we carry out all the calculations explicitly in the model \eqref{eq:Lag} with operators defined in Eq.~\eqref{eq:dLag}, 
In what follows, we present all our results in terms of rescaled couplings and for the rescaled operators $O'$ \eqref{eq:dim3_ops_rescaled}.
All expressions are available in computer-readable form as Supplemental
Material.

\begin{figure}
\begin{align*}
	\begin{pmatrix}
		\left[
			\begin{tikzpicture}[scale=0.9]
        \begin{pgfonlayer}{nodelayer}
		\node [onepi] (1pi) at (-4.2,1) {};
                \node [style=none,label={[xshift=5pt,yshift=-1pt]$O_1$}] (Op) at (-7, 1) {};
                \node [style=Op] (0) at (-6, 1) {};
                \node [style=none] (9) at (-3, -0.5) {};
                \node [style=none] (10) at (-3, 2.5) {};
        \end{pgfonlayer}
        \begin{pgfonlayer}{edgelayer}
                %\draw [style=scalar] (0) to (3.center);
		\draw [style={fermion,PR}] (1pi.center) to[out=-135,in=-45] (0.center);
		\draw [style={fermion,PL}] (0.center) to[out=45,in=135] (1pi.center);
                \draw [style=fermion,PR] (9) to (1pi.center);
                \draw [style=fermion,PL] (1pi) to (10.center);
        \end{pgfonlayer}
        \end{tikzpicture}
	+
\begin{tikzpicture}[scale=0.9]
        \begin{pgfonlayer}{nodelayer}
		\node [onepi] (1pi) at (-4.2,1) {};
                \node [style=none,label={[xshift=5pt,yshift=-1pt]$O_1$}] (Op) at (-7, 1) {};
                \node [style=Op] (0) at (-6, 1) {};
                %\node [style=none] (11) at (-3, 1) {};
		% This is for O1
			\node [style=ffh,label={$y$}] (11) at (-3, 1) {};
			\node [style=none] (psiL) at (-2,2.5) {};
			\node [style=none] (psiR) at (-2,-0.5) {};
		\node [style=none] (bb) at (-1.5,1) {}; % just reserve space
        \end{pgfonlayer}
        \begin{pgfonlayer}{edgelayer}
                %\draw [style=scalar] (0) to (3.center);
		\draw [style={fermion,PR}] (1pi.center) to[out=-135,in=-45] (0.center);
		\draw [style={fermion,PL}] (0.center) to[out=45,in=135] (1pi.center);
                \draw [style=scalar] (11) to (1pi.center);
			\draw[style=fermion,PL] (11.center) to (psiL);
			\draw[style=fermion,PR] (psiR) to (11.center);
        \end{pgfonlayer}
\end{tikzpicture}
\right]
	&
\left[
	\begin{tikzpicture}[scale=0.9]
        \begin{pgfonlayer}{nodelayer}
		\node [onepi] (1pi) at (-4.2,1) {};
                \node [style=none,label={[xshift=5pt,yshift=-1pt]$O_1$}] (Op) at (-7, 1) {};
                \node [style=Op] (0) at (-6, 1) {};
                \node [style=none] (c) at (-3, -0.5) {};
                \node [style=none] (a) at (-3, 2.5) {};
                \node [style=none] (b) at (-3, 1) {};
        \end{pgfonlayer}
        \begin{pgfonlayer}{edgelayer}
                %\draw [style=scalar] (0) to (3.center);
		\draw [style={fermion,PR}] (1pi.center) to[out=-135,in=-45] (0.center);
		\draw [style={fermion,PL}] (0.center) to[out=45,in=135] (1pi.center);
                \draw [style=scalar] (c) to (1pi.center);
                \draw [style=scalar] (a) to (1pi.center);
                \draw [style=scalar] (b) to (1pi.center);
        \end{pgfonlayer}
        \end{tikzpicture}
	+
\begin{tikzpicture}[scale=0.9]
        \begin{pgfonlayer}{nodelayer}
		\node [onepi] (1pi) at (-4.2,1) {};
                \node [style=none,label={[xshift=5pt,yshift=-1pt]$O_1$}] (Op) at (-7, 1) {};
                \node [style=Op] (0) at (-6, 1) {};
			\node [style=h4,label={[xshift=-2pt]$v$}] (11) at (-3, 1) {};
			\node [style=none] (a) at (-2,2.5) {};
			\node [style=none] (b) at (-2,1) {};
			\node [style=none] (c) at (-2,-0.5) {};
		\node [style=none] (bb) at (-1.5,1) {}; % just reserve space
        \end{pgfonlayer}
        \begin{pgfonlayer}{edgelayer}
                %\draw [style=scalar] (0) to (3.center);
		\draw [style={fermion,PR}] (1pi.center) to[out=-135,in=-45] (0.center);
		\draw [style={fermion,PL}] (0.center) to[out=45,in=135] (1pi.center);
                \draw [style=scalar] (11) to (1pi.center);
  	       		\draw [style=scalar] (11) to (a);
                	\draw [style=scalar] (11) to (b);
                	\draw [style=scalar] (11) to (c);
        \end{pgfonlayer}
\end{tikzpicture}
\right]
	&
	\left[
	\begin{tikzpicture}[scale=0.9]
        \begin{pgfonlayer}{nodelayer}
		\node [onepi] (1pi) at (-4.2,1) {};
                \node [style=Op] (0) at (-6, 1) {};
                \node [style=none,label={[xshift=5pt,yshift=-1pt]$O_1$}] (Op) at (-7, 1) {};
                \node [style=none] (9) at (-3, -0.5) {};
                \node [style=none] (10) at (-3, 2.5) {};
                \node [style=none] (11) at (-3, 1) {};
        \end{pgfonlayer}
        \begin{pgfonlayer}{edgelayer}
                %\draw [style=scalar] (0) to (3.center);
		\draw [style={fermion,PR}] (1pi.center) to[out=-135,in=-45] (0.center);
		\draw [style={fermion,PL}] (0.center) to[out=45,in=135] (1pi.center);
                \draw [style=scalar] (9) to (1pi.center);
                \draw [style=scalar] (10) to (1pi.center);
                \draw [style=scalar] (11) to (1pi.center);
        \end{pgfonlayer}
        \end{tikzpicture}
	+
\begin{tikzpicture}[scale=0.9]
        \begin{pgfonlayer}{nodelayer}
		\node [onepi] (1pi) at (-4.2,1) {};
                \node [style=none,label={[xshift=5pt,yshift=-1pt]$O_1$}] (Op) at (-7, 1) {};
                \node [style=Op] (0) at (-6, 1) {};
			\node [style=h4,label={[xshift=-2pt]$u$}] (11) at (-3, 1) {};
			\node [style=none] (a) at (-2,2.5) {};
			\node [style=none] (b) at (-2,1) {};
			\node [style=none] (c) at (-2,-0.5) {};
		\node [style=none] (bb) at (-1.5,1) {}; 
        \end{pgfonlayer}
        \begin{pgfonlayer}{edgelayer}
                %\draw [style=scalar] (0) to (3.center);
		\draw [style={fermion,PR}] (1pi.center) to[out=-135,in=-45] (0.center);
		\draw [style={fermion,PL}] (0.center) to[out=45,in=135] (1pi.center);
                \draw [style=scalar] (11) to (1pi.center);
  	       		\draw [style=scalar] (11) to (a);
                	\draw [style=scalar] (11) to (b);
                	\draw [style=scalar] (11) to (c);
        \end{pgfonlayer}
\end{tikzpicture}
\right]
\\[1cm]
\left[
\begin{tikzpicture}[scale=0.9]
        \begin{pgfonlayer}{nodelayer}
		\node [onepi] (1pi) at (-4.3,1) {};
                \node [style=none,label={[xshift=5pt,yshift=-1pt]$O_2$}] (Op) at (-7, 1) {};
                \node [style=Op] (0) at (-6, 1) {};
                \node [style=none] (9) at (-3, -0.5) {};
                \node [style=none] (10) at (-3, 2.5) {};
        \end{pgfonlayer}
        \begin{pgfonlayer}{edgelayer}
                %\draw [style=scalar] (0) to (3.center);
		\draw [style=scalar] (0.center) to[out=-45,in=-135] (1pi.center);
		\draw [style=scalar] (0.center) to[out=45,in=135] (1pi.center);
		\draw [style=scalar] (0.center) to (1pi.center);
                \draw [style=fermion,PR] (9) to (1pi.center);
                \draw [style=fermion,PL] (1pi) to (10.center);
        \end{pgfonlayer}
        \end{tikzpicture}
	+
\begin{tikzpicture}[scale=0.9]
        \begin{pgfonlayer}{nodelayer}
		\node [onepi] (1pi) at (-4.2,1) {};
                \node [style=none,label={[xshift=5pt,yshift=-1pt]$O_2$}] (Op) at (-7, 1) {};
                \node [style=Op] (0) at (-6, 1) {};
                \node [style=none] (11) at (-3, 1) {};
			\node [style=ffh,label={$y$}] (11) at (-3, 1) {};
			\node [style=none] (psiL) at (-2,2.5) {};
			\node [style=none] (psiR) at (-2,-0.5) {};
		\node [style=none] (bb) at (-1.5,1) {}; % just reserve space
        \end{pgfonlayer}
        \begin{pgfonlayer}{edgelayer}
                %\draw [style=scalar] (0) to (3.center);
		\draw [style=scalar] (0.center) to[out=-45,in=-135] (1pi.center);
		\draw [style=scalar] (0.center) to[out=45,in=135] (1pi.center);
		\draw [style=scalar] (0.center) to (1pi.center);
                \draw [style=scalar] (11) to (1pi.center);
		% This is for O1
			\draw[style=fermion,PL] (11.center) to (psiL);
			\draw[style=fermion,PR] (psiR) to (11.center);
        \end{pgfonlayer}
\end{tikzpicture}
\right]
	&
	\left[
	\begin{tikzpicture}[scale=0.9]
        \begin{pgfonlayer}{nodelayer}
		\node [onepi] (1pi) at (-4.2,1) {};
                \node [style=none,label={[xshift=5pt,yshift=-1pt]$O_2$}] (Op) at (-7, 1) {};
                \node [style=Op] (0) at (-6, 1) {};
                \node [style=none] (9) at (-3, -0.5) {};
                \node [style=none] (10) at (-3, 2.5) {};
                \node [style=none] (11) at (-3, 1) {};
        \end{pgfonlayer}
        \begin{pgfonlayer}{edgelayer}
                %\draw [style=scalar] (0) to (3.center);
		\draw [style=scalar] (0.center) to[out=-45,in=-135] (1pi.center);
		\draw [style=scalar] (0.center) to[out=45,in=135] (1pi.center);
		\draw [style=scalar] (0.center) to (1pi.center);
                \draw [style=scalar] (9) to (1pi.center);
                \draw [style=scalar] (10) to (1pi.center);
                \draw [style=scalar] (11) to (1pi.center);
        \end{pgfonlayer}
        \end{tikzpicture}
	+
\begin{tikzpicture}[scale=0.9]
        \begin{pgfonlayer}{nodelayer}
		\node [onepi] (1pi) at (-4.2,1) {};
                \node [style=none,label={[xshift=5pt,yshift=-1pt]$O_2$}] (Op) at (-7, 1) {};
                \node [style=Op] (0) at (-6, 1) {};
                \node [style=none] (11) at (-3, 1) {};
			\node [style=h4,label={[xshift=-2pt]$v$}] (11) at (-3, 1) {};
			\node [style=none] (a) at (-2,2.5) {};
			\node [style=none] (b) at (-2,1) {};
			\node [style=none] (c) at (-2,-0.5) {};
		\node [style=none] (bb) at (-1.5,1) {}; % just reserve space
        \end{pgfonlayer}
        \begin{pgfonlayer}{edgelayer}
                %\draw [style=scalar] (0) to (3.center);
		\draw [style=scalar] (0.center) to[out=-45,in=-135] (1pi.center);
		\draw [style=scalar] (0.center) to[out=45,in=135] (1pi.center);
		\draw [style=scalar] (0.center) to (1pi.center);
                \draw [style=scalar] (11) to (1pi.center);
                	\draw [style=scalar] (11) to (a);
                	\draw [style=scalar] (11) to (b);
                	\draw [style=scalar] (11) to (c);
        \end{pgfonlayer}
\end{tikzpicture}
\right]
	&
	\left[
	\begin{tikzpicture}[scale=0.9]
        \begin{pgfonlayer}{nodelayer}
		\node [onepi] (1pi) at (-4.2,1) {};
                \node [style=Op] (0) at (-6, 1) {};
                \node [style=none,label={[xshift=5pt,yshift=-1pt]$O_2$}] (Op) at (-7, 1) {};
                \node [style=none] (9) at (-3, -0.5) {};
                \node [style=none] (10) at (-3, 2.5) {};
                \node [style=none] (11) at (-3, 1) {};
        \end{pgfonlayer}
        \begin{pgfonlayer}{edgelayer}
                %\draw [style=scalar] (0) to (3.center);
		\draw [style=scalar] (0.center) to[out=-45,in=-135] (1pi.center);
		\draw [style=scalar] (0.center) to[out=45,in=135] (1pi.center);
		\draw [style=scalar] (0.center) to (1pi.center);
                \draw [style=scalar] (9) to (1pi.center);
                \draw [style=scalar] (10) to (1pi.center);
                \draw [style=scalar] (11) to (1pi.center);
        \end{pgfonlayer}
        \end{tikzpicture}
	+
\begin{tikzpicture}[scale=0.9]
        \begin{pgfonlayer}{nodelayer}
		\node [onepi] (1pi) at (-4.2,1) {};
                \node [style=none,label={[xshift=5pt,yshift=-1pt]$O_2$}] (Op) at (-7, 1) {};
                \node [style=Op] (0) at (-6, 1) {};
                \node [style=none] (11) at (-3, 1) {};
			\node [style=h4,label={[xshift=-2pt]$u$}] (11) at (-3, 1) {};
			\node [style=none] (a) at (-2,2.5) {};
			\node [style=none] (b) at (-2,1) {};
			\node [style=none] (c) at (-2,-0.5) {};
		\node [style=none] (bb) at (-1.5,1) {}; % just reserve space
        \end{pgfonlayer}
        \begin{pgfonlayer}{edgelayer}
                %\draw [style=scalar] (0) to (3.center);
		\draw [style=scalar] (0.center) to[out=-45,in=-135] (1pi.center);
		\draw [style=scalar] (0.center) to[out=45,in=135] (1pi.center);
		\draw [style=scalar] (0.center) to (1pi.center);
                \draw [style=scalar] (11) to (1pi.center);
                	\draw [style=scalar] (11) to (a);
                	\draw [style=scalar] (11) to (b);
                	\draw [style=scalar] (11) to (c);
        \end{pgfonlayer}
\end{tikzpicture}
\right]
\\[1cm]
\left[
\begin{tikzpicture}[scale=0.9]
        \begin{pgfonlayer}{nodelayer}
		\node [onepi] (1pi) at (-4.3,1) {};
                \node [style=none,label={[xshift=5pt,yshift=-1pt]$O_3$}] (Op) at (-7, 1) {};
                \node [style=Op] (0) at (-6, 1) {};
                \node [style=none] (9) at (-3, -0.5) {};
                \node [style=none] (10) at (-3, 2.5) {};
        \end{pgfonlayer}
        \begin{pgfonlayer}{edgelayer}
                %\draw [style=scalar] (0) to (3.center);
		\draw [style=scalar] (0.center) to[out=-45,in=-135] (1pi.center);
		\draw [style=scalar] (0.center) to[out=45,in=135] (1pi.center);
		\draw [style=scalar] (0.center) to (1pi.center);
                \draw [style=fermion,PR] (9) to (1pi.center);
                \draw [style=fermion,PL] (1pi) to (10.center);
        \end{pgfonlayer}
        \end{tikzpicture}
	+
\begin{tikzpicture}[scale=0.9]
        \begin{pgfonlayer}{nodelayer}
		\node [onepi] (1pi) at (-4.2,1) {};
                \node [style=none,label={[xshift=5pt,yshift=-1pt]$O_3$}] (Op) at (-7, 1) {};
                \node [style=Op] (0) at (-6, 1) {};
                \node [style=none] (11) at (-3, 1) {};
			\node [style=ffh,label={$y$}] (11) at (-3, 1) {};
			\node [style=none] (psiL) at (-2,2.5) {};
			\node [style=none] (psiR) at (-2,-0.5) {};
		\node [style=none] (bb) at (-1.5,1) {}; % just reserve space
        \end{pgfonlayer}
        \begin{pgfonlayer}{edgelayer}
                %\draw [style=scalar] (0) to (3.center);
		\draw [style=scalar] (0.center) to[out=-45,in=-135] (1pi.center);
		\draw [style=scalar] (0.center) to[out=45,in=135] (1pi.center);
		\draw [style=scalar] (0.center) to (1pi.center);
                \draw [style=scalar] (11) to (1pi.center);
			\draw[style=fermion,PL] (11.center) to (psiL);
			\draw[style=fermion,PR] (psiR) to (11.center);
        \end{pgfonlayer}
\end{tikzpicture}
\right]
	&
	\left[
	\begin{tikzpicture}[scale=0.9]
        \begin{pgfonlayer}{nodelayer}
		\node [onepi] (1pi) at (-4.2,1) {};
                \node [style=none,label={[xshift=5pt,yshift=-1pt]$O_3$}] (Op) at (-7, 1) {};
                \node [style=Op] (0) at (-6, 1) {};
                \node [style=none] (9) at (-3, -0.5) {};
                \node [style=none] (10) at (-3, 2.5) {};
                \node [style=none] (11) at (-3, 1) {};
        \end{pgfonlayer}
        \begin{pgfonlayer}{edgelayer}
                %\draw [style=scalar] (0) to (3.center);
		\draw [style=scalar] (0.center) to[out=-45,in=-135] (1pi.center);
		\draw [style=scalar] (0.center) to[out=45,in=135] (1pi.center);
		\draw [style=scalar] (0.center) to (1pi.center);
                \draw [style=scalar] (9) to (1pi.center);
                \draw [style=scalar] (10) to (1pi.center);
                \draw [style=scalar] (11) to (1pi.center);
        \end{pgfonlayer}
        \end{tikzpicture}
	+
\begin{tikzpicture}[scale=0.9]
        \begin{pgfonlayer}{nodelayer}
		\node [onepi] (1pi) at (-4.2,1) {};
                \node [style=none,label={[xshift=5pt,yshift=-1pt]$O_3$}] (Op) at (-7, 1) {};
                \node [style=Op] (0) at (-6, 1) {};
                \node [style=none] (11) at (-3, 1) {};
			\node [style=h4,label={[xshift=-2pt]$v$}] (11) at (-3, 1) {};
			\node [style=none] (a) at (-2,2.5) {};
			\node [style=none] (b) at (-2,1) {};
			\node [style=none] (c) at (-2,-0.5) {};
		\node [style=none] (bb) at (-1.5,1) {}; % just reserve space
        \end{pgfonlayer}
        \begin{pgfonlayer}{edgelayer}
                %\draw [style=scalar] (0) to (3.center);
		\draw [style=scalar] (0.center) to[out=-45,in=-135] (1pi.center);
		\draw [style=scalar] (0.center) to[out=45,in=135] (1pi.center);
		\draw [style=scalar] (0.center) to (1pi.center);
                \draw [style=scalar] (11) to (1pi.center);
                	\draw [style=scalar] (11) to (a);
                	\draw [style=scalar] (11) to (b);
                	\draw [style=scalar] (11) to (c);
        \end{pgfonlayer}
\end{tikzpicture}
\right]
	&
	\left[
	\begin{tikzpicture}[scale=0.9]
        \begin{pgfonlayer}{nodelayer}
		\node [onepi] (1pi) at (-4.2,1) {};
                \node [style=Op] (0) at (-6, 1) {};
                \node [style=none,label={[xshift=5pt,yshift=-1pt]$O_3$}] (Op) at (-7, 1) {};
                \node [style=none] (9) at (-3, -0.5) {};
                \node [style=none] (10) at (-3, 2.5) {};
                \node [style=none] (11) at (-3, 1) {};
        \end{pgfonlayer}
        \begin{pgfonlayer}{edgelayer}
                %\draw [style=scalar] (0) to (3.center);
		\draw [style=scalar] (0.center) to[out=-45,in=-135] (1pi.center);
		\draw [style=scalar] (0.center) to[out=45,in=135] (1pi.center);
		\draw [style=scalar] (0.center) to (1pi.center);
                \draw [style=scalar] (9) to (1pi.center);
                \draw [style=scalar] (10) to (1pi.center);
                \draw [style=scalar] (11) to (1pi.center);
        \end{pgfonlayer}
        \end{tikzpicture}
	+
\begin{tikzpicture}[scale=0.9]
        \begin{pgfonlayer}{nodelayer}
		\node [onepi] (1pi) at (-4.2,1) {};
                \node [style=none,label={[xshift=5pt,yshift=-1pt]$O_3$}] (Op) at (-7, 1) {};
                \node [style=Op] (0) at (-6, 1) {};
                \node [style=none] (11) at (-3, 1) {};
			\node [style=h4,label={[xshift=-2pt]$u$}] (11) at (-3, 1) {};
			\node [style=none] (a) at (-2,2.5) {};
			\node [style=none] (b) at (-2,1) {};
			\node [style=none] (c) at (-2,-0.5) {};
		\node [style=none] (bb) at (-1.5,1) {}; % just reserve space
        \end{pgfonlayer}
        \begin{pgfonlayer}{edgelayer}
                %\draw [style=scalar] (0) to (3.center);
		\draw [style=scalar] (0.center) to[out=-45,in=-135] (1pi.center);
		\draw [style=scalar] (0.center) to[out=45,in=135] (1pi.center);
		\draw [style=scalar] (0.center) to (1pi.center);
                \draw [style=scalar] (11) to (1pi.center);
                	\draw [style=scalar] (11) to (a);
                	\draw [style=scalar] (11) to (b);
                	\draw [style=scalar] (11) to (c);
        \end{pgfonlayer}
\end{tikzpicture}
\right]
\end{pmatrix}
\end{align*}
\caption{Green's functions used to extract anomalous dimensions $\tilde \gamma_O$ of the dimensions-3 operators with the account of equations of motion, expressing $O_4$ as a linear combination of $O_{1-3}$.}
\label{fig:ZO3x3}
\end{figure}

\section{Results}\label{sec:Res}
In this section, we summarize our results for $\beta$-functions for (in)finite $N_c$ and anomalous dimensions. Moreover, we determine fixed points and universal scaling dimensions up to the third nontrivial order in the Veneziano parameter also for both cases. 

\begin{table}
\begin{center}
	\begin{align*}
\begin{array}{|c|rrr|r|}
	\hline
  & \hfill \epsilon ^3 \hfill& \hfill \epsilon ^4 \hfill & \hfill \epsilon ^5 \hfill  & \hfill \epsilon ^6 \hfill\\
\hline
	\beta _g^{{(1)}} & 0.277419 & 0.94969 & 8.85345 & \textcolor{myred}{13.7629} \\
	\beta _g^{{(2)}} & -0.277419 & -3.42714 & -24.4312 & \textcolor{myred}{-120.131} \\
	\beta _g^{{(3)}} & 0 & 2.47745 & 19.7319 & \textcolor{myred}{209.597} \\
	\beta _g^{{(4)}} & 0 & 0 & \textcolor{myblue}{-4.15417} & \textcolor{myred}{-37.2974} \\
\hline
 \beta _g & 0 & 0 & 0 & \textcolor{myred}{65.9311} \\
	\hline
\end{array}
\quad 
\begin{array}{|c|rr|r|}
\hline
  & \hfill \epsilon ^3 \hfill& \hfill \epsilon ^4 \hfill & \hfill \epsilon ^5 \hfill  \\
\hline
	\beta _y^{{(1)}} & 0.49336 & 1.8551 & \textcolor{myred}{10.4922} \\
	\beta _y^{{(2)}} & -0.49336 & -3.11773 & \textcolor{myred}{-28.4972} \\
	\beta _y^{{(3)}} & 0 & \textcolor{myblue}{1.26263} & \textcolor{myred}{6.86205} \\
	\beta _y^{{(4)}} & 0 & 0 & \hfill \textcolor{myred}{?} \hfill \\
\hline
 \beta _y & 0 & 0 &  \textcolor{myred}{-11.143} \\
 \hline
\end{array}
\end{align*}
\begin{align*}
\begin{array}{|c|rr|r|}
\hline
  & \hfill \epsilon ^3 \hfill& \hfill \epsilon ^4 \hfill & \hfill \epsilon ^5 \hfill  \\
\hline
 \beta _u^{{(1)}} & -0.258097 & -2.26905 & \textcolor{myred}{-9.06296}  \\
 \beta _u^{{(2)}} & 0.258097 & 2.52154 & \textcolor{myred}{22.1462}  \\
 \beta _u^{{(3)}} & 0 & \textcolor{myblue}{-0.252485} & \textcolor{myred}{-3.72929}  \\
 \beta _u^{{(4)}} & 0 & 0 & \hfill \textcolor{myred}{?} \hfill \\
\hline
 \beta _u & 0 & 0 & \textcolor{myred}{9.3539}  \\
 \hline
\end{array}
\quad
\begin{array}{|c|rr|r|}
\hline
  & \hfill \epsilon ^3 \hfill& \hfill \epsilon ^4 \hfill & \hfill \epsilon ^5 \hfill  \\
\hline
 \beta _v^{{(1)}} & -0.992548 & -9.26177 & \textcolor{myred}{-24.2444}  \\
 \beta _v^{{(2)}} & 0.992548 & 9.04913 & \textcolor{myred}{82.5704}  \\
 \beta _v^{{(3)}} & 0 & \textcolor{myblue}{0.212636} & \textcolor{myred}{-10.4633}  \\
 \beta _v^{{(4)}} & 0 & 0 & \hfill \textcolor{myred}{?} \hfill \\
\hline
 \beta _v & 0 & 0 & \textcolor{myred}{47.8627}  \\
 \hline
\end{array}
\end{align*}
\end{center}
\caption{The model beta-functions at the fixed point as series in $\epsilon$. Cancellations between contributions from different loop orders are presented up to $\mathcal{O}(\epsilon^5)$ for $\beta_g$ and up to $\mathcal{O}(\epsilon^4)$ for $\beta_{y,u,v}$. Higher-order terms marked by the red color are not reliably  
	calculated and are not summed up to zero, but (the sum) can be used to estimate the size of typical high-order contribution. 
}
\label{tab:shifts}
\end{table}

\subsubsection{Fixed points}
As it is well known in perturbation theory the $\beta$-functions can be given as
\begin{align}
    \beta_x \equiv \frac{d \alpha_x}{d \ln \mu} = \beta_x^{(1)} +  \beta_x^{(2)} +  \beta_x^{(3)} + ...,
    \label{eq:beta_expansion}
\end{align}
where $\beta_x^{(n)}$ denotes the $n$-th loop contribution, and $x = \{g,y,u,v\}$.  In Ref.~\cite{Litim:2023tym} they were found for the first time and provided both in the Veneziano limit and with finite-$N_c$ corrections. We recomputed them independently and provide the lengthy expressions in Appendix \ref{sec:beta_func} and as Supplemental
Material.

Using the 433-order $\beta$ functions and solving $\beta_i(\alpha_j^*)=0$ systematically, we determine interacting fixed points up to complete third order in the small parameter $\epsilon$. We expressed fixed points as a series expansion
\begin{align}
    \alpha_x^* & = c_{x}^{(1)}f_{x}^{(1)}(N_c)\epsilon + c_{x}^{(2)}f_{x}^{(2)}(N_c)\epsilon^2 + c_{x}^{(3)}f_{x}^{(3)}(N_c)\epsilon^3 + O(\epsilon^4),
    \label{eq:fp_exp}
\end{align}
where $x=\{g,y,u,v\}$. 
We have 6 possible solutions with fixed points, however, following Refs.~\cite{Litim:2014uca,Bond:2017lnq,Bond:2021tgu}, we choose  a fully interacting UV fixed point ($\alpha_{g,y,u,v}^*\neq 0$) that exhibits asymptotically safe behavior. 

The coefficients $c_{x}^{(i)}$ of series expansion can be found by solving the beta functions in this limit order by order for the stationary point. We recomputed the exact expressions for $c_x^{(i)}$ and found agreement with Ref.~\cite{Litim:2023tym}
\begin{align}
    c_{g}^{(1)}&=\frac{26  }{57} , \qquad c_{g}^{(2)}=-\frac{23 \left(13068 \sqrt{23}-75245\right) }{370386},\\
    c_{g}^{(3)}&=\frac{\left(5025189312 \zeta_3+353747709269-71703657432 \sqrt{23}\right) }{2406768228},\\
    c_{y}^{(1)}&=\frac{4 }{19}, \qquad c_{y}^{(2)}=\frac{\left(43549-6900 \sqrt{23}\right) }{20577},\\
    c_{y}^{(3)}&=\frac{\left(29734848 \zeta_3+2893213181-580847448 \sqrt{23}\right) }{44569782},\\
c_{u}^{(1)}&=\frac{\sqrt{23}-1}{19}, \qquad c_{u}^{(2)}=  \frac{365825 \sqrt{23}-1476577}{631028},\\
c_{u}^{(3)}&=\left(\frac{5173524931447 \sqrt{23}-24197965967251}{282928976136}-\frac{416 \left(\sqrt{23}-12\right) \zeta_3}{6859}\right),\\
c_{v}^{(1)}&=\frac{\sqrt{2 \left(10+3 \sqrt{23}\right)}-2 \sqrt{23}} {19}, \\ c_{v}^{(2)}&=\frac{268229312-68836310 \sqrt{23}+46652027 \sqrt{2 \left(10+3 \sqrt{23}\right)}-9153184 \sqrt{46 \left(10+3 \sqrt{23}\right)} 
   }{67519996},\\
   c_{v}^{(3)}&= +\frac{1}{3239253847781064}\left[498710025259776 \zeta_3-105786975055104 \sqrt{23} \zeta_3\right. {}\nonumber\\ 
			& \hspace{0.5cm} \left.+6742118880147456 \sqrt{2 \left(10+3 \sqrt{23}\right)}
   \zeta_3-1374312631206528 \sqrt{46 \left(10+3 \sqrt{23}\right)} \zeta_3\right.{}\nonumber\\ 
			& \hspace{0.5cm}\left.+479791813615522776-103064713697904086 \sqrt{23}+74641138195038841 \sqrt{2 \left(10+3
   \sqrt{23}\right)}\right.{}\nonumber\\ 
			& \hspace{0.5cm}\left.-15585870376520334 \sqrt{46 \left(10+3 \sqrt{23}\right)}\right]
\end{align}

For convenience, we provide the numerical results in the Veneziano limit \cite{Litim:2023tym}
\begin{align}
    \alpha_g^* &= 0.456 \epsilon + 0.781 \epsilon^2 + 6.610 \epsilon^3 \textcolor{myred}{+ 24.137}  \epsilon^4,
    \label{eq:ag_fp}\\
    \alpha_y^* &=0.211 \epsilon + 0.508 \epsilon^2 + 3.322 \epsilon^3 \textcolor{myred}{+ 15.212}  \epsilon^4,\label{eq:ay_fp}\\
    \alpha_u^* &=  0.200 \epsilon + 0.440 \epsilon^2 + 2.693 \epsilon^3 \textcolor{myred}{+ 12.119}  \epsilon^4,\label{eq:au_fp}\\
    -\alpha_v^* & = 0.137 \epsilon + 0.632 \epsilon^2 + 4.313 \epsilon^3 \textcolor{myred}{+ 24.147 } \epsilon^4,
    \label{eq:av_fp}
\end{align}
where we also include subleading terms $\mathcal{O}(\epsilon^4)$ terms that will be modified when the $544$-result will be available \cite{Litim:2023tym}.
In addition,  Table \ref{tab:shifts} shows various $\beta^{(l)}_x$ evaluated at the fixed point $(\alpha_g^*, \alpha_y^*, \alpha_u^*, \alpha_v^*)$ given above. One sees the typical size of the coefficients that cancel at each order of $\epsilon$-expansion up to $\mathcal{O}(\epsilon^5)$ in $\beta_g$, and up to $\mathcal{O}(\epsilon^4)$ in other beta functions. Subleading terms that do not add up to zero due to missing high-order corrections are also indicated.

The functions $f_{x}^{(i)}$ first introduced in Ref.~\cite{Bond:2021tgu} capture the dependence on $N_c$ and can be computed in the same manner as the $c_{x}^{(i)}$. It should be noted that $\lim_{N_c\to \infty} f_{x}^{(i)}\equiv 1$, which  means that the FPs found for in(finite) cases are connected continuously. Therefore, for the infinite-$N_c$ case we have only $c_{x}^{(i)}$, which are exact numbers. However, the full form of $f_{x}^{(i)}$ is complicated\footnote{Available by demand from the authors.}. Because of this, following the ideas of \cite{Litim:2020jvl}, we fitted all the finite-$N_c$ corrections in the range $N_c \in [3,100]$ as ratios of two second-order polynomial up to $\epsilon^3$. The results can be found in Appendix \ref{sec:other_fits}. 
Then, we carried out another fit when ratios of two  fourth-order polynomials were considered and obtained the following expressions:
\begin{align}
    f_{g}^{(1)}&=\frac{N_c^2}{N_c^2-\frac{110}{19}}, \qquad f_{g}^{(2)}=\frac{N_c^4-0.534N_c^2+2.485}{N_c^4-13.106N_c^2+43.594}, \qquad f_{g}^{(3)}=\frac{N_c^4+8.103N_c^2+40.270}{N_c^4-15.278N_c^2+59.731},
    \label{eq:new_fit_fg}\\
    f_{y}^{(1)}&=\frac{N_c^2-1}{N_c^2-\frac{110}{19}}, \qquad f_{y}^{(2)}=\frac{N_c^4-0.976N_c^2+1.185}{N_c^4-12.691N_c^2+40.681}, \qquad f_{y}^{(3)}=\frac{N_c^4+6.520N_c^2+32.220}{N_c^4-15.226N_c^2+59.311},
    \label{eq:new_fit_fy}\\
    f_{u}^{(1)}&=\frac{N_c^4-1.047N_c^2+0.047}{N_c^4-5.863N_c^2+0.428}, \qquad f_{u}^{(2)}=\frac{N_c^4-1.049N_c^2+1.545}{N_c^4-12.779N_c^2+41.292}, \qquad f_{u}^{(3)}=\frac{N_c^4+7.239N_c^2+35.870}{N_c^4-15.284N_c^2+59.770},
    \label{eq:new_fit_fu}\\
    f_{v}^{(1)}&=\frac{N_c^4-1.029N_c^2+0.029}{N_c^4-5.878N_c^2+0.511}, \qquad f_{v}^{(2)}=\frac{N_c^4-1.718N_c^2+1.0112}{N_c^4-12.330N_c^2+38.240}, \qquad f_{v}^{(3)}=\frac{N_c^4+4.258N_c^2+19.300}{N_c^4-14.946N_c^2+56.311}.
    \label{eq:new_fit_fv}
\end{align}
which turn out to be more accurate. . 
The maximal value of the mean squared error (MSE) for the $f_x^{(i)}$ with our fourth (second) order polynomial fits in each order of $\epsilon$ is as follows: $10^{-12}$ ($4 \cdot 10^{-11}$), $ 10^{-9}$ ($8 \cdot 10^{-5}$), $8\cdot 10^{-5}$ ($4 \cdot 10^{-3}$).
While the fit with the second-order polynomials seems to provide good approximation we restrict ourselves to a more precise  four-order result.

At the end, we carry our numerical comparison of the two types of fits and exact results. The green line in Fig.\ref{fig:afp_Nc} corresponds to the second order fit  (see Appendix \ref{sec:other_fits}), and purple line is our fourth order fit \eqref{eq:new_fit_fg},\eqref{eq:new_fit_fy},\eqref{eq:new_fit_fu},\eqref{eq:new_fit_fv}. One can see that the green line ``misses'' exact points in Fig.\ref{fig:afp_Nc} in the presented range of $N_c$, while the purple line goes through the dots.  
\begin{figure}[h]
		\begin{tabular}{cc}
			\includegraphics[width=0.44\textwidth]{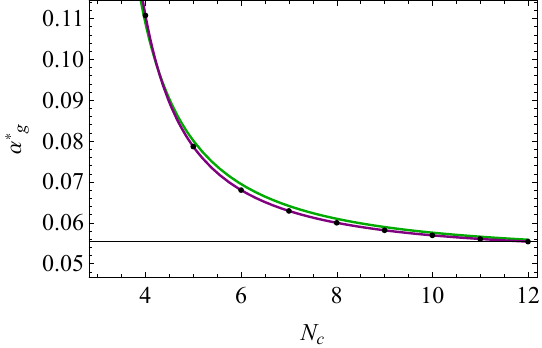}&
			\includegraphics[width=0.44\textwidth]{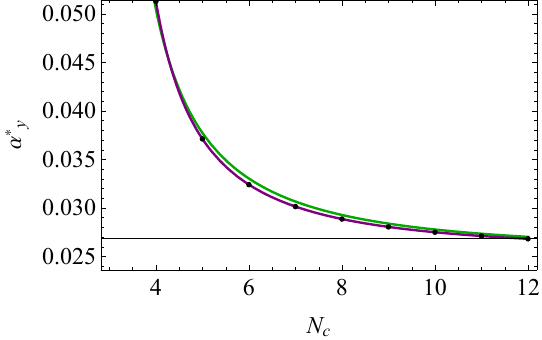}\\
			\includegraphics[width=0.44\textwidth]{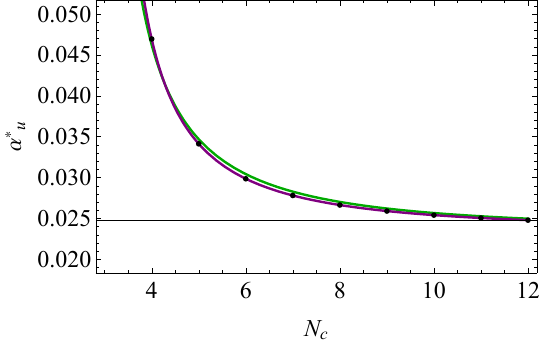}&
			\includegraphics[width=0.44\textwidth]{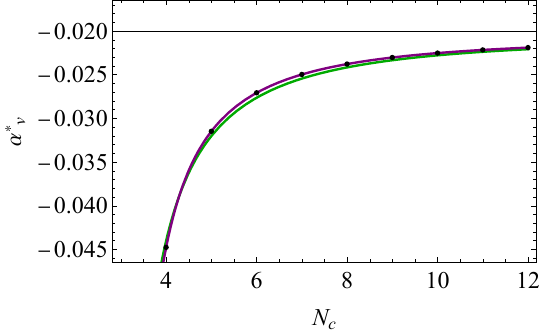}
	\end{tabular}
	\caption{The dependence of couplings from the number of colors $N_c$ for fixed  $\epsilon = 0.09$. Here the purple line is our  fourth-order polynomial fit \eqref{eq:new_fit_fg},\eqref{eq:new_fit_fy},\eqref{eq:new_fit_fu},\eqref{eq:new_fit_fv}, the green line is 2nd order fit \eqref{eq:fit_2nd_order}. The black dots correspond to exact values. It should be noted that we omit the point $N_c=3$, since it lies higher in the FP's scale, therefore the difference between the fits is not clearly visible. }
\label{fig:afp_Nc}
\end{figure}

\subsubsection{Critical exponents}
Let us study the universal critical exponents. The latter can be obtained as the eigenvalues of the stability matrix
\begin{align}
    M_{ij} = \frac{\partial\beta_i}{\partial\alpha_j}\vert_{\alpha=\alpha^*},
\end{align}
which again we expand as a power series 
\begin{align}
    \theta_j = c_{\theta_j}^{(1)}f_{\theta_j}^{(1)}\epsilon +  c_{\theta_j}^{(2)}f_{\theta_j}^{(2)}\epsilon^2 +c_{\theta_j}^{(3)}f_{\theta_j}^{(3)}\epsilon^3+...
    \label{eq:theta_expansion}
\end{align}
and find perfect agreement with \cite{Litim:2023tym} (note that for $\theta_1$ expansion starts at $\epsilon^2$ so $c^{(1)}_{\theta_1} = 0$): 
\begin{align}
	c^{(2)}_{\theta_1}&=-\frac{104}{171},\qquad c^{(3)}_{\theta_1} = \frac{2296}{3249}, \qquad c^{(4)}_{\theta_1} = \frac{\left(43551640704 \zeta_3+1405590649319-281341851912 \sqrt{23}\right)}{15643993482}, \\
    c^{(1)}_{\theta_2}&=\frac{52}{19},\quad c^{(2)}_{\theta_2} = \frac{\left(136601719-22783308 \sqrt{23}\right)}{4094823}, \quad c^{(3)}_{\theta_2} = \frac{5 \left(547695099865475491 - 111718308712462080 \sqrt{23}\right)}{2692813775855538},\\
    c^{(1)}_{\theta_3}&=\frac{8}{19} \sqrt{20+6 \sqrt{23}},\qquad c^{(2)}_{\theta_3} = \frac{2 \left(45155739-9153184 \sqrt{23}\right) \sqrt{20+6 \sqrt{23}} }{16879999}, \\
    c^{(3)}_{\theta_3} &= \frac{\sqrt{20+6 \sqrt{23}} \left(-918044509824 \left(1497 \sqrt{23}-7558\right) \zeta_3+73205713038142585-15289473238519518 \sqrt{23}\right)}{404906730972633},\\
    c^{(1)}_{\theta_4}&=\frac{16 \sqrt{23}}{19},\qquad c^{(2)}_{\theta_4} =\frac{4 \left(68248487 \sqrt{23}-255832864\right)} {31393643}, \\ c^{(3)}_{\theta_4} &=\frac{2 \left(37418532792608300581 \sqrt{23}-174067504271892880236\right)}{278706225801048183}.
\end{align}
Numerical evaluation of these coefficients gives in the Veneziano limit \cite{Litim:2023tym}
\begin{align}
  -\theta_1 &= 0.608 \epsilon^2 - 0.707 \epsilon^3 - 6.947 \epsilon^4 \textcolor{myred}{- 4.825 \epsilon^5} ,\label{eq:th1}\\
   \theta_2 &= 2.737 \epsilon + 6.676 \epsilon^2 + 22.120 \epsilon^3 \textcolor{myred}{+ 102.55 \epsilon^4} ,\\
   \theta_3 &= 2.941 \epsilon + 1.041 \epsilon^2 + 5.137 \epsilon^3 \textcolor{myred}{- 62.340 \epsilon^4} ,\label{eq:th3}\\
   \theta_4 &= 4.039 \epsilon + 9.107 \epsilon^2 + 38.646 \epsilon^3 \textcolor{myred}{+ 87.016 \epsilon^4}    \label{eq:theta_inf_Nc}
\end{align}
and for not too large $\epsilon$ (see below) there is only one relevant direction corresponding to $\theta_1 < 0$ giving rise to an asymptotically-safe scenario.  In Eq.~\eqref{eq:theta_inf_Nc} we again highlight terms that are not determined precisely in the 433 approximation and will be modified by high-order terms.

Finite-$N_c$ corrections are incorporated in the factors, which we approximate as\footnote{It is interesting that our fourth-order fit for $f^{(3)}_{\theta_1}$ gives ``exact result'', which we were not able to reproduce analytically. Yet numerical comparison up to 10000 digits shows no difference.}: 
\begin{align}
    f_{\theta_1}^{(2)} & =\frac{N_c^2 }{N_c^2 - \frac{110}{19}}, & 
    f_{\theta_1}^{(3)} & =\frac{N_c^2 (N_c^2-\frac{326}{287})}{(N_c^2 - \frac{110}{19})^2}, &
	f_{\theta_1}^{(4)} & = \frac{N_c^4 + 10.21 N_c^2 + 49.2}{N_c^4 - 15.23 N_c^2 + 59.44} ,\\
	f_{\theta_2}^{(1)} & =\frac{N_c^2 - 1}{N_c^2 - \frac{110}{19}}, & 
	f_{\theta_2}^{(2)} & = \frac{N_c^4 - 0.8198 N_c^2 + 0.33}{N_c^4 - 12.667 N_c^2 + 40.51}, &
	f_{\theta_2}^{(3)} & = \frac{N_c^4 + 8.568 N_c^2 + 44.8}{N_c^4-15.70 N_c^2 + 63.07} ,\\
	f_{\theta_3}^{(1)} & = \frac{N_c^4 - 0.3883 N_c^2 - 0.6036}{N_c^4-5.265 N_c^2 - 3.04}, & 
	f_{\theta_3}^{(2)} & = \frac{N_c^4 - 1.580 N_c^2 + 3.753}{N_c^4- 12.767 N_c^2 + 41.189}, &
	f_{\theta_3}^{(3)} & = \frac{N_c^4 + 11.09 N_c^2 + 48.6}{N_c^4 - 15.42 N_c^2 + 60.86}, \\
	f_{\theta_4}^{(1)} & = \frac{N_c^4 -0.7726 N_c^2 - 0.2459}{N_c^4-5.392 N_c^2 - 2.30}, & 
	f_{\theta_4}^{(2)} & = \frac{N_c^4 + 16.58 N_c^2 - 11}{N_c^4 - 13.33 N_c^2 +45.33}, &
	f_{\theta_4}^{(3)} & = \frac{N_c^4 + 61.95 N_c^2 + 357}{N_c^4 - 15.64 N_c^2 + 62.73}.
\end{align}
The maximal value of MSE for the given (second-order result from the appendix) approximations are
$10^{-19}$ ($10^{-10}$), 
$10^{-10}$ ($10^{-3}$), and
$10^{-7}$ ($0.15$) for first, second and third nontrivial orders in $\epsilon$.

\subsubsection{Other anomalous dimensions}\label{sec:other_anom}

In this subsection we provide results for the scalar and fermion anomalous dimensions. In addition, we consider the operator $\Tr(H^\dagger H)$ coupled to $m^2$ in \eqref{eq:Lag} together with the anomalous dimensions corresponding to dimension-three eigenoperators discussed earlier. 
The full expressions for anomalous dimensions beyond the Veneziano limit can be found in Appendix \ref{sec:ADM_3} and are also available in the Supplemental Material. 

At the fixed point the anomalous dimensions are given as a series expansion in $\epsilon$. 
In the Veneziano limit we have gauge-independent coefficients
\begin{align}
    c_{H}^{(1)} & = \frac{4}{19}, & 
	c_{H}^{(2)} & =  \frac{14567}{6859}-\frac{2376 \sqrt{23}}{6859}, &% \frac{14567-2376 \sqrt{23}}{13718}, &
	c_{H}^{(3)} & =  \frac{8816623159}{133709346}-\frac{753675598}{2476099 \sqrt{23}},
\end{align}
for the scalar fields, and   
\begin{align}
    c_{\psi}^{(1)} & = \frac{11}{19}, \qquad 
	c_{\psi}^{(2)}  = \frac{3738501-683100 \sqrt{23}}{740772},  \nonumber\\
	c_{\psi}^{(3)} & =  \frac{2879380764 \zeta (3)+780746553081-158608932408
   \sqrt{23}}{4813536456}, 
	\end{align}
which is valid in the Landau gauge $\xi=\xi^*=0$ (see also Ref.~\cite{Litim:2023tym}).
The gauge-independent result for $m^2$ can be cast into
	\begin{align}
    c_{m^2}^{(1)} & = \frac{4}{19}\sqrt{2(10+3\sqrt{23})}, \qquad 
	c_{m^2}^{(2)}  = \frac{\left(45155739-9153184 \sqrt{23}\right) \sqrt{20+6\sqrt{23}}}{16879999}, \nonumber\\
	c_{m^2}^{(3)} & =  \frac{\sqrt{20+6 \sqrt{23}} \left(-918044509824 \left(1497
   \sqrt{23}-7558\right) \zeta
   (3)+73205713038142585-15289473238519518
   \sqrt{23}\right)}{809813461945266}.
\end{align}
while gluon ($G$) and ghost ($c$) field anomalous dimensions in the Landau gauge\footnote{The result $\gamma_G = - 2 \gamma_c$ in the Landau gauge is the consequence of the relation for the renormalization constant for gluon-ghost-ghost vertex $Z_{Gcc} = Z_{g} \cdot Z_{G}^{1/2} \cdot Z_c$ that leads to $\gamma_{Gcc} = (\gamma_G + 2 \gamma_c) - \beta_{\alpha_g}/(2 \alpha_g)$. Since $\beta_{\alpha_g}(\alpha^*) = 0$ and in the Landau gauge $Z_{Gcc} = 1$, we have $\gamma_{Gcc} = 0$ and $\gamma_G + 2 \gamma_c = 0$.} are given as
\begin{align}
	c_{G}^{(1)} & = -2c_{c}^{(1)}=\frac{13}{19}, \qquad 
	c_{G}^{(2)}  = -2c_{c}^{(2)}=\frac{188725-33396 \sqrt{23}}{27436}, \nonumber\\
	c_{G}^{(3)} & = -2c_{c}^{(3)}= \frac{288694588 \zeta
		(3)+38849548925-7884584928
	\sqrt{23}}{178279128}.
\end{align}
Substituting the coefficients into the power series, we have in the Veneziano limit and (for $\xi=0$) \cite{Litim:2023tym}
\begin{align}
    \gamma_H & = 0.2105\epsilon + 0.4625 \epsilon^2 + 2.4711 \epsilon^3,
    \label{eq:gamma_scalar}\\
    \gamma_\psi & =  0.5789\epsilon  
    + 0.6243\epsilon^2 + 4.8916\epsilon^3,
    \label{eq:gamma_fermion}\\
    \gamma_{m^2} & = 1.4703\epsilon + 0.5207\epsilon^2 + 2.5684 \epsilon^3,
	\label{eq:gamma_mphi}\\
    \gamma_{G} & = -2\gamma_{c}= 0.6842\epsilon + 1.0411\epsilon^2 + 7.7599 \epsilon^3
	\label{eq:gamma_gauge_ghost}.
\end{align}
The factors that account for the finite-$N_c$ corrections are again approximated as
\begin{align}
	f_{H}^{(1)} & =\frac{N_c^2 - 1}{N_c^2 - \frac{110}{19}}, & 
	f_{H}^{(2)} & = \frac{N_c^4 -0.819 N_c^2 + 1.265}{N_c^4-12.747 N_c^2 + 41.073}, &
	f_{H}^{(3)} & = \frac{N_c^4 +8.310 N_c^2 + 41.460}{N_c^4-15.354 N_c^2 + 60.340},\\
    f_{\psi}^{(1)} & = \frac{N_c^2 - 1}{N_c^2 - \frac{110}{19}}, & 
	f_{\psi}^{(2)} & = \frac{N_c^4 +0.459 N_c^2 + 3.821}{N_c^4-13.218 N_c^2 + 44.414}, &
	f_{\psi}^{(3)} & = \frac{N_c^4 +10.100 N_c^2 + 51.904}{N_c^4-15.483 N_c^2 + 61.368},\\
    f_{m^2}^{(1)} & =\frac{N_c^4 - 2.678 N_c^2+1.677 }{N_c^4- 7.468 N_c^2 +9.715}, 
    & 
	f_{m^2}^{(2)} & = \frac{N_c^4+ 9.603 N_c^2+11.895 }{N_c^4- 13.542 N_c^2 +46.829}, &
	f_{m^2}^{(3)} & =  \frac{N_c^4 + 47.739 N_c^2 + 253.184}{N_c^4-15.868 N_c^2 + 64.530},
	\\
    f_{G(c)}^{(1)} & =\frac{N_c^2 }{N_c^2 - \frac{110}{19}}, & 
	f_{G(c)}^{(2)} & = \frac{N_c^4 -0.522N_c^2 + 3.152}{N_c^4-13.217 N_c^2 + 44.395}, &
	f_{G(c)}^{(3)} & = \frac{N_c^4 +9.086 N_c^2 + 48.136}{N_c^4-15.440 N_c^2 + 61.028}.
\end{align}
where the maximal MSE is no worse than $3\cdot 10^{-7}$ in comparison to $10^{-1}$ for the approximations given in Eq.~\eqref{eq:fit_2nd_order_fields_and_mm}. 

Finally, the eigenvalues for the dimension-three operator $4\times4$ anomalous dimension matrix are given by
\begin{align}
	\gamma_j & = c_{\gamma_j}^{(1)}f_{\gamma_j}^{(1)}\epsilon +  c_{\gamma_j}^{(2)}f_{\gamma_j}^{(2)}\epsilon^2 +c_{\gamma_j}^{(3)}f_{\gamma_j}^{(3)}\epsilon^3+...
\end{align}
with
\begin{align}
	c_{\gamma_1}^{(1)} & = - c_{\gamma_2}^{(1)} = c_{H}^{(1)} & 
	c_{\gamma_1}^{(2)} & = - c_{\gamma_2}^{(2)} = c_{H}^{(2)} & 
	c_{\gamma_1}^{(3)} & = - c_{\gamma_2}^{(3)} = c_{H}^{(3)} \\ 
	c_{\gamma_3}^{(1)} & =  \frac{4}{19} \left(1+2 \sqrt{23}\right), & 
	c_{\gamma_3}^{(2)} & =  \frac{606162}{6859 \sqrt{23}}-\frac{99745}{6859}, & 
	c_{\gamma_3}^{(3)} & = \frac{4207372301377}{ 1537657479\sqrt{23}}-\frac{73545557081}{133709346}. 
\end{align}
\begin{align}
	c_{\gamma_4}^{(1)} & = \frac{4}{19} \left(1+\sqrt{20+6 \sqrt{23}}\right), \\
	c_{\gamma_4}^{(2)} & = \frac{14567}{6859}
	-\frac{2376 \sqrt{23}}{6859}
	+\frac{\sqrt{2(1475668498887 \sqrt{23}-7061359720318)}}{6859\sqrt{2461}} 
,\\
	c_{\gamma_4}^{(3)} & = 
	\frac{8816623159}{133709346}-\frac{753675598}{2476099 \sqrt{23}}
+ \frac{832 \sqrt{2 \left(931899 \sqrt{23}-4436554\right)} \zeta_3}{6859\sqrt{107}} \\
		  & -\frac{\sqrt{3990932506333661629040635839771 \sqrt{23}-19139697164494183329626575881170}}{164529350253\sqrt{4922}}
\end{align}
Evaluating the coefficients in the Veneziano limit, we obtain
\begin{align}
	\gamma_H = \gamma_1 = - \gamma_2 & = 
	0.21053\epsilon  
	+0.46247\epsilon ^2 
	 + 2.47105\epsilon ^3 , \\
	\gamma_3 & = 
	2.22982\epsilon 
	+3.88519\epsilon ^2 
	+ 20.5012\epsilon ^3 , \\
	\gamma_4 & = 
	1.68082\epsilon   
	+0.98321\epsilon ^2 
	+5.03949\epsilon ^3.
\end{align}
One can see that all but $\gamma_2$ is positive for $\epsilon>0$ and does not pose a problem to unitarity. Moreover, $\gamma_2 = - \gamma_H$ corresponds to the linear combination of operators\footnote{This fact can be deduced by considering the renormalization of local operator 
$[\delta S/\delta H]_R = Z_H^{1/2} \cdot (\delta S/\delta H)_0$.} $O_{1-4}$ that vanish due to the equations of motion, while $\gamma_1 = \gamma_H$ corresponds to a linear combination of $O_{1-3}$ that becomes a descendant of $\Tr(H) + \hc$ when EOMs are imposed. 

As for the finite-$N_c$ factors, we provide the following approximate expressions 

\begin{align}
	f_{\gamma_1,\gamma_2}^{(1)} & = f_{H}^{(1)}, &
	f_{\gamma_1,\gamma_2}^{(2)} & = f_{H}^{(2)}, & 
	f_{\gamma_1,\gamma_2}^{(3)} & = f_{H}^{(3)}, \\ 
	f_{\gamma_3}^{(1)} & = \frac{N_c^4 - 0.751 N_c^2 - 0.265}{N_c^4-5.392 N_c^2 - 2.300}, & 
	f_{\gamma_3}^{(2)} & = \frac{N_c^4 + 5.478 N_c^2 - 0.150}{N_c^4-13.400 N_c^2 + 45.771}, &
	f_{\gamma_3}^{(3)} & = \frac{N_c^4 + 27.974 N_c^2 + 149.100}{N_c^4-15.762 N_c^2 + 63.660}, \\
	f_{\gamma_4}^{(1)} & = \frac{N_c^4 - 0.397 N_c^2 - 0.596}{N_c^4-5.265 N_c^2 - 3.036}, & 
	f_{\gamma_4}^{(2)} & = \frac{N_c^4 - 0.923 N_c^2 + 3.520}{N_c^4-12.924 N_c^2 + 42.288}, &
	f_{\gamma_4}^{(3)} & = \frac{N_c^4 + 10.300 N_c^2 + 50.500}{N_c^4-15.454 N_c^2 + 61.140}. 
\end{align}
The maximal value of MSE in these fits are $10^{-9}$, while the corresponding value for the ratio of two second-order polynomials \eqref{eq:gamma_2nd_order} reaches $4\cdot 10^{-2}$. 

\section{Conformal window}\label{sec:conf_window}
In this section we investigate the size of the UV conformal window for the asymptotically safe theory with action equation ~\eqref{eq:Lag} using perturbation theory.

\subsection{Constraints on the UV conformal window} 

We can find the UV conformal window directly from the expressions for fixed points and scaling exponents given in the previous sections. In the case when we retain only first three non-vanishing powers of $\epsilon$, we will call the bound on $\epsilon$ strict. 
The reason for this is that the higher-order (``subleading'') coefficients in the power expansion \eqref{eq:fp_exp},\eqref{eq:theta_expansion} are not (yet) accurately determined due to the absence of higher loop terms in $\beta$-functions. This scheme is dictated first by the following constraints
\begin{itemize}
    \item perturbativity for couplings $0<|\alpha^*|\lesssim 1$ \cite{Weinberg:1978kz};
    \item vacuum stability $\alpha_u^* >0$ and  $\alpha_u^*+\alpha_v^* >0$ \cite{Paterson:1980fc};
    \item no fixed point merger \cite{Bond:2021tgu} (the collision of the UV fixed point with an IR fixed point corresponds to $\theta = 0$).
\end{itemize}

The second strategy employs the approximation, where we retain subleading terms in $\epsilon$,  so we refer to its bounds as subleading. There we also take into account all the above mentioned constraints from couplings, vacuum stability and critical exponents.

\subsection{Investigation of the UV conformal window}

The UV conformal window can be investigated using the above mentioned constraints. To do so, we first can equate the perturbative expressions for FP couplings to unity, ($\alpha_u^* + \alpha_v^*$) and scaling exponents to zero and choose the smallest positive solution for $\epsilon$. Second, we can find the Padé  approximants for these constraints and make the same manipulations. We represent our results as $\alpha^* = \epsilon P_{ij}$ and $\theta = \epsilon^2 P_{ij}$, where $P_{ij}$ are Padé  approximants and $i+j=2(3)$ for the strict (subleading) case. However,  we cannot confidently trust the obtained results,  because these approximants contain non-physical poles. Nevertheless, they give tighter constraints on the conformal window and we provide all approximants that can be constructed from available series together with the corresponding bounds.

\begin{itemize}
    \item From perturbative expansion of couplings \eqref{eq:ag_fp}-\eqref{eq:av_fp}, we note that the tightest bound on $\epsilon_{strict}$  arises from the gauge coupling.  
    First, equating the gauge coupling to unity, we can find $\epsilon_{strict}$ and $\epsilon_{subl}$
    \begin{align}
        \epsilon_{strict} \approx  0.457,\qquad
        \epsilon_{subl} \approx  0.363.
        \label{eq:strong_coup}
    \end{align}
    Second, we can construct the Padé  approximants
    \begin{equation}
        \begin{aligned}
        P_{11}^{\alpha_g^*} &= \frac{0.456-3.081\epsilon}{1-8.467\epsilon},\\
        P_{12}^{\alpha_g^*} &= \frac{0.456-0.328\epsilon}{1-2.431\epsilon-10.331\epsilon^2}, \qquad P_{21}^{\alpha_g^*} = \frac{0.456-0.885\epsilon+3.760\epsilon^2}{1-3.651\epsilon}
    \end{aligned}
    \end{equation}
    where the first line correspond to the strict case, the second to subleading, and find the bounds
    \begin{equation}
        \begin{aligned}
        \epsilon_{strict_{11}} &\approx  0.117,\qquad
        \epsilon_{subl_{12}} &\approx  0.203, \qquad \epsilon_{subl_{21}}\approx 0.243.
        \label{eq:strong_coup_Padé }
    \end{aligned}
    \end{equation}
    
    \item In the same manner we can investigate the bounds arising from the vacuum stability condition \cite{Paterson:1980fc}:
    \begin{align}
        \epsilon_{strict} \approx  0.146,\qquad
        \epsilon_{subl} \approx  0.116.
        \label{eq:vacuum_inst}
    \end{align}
    To provide more stronger constrains for $\epsilon$, we use the Padé  approximants:
    \begin{equation}
        \begin{aligned}
        P_{11}^{\alpha_u^*+\alpha_v^*} &= \frac{0.0625-0.719\epsilon}{1-8.438\epsilon}, \\
        P_{12}^{\alpha_u^*+\alpha_v^*} &= \frac{0.625-0.673\epsilon}{1-7.695\epsilon+2.281\epsilon^2}, \qquad P_{21}^{\alpha_u^*+\alpha_v^*} = \frac{0.625-0.656\epsilon-0.194\epsilon^2}{1-7.425\epsilon}.
    \end{aligned}
    \end{equation}
    The UV conformal window in this case is constrained as
    \begin{equation}
        \begin{aligned}
        \epsilon_{strict_{11}} &\approx  0.087,\qquad
        \epsilon_{subl_{12}} &\approx  0.09287, \qquad \epsilon_{subl_{21}}\approx 0.09272.
        \label{eq:vacuum_inst_Padé }
    \end{aligned}
    \end{equation}
   
    \item After the calculation of scaling exponents \eqref{eq:theta_inf_Nc},
    we notice the behavior of $\theta_{2,3,4}$ with same-sign corrections at every order. However, the sign of leading $\epsilon^2$ coefficient for the relevant scaling exponent $\theta_{1}$  differs from other loop terms, which is the indication of possible FP merger.  
    Thus, we can extract the constraints from the relevant scaling exponent, solving $\theta_1 = 0$:
    \begin{align}
        \epsilon_{strict} \approx  0.249,\qquad
        \epsilon_{subl} \approx  0.234.
        \label{eq:fp_merger}
    \end{align}
    And using the Padé  approximation:
    \begin{equation}
        \begin{aligned}
         P_{11}^{\theta_1} &= \frac{0.608-6.681\epsilon}{1-9.826\epsilon},\\
         P_{12}^{\theta_1} &= \frac{0.608-1.717\epsilon}{1-0.661\epsilon+9.495\epsilon^2}, \qquad
         P_{21}^{\theta_1} = \frac{0.608-1.129\epsilon-6.455\epsilon^2}{1-0.695\epsilon},
     \end{aligned}
    \end{equation}
    we get
    \begin{equation}
        \begin{aligned}
        \epsilon_{strict_{11}} &\approx  0.091,\qquad
        \epsilon_{subl_{12}} &\approx  0.354, \qquad \epsilon_{subl_{21}}\approx 0.232.
        \label{eq:fp_merger_Padé }
    \end{aligned}
    \end{equation}
 \end{itemize}
It should be noted, that at 433 order we have only one fixed point merger. However, if the subleading tendency continues in the 544 order, it will lead to the additional  $\theta_3$ merger  \eqref{eq:th3}, which we do not study in this paper. 

\begin{figure}[h]
		\begin{tabular}{ccc}
		\includegraphics[width=0.31\textwidth]{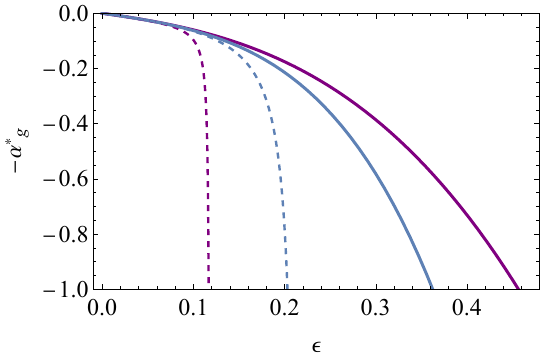}&
        \includegraphics[width=0.31\textwidth]{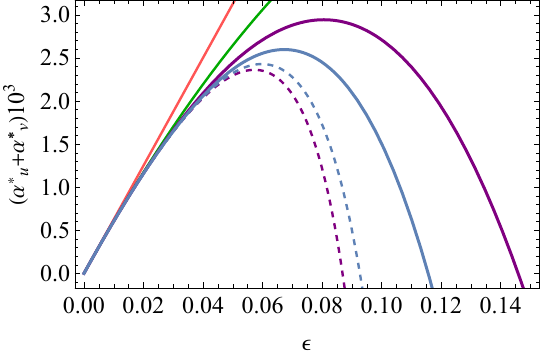}&
        \includegraphics[width=0.405\textwidth]{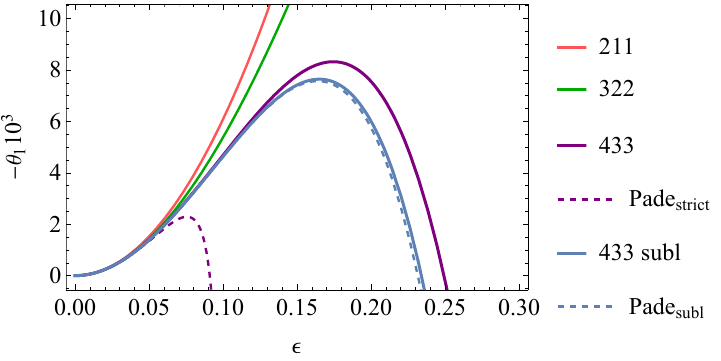}
	\end{tabular}
	\caption{$\epsilon_{strict, subl}$ bounds for 433 order (full lines) \eqref{eq:strong_coup},\eqref{eq:vacuum_inst},\eqref{eq:fp_merger} and their tighter Padé  ressummations (dashed lines) \eqref{eq:strong_coup_Padé },\eqref{eq:vacuum_inst_Padé },\eqref{eq:fp_merger_Padé }. The 211 and 322 orders also illustrated.}
\label{fig:UV_bound}
\end{figure}

Moreover, we illustrate our results in the Fig.\ref{fig:UV_bound}.
Here we show up the bounds for strict and subleading approaches (full lines) \eqref{eq:strong_coup},\eqref{eq:vacuum_inst},\eqref{eq:fp_merger} and their Padé  approximants (dashed lines) that provide tightest constraints\eqref{eq:strong_coup_Padé },\eqref{eq:vacuum_inst_Padé },\eqref{eq:fp_merger_Padé } at 433 order. In addition, we add 211 and 322 orders for comparison. 
From these figures and from \eqref{eq:strong_coup},\eqref{eq:strong_coup_Padé },\eqref{eq:vacuum_inst},\eqref{eq:vacuum_inst_Padé },\eqref{eq:fp_merger}, and \eqref{eq:fp_merger_Padé }, we can deduce that the $\epsilon_{subl}$ bound  is systematically tighter than the $\epsilon_{strict}$ bound, which arise from the last summands \eqref{eq:ag_fp}, \eqref{eq:th1}. The same situation was obtained at 322 order, see Ref.\cite{Bond:2017tbw}.

\begin{figure}[h]
		\begin{tabular}{cc}
\includegraphics[width=0.425\textwidth]{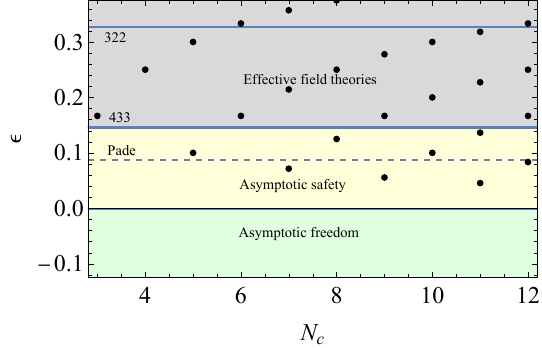}&
\includegraphics[width=0.58\textwidth]{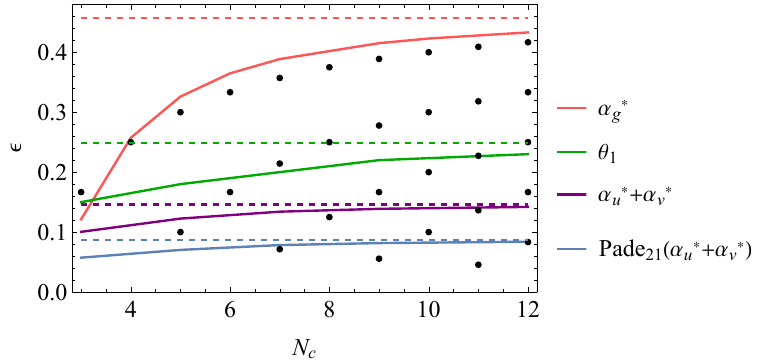}
	\end{tabular}
	\caption{Left panel: conformal window with asymptotic safety (yellow band), also showing regimes with asymptotic freedom (green) and effective theories (grey). Dots indicate the first few integer solutions \eqref{eq:AS_QFT} at 433 order. Moreover, this panel contains the new upper bound on $\epsilon$ at 433 order \eqref{eq:vacuum_inst}, and the tighter Padé  approximant bound (dashed line) \eqref{eq:vacuum_inst_Padé }. Previous upper bounds at 322 \cite{Bond:2017tbw} order also indicated. The right panel compares different schemes \eqref{eq:strong_coup},\eqref{eq:vacuum_inst},\eqref{eq:fp_merger} as given by $\epsilon_{strict}$. The dashed lines represent the asymptotic value; the full lines represent the upper boundary for the $\epsilon$ as functions of $N_c$. }
\label{fig:eps_Nc}
\end{figure}

At the end, we demonstrate   the size of the UV conformal window in Fig.\ref{fig:eps_Nc}. The left panel includes $\epsilon_{strict}$ bounds arising from vacuum stability \eqref{eq:vacuum_inst},\eqref{eq:vacuum_inst_Padé }. For comparison we have also indicated the previous bound at 322 order \cite{Bond:2017tbw}.
The right panel illustrates the boundaries for finite values of $N_c$. From this picture we can see that all constraints \eqref{eq:strong_coup},\eqref{eq:vacuum_inst},\eqref{eq:fp_merger} share roughly the same rate of convergence. In addition, it is clear, that finite-$N_c$ corrections contract the conformal window (full lines) in comparison to infinite  results (dashed lines).

Finally, using the tightest bound on the conformal window which is given by the vacuum instability $\epsilon_{strict} \approx 0.146$ \eqref{eq:vacuum_inst}, we obtain the smallest pair of integer values %(see Tab.\ref{tab:Nc_Nf} and Fig.\ref{fig:eps_Nc}) 
for ($N_c$, $N_f$) compatible with asymptotic safety, which are indicated by black dots in Fig.\ref{fig:eps_Nc}. The first few integer solutions are
\begin{align}
    (N_c,N_f)=(5, 28), (7, 39), (8, 45), (9, 50), (10, 56), (11,61), (11,62),(12,67)...
    \label{eq:AS_QFT}
\end{align}

\subsection{(De)stabilizing fluctuations}

Following the ideas of Refs.\cite{Bond:2017tbw,Bond:2021tgu}, let us consider into which direction the higher loop corrections shift the beta functions. To do this we substitute the fixed points to order $O(\epsilon)^3$ back to the beta functions  and take series expansion in $\epsilon$ up to the first non-vanishing order. Keeping only highest available terms in RG functions, we obtain   
(c.f., also Table \ref{tab:shifts})
\begin{equation}
       \begin{aligned}
 \beta_g^{(4)}|_{322}&=-4.154\left(\frac{55.257+N_c^2}{-8.708+N_c^2}\right)\epsilon^5,\\
\beta_y^{(3)}|_{322}&=1.263\left(\frac{33.923+N_c^2}{-8.442+N_c^2}\right)\epsilon^4,\\
\beta_u^{(3)}|_{322}&=-0.252\left(\frac{19.213+N_c^2}{-8.304+N_c^2}\right)\epsilon^4,\\
    \beta_v^{(3)}|_{322}&=0.213\left(\frac{83.582+N_c^2}{-8.514+N_c^2}\right)\epsilon^4.
    \label{eq:leading_shifts}
\end{aligned}
\end{equation}

Negative shifts to the beta functions ($\Delta\beta < 0$) are supposed to stabilize the UV fixed point \cite{Bond:2017tbw,Bond:2021tgu}, while $\Delta\beta > 0$, conversely,  shift the zero towards larger values, which could even destabilize it. Thus, the obtained leading shifts for finite $N_c$ gives us a qualitative picture of the trend from higher-order loop contributions.  It should be noted that these shifts have changed their signs compared to the previous results obtained in Refs.\cite{Bond:2017tbw,Bond:2021tgu}.

In the same manner, we insert the fixed points up to $\epsilon^3$ to the full available beta functions and find the subleading shifts for finite $N_c$ (see Table \ref{tab:shifts} with the Veneziano limit):
\begin{equation}
    \begin{aligned}
    \beta_g|_{433}&=65.931\left(\frac{183.114+N_c^2}{-8.948+N_c^2}\right)\epsilon^6,\\
    \beta_y|_{433}&=-11.143\left(\frac{75.486+N_c^2}{-8.946+N_c^2}\right)\epsilon^5,\\
    \beta_u|_{433}&=9.354\left(\frac{64.415+N_c^2}{-8.877+N_c^2}\right)\epsilon^5,\\
    \beta_v|_{433}&=47.863\left(\frac{51.740+N_c^2}{-7.993+N_c^2}\right)\epsilon^5.
    \label{eq:subleading_shifts}
\end{aligned}
\end{equation}
We see that the results for $\beta_{g,y,u}$ change signs as compared to \eqref{eq:leading_shifts}. However, they retained their behavior as in the case of Refs.\cite{Bond:2017tbw,Bond:2021tgu}. Therefore, we can expect, that in  the 544 order we will obtain similar results.  

\section{Conclusion}\label{sec:Concl}

The availability of interacting UV fixed points in particle physics presents numerous prospects for constructing models, see, e.g., Ref.~\cite{Bond:2017wut}. However, comprehending the size of the corresponding conformal window is equally crucial for any real-world applications. In this paper we investigated the Litim-Sannino model with action \eqref{eq:Lag} at the 433 order. Extending the findings of \cite{Litim:2014uca,Bond:2017tbw,Bond:2021tgu} and confirming the results of \cite{Litim:2023tym}, we have performed a full search for interacting fixed points and  computed scaling exponents and  anomalous dimensions up to third order in the small parameter $\epsilon$ \eqref{eq:eps_def}. It was also noted that their full expressions for finite $N_c$ are complicated, therefore, we performed fourth(second) order fits and provide the corresponding approximate expressions together. By comparing means square errors, we conclude that the ratio of fourth-order polynomials provide better approximation in certain cases; see, e.g., Fig.\ref{fig:afp_Nc}.

Moreover,  we studied the size of the conformal window imposing conditions on the fixed point values of the couplings and scaling exponents. We used different approximation orders and estimate the effect of subleading corrections. We compared various restrictions coming from the perturbativity of the strong coupling \eqref{eq:strong_coup},  vacuum instability \eqref{eq:vacuum_inst} and possible fixed point merger \eqref{eq:fp_merger}. Despite their qualitatively different origins, constraints are quantitatively similar, with vacuum stability offering the tightest bound \eqref{eq:vacuum_inst}. In addition, following Ref.~\cite{Litim:2023tym} we tried to resum the $\epsilon$ series by means of Padé  approximation and find that it gives even stronger constraints \eqref{eq:strong_coup_Padé },\eqref{eq:vacuum_inst_Padé },\eqref{eq:fp_merger_Padé }. However, the presence of non-physical poles in the approximants used to derive the bounds undermines our confidence in their accuracy. Perhaps, this should be explored using other types of approximations and when high-order contributions in the 544-scheme will be available. We summarize our results in Fig.\ref{fig:eps_Nc}, where we illustrated the asymptotic safety regime together with the size of the UV conformal window. At the end, the asymptotically safe quantum field theories, which lie within the allowed conformal window were also demonstrated in Eq.\eqref{eq:AS_QFT} and Fig.\ref{fig:eps_Nc}.

Furthermore, we notice that the authors of Ref.~\cite{Litim:2023tym} mentioned 
that the anomalous dimension $\gamma_{m_\psi}$ of the fermion mass, which they computed in their paper, is \emph{negative} to the leading order in $\epsilon$, while all the next-to-leading order terms are positive. Because of this, $\gamma_{m_\psi}$ can become negative and  potentially pose a threat to unitarity \cite{Litim:2023tym}. We argue that this negative leading-order result is nothing else but the leading-order contribution to $(-\gamma_H)$. Positive  higher-order terms computed in Ref.~\cite{Litim:2023tym} cannot be trusted when evaluating scaling dimensions, since at the two-loop level the mixing comes into play. 
We account for this mixing in the present study and compute the anomalous-dimension matrix eigenvalues, one of which should be $(-\gamma_H)$ at any loop.  
But this is not the end of the story. We also argue that if EOMs are taken into account the anomalous dimension matrix of dimension-three operators is modified $\gamma_O \to \tilde \gamma_O$ such that $\tilde\gamma_O$ has the same eigenvalues as $\gamma_O$ but with the flipped sign of the ``dangerous'' eigenvalue $(-\gamma_H) \to \gamma_H$. This can be anticipated and represents correct scaling of the operator that enters EOMs alongside with $O_4 \propto \partial^2 \Tr(H)$. Our analysis shows that all eigenvalues of the reduced anomalous dimension matrix are positive so dimension-three operators that break $G$ flavor symmetry do not spoil unitarity. 

We believe that our findings can be used in a more elaborate analysis of the asymptotic safety in the Veneziano limit and beyond the latter. It is interesting how the results will be modified when the 544 order beta functions will be available, e.g., in connection with possible additional FP merger due to potentially negative contribution of $\mathcal{O}(\epsilon^4)$ to $\theta_3$.  

We thank Oleg Antipin,  Nikita Lebedev, Georgy Kalagov, Dmitry Kazakov, Lukas Mizisin, and Vitaly Velizhanin for fruitful discussions. We also appreciate the correspondence with Tom Steudtner on Ref.~\cite{Litim:2023tym} at the early stage of our study. The work of a A.I.M was supported by the JINR AYSS Foundation, Project No 24-301-02.
\appendix

\section{Results}
In this appendix we provide complete expressions for the renormalization-group functions beyond the Veneziano limit. We follow \cite{Litim:2023tym} and introduce convenient abbreviations $\rc \equiv N_c^{-2}$, $\rf\equiv \left[(\frac{11}{2}+\epsilon)N_c \right]^{-2}$.
\subsection{$\beta$-functions}\label{sec:beta_func}
The beta functions for rescaled coupling \eqref{eq:dim4_couplings_VL} read 

\begin{align}
    \beta_g^{(1)}\alpha_g^{-2}&= \frac{4\epsilon}{3},
\\
    \beta_g^{(2)}\alpha_g^{-2}&= \left[25+\frac{26\epsilon}{3}-(11+2\epsilon)\rc\right]\alpha_g - \frac{1}{2}\alpha_y  \left(11+2\epsilon\right)^2,\\
\beta_g^{(3)}\alpha_g^{-2}&=\left[\frac{6309+954\epsilon-224\epsilon^2}{54}+\frac{11(11+2\epsilon)(\epsilon-3)}{18}\rc - \frac{11+2\epsilon}{4}\rc^2\right]\alpha_g^2 -\frac{3}{8}\left(9-\rc\right)(11+2\epsilon)^2\alpha_g\alpha_y {}\nonumber\\ 
			& \hspace{0.5cm}
   +\frac{1}{4}(11+2\epsilon)^2(3\epsilon+20)\alpha_y^2,
\end{align}

\begin{align}
   \beta_g^{(4)}\alpha_g^{-2}&= \alpha_g^2 \alpha_y
   \left[\rc^2 \left(\left(18 \zeta_3-\frac{3}{4}\right)
   \epsilon ^2+\left(198 \zeta_3-\frac{33}{4}\right) \epsilon
   +\frac{1089 \zeta_3}{2}-\frac{363}{16}\right)\right.{}\nonumber\\ 
			& \hspace{0.5cm}\left.+\rc
   \left(\left(-54 \zeta_3-\frac{1184}{9}\right) \epsilon
   ^2+\left(-594 \zeta_3-\frac{45749}{72}\right) \epsilon
   -\frac{3267 \zeta_3}{2}-\frac{161 \epsilon
   ^3}{18}-\frac{24079}{24}\right)\right.{}\nonumber\\ 
			& \hspace{0.5cm}\left.+\left(36 \zeta_3+\frac{8017}{36}\right) \epsilon ^2+\left(396 \zeta_3+\frac{38797}{72}\right) \epsilon +1089 \zeta_3+\frac{379
   \epsilon ^3}{18}-\frac{12947}{48}\right]{}\nonumber\\ 
			& \hspace{0.5cm}+\alpha_g^3
   \left[\rc^3 \left(\frac{23 \epsilon
   }{4}+\frac{253}{8}\right)+\rc^2
   \left(\left(\frac{623}{27}-\frac{488 \zeta_3}{9}\right) \epsilon
   ^2+\left(\frac{29753}{108}-\frac{5456 \zeta_3}{9}\right)
   \epsilon -1694 \zeta_3+\frac{19613}{24}\right)\right.{}\nonumber\\ 
			& \hspace{0.5cm}\left.+\rc
   \left(\left(\frac{128 \zeta_3}{9}+\frac{7495}{243}\right)
   \epsilon ^2+\left(\frac{2504 \zeta_3}{9}+\frac{71765}{324}\right) \epsilon +396 \zeta_3+\frac{154
   \epsilon ^3}{243}+\frac{30047}{72}\right)\right.{}\nonumber\\ 
			& \hspace{0.5cm}\left.+\left(-\frac{56 \zeta
   _{3}}{3}-\frac{21598}{243}\right) \epsilon ^2+\left(-\frac{1808
   \zeta_3}{9}-\frac{123473}{324}\right) \epsilon -550 \zeta
   _{3}-\frac{260 \epsilon
   ^3}{243}-\frac{14731}{72}\right]{}\nonumber\\ 
			& \hspace{0.5cm}+\alpha_u^2 \alpha_y \left[12 \rc+12 \epsilon ^2+132 \epsilon +363\right]+48
   \alpha_u \alpha_v \alpha_y
   \rc{}\nonumber\\ 
			& \hspace{0.5cm}+\alpha_u \alpha_y^2 \left[-\rc
   (10 \epsilon +55)-10 \epsilon ^3-165 \epsilon ^2-\frac{1815
   \epsilon }{2}-\frac{6655}{4}\right]-\alpha_v
   \alpha_y^2 \rc (20 \epsilon +110){}\nonumber\\ 
			& \hspace{0.5cm}+12 \alpha_v^2 \alpha_y \rc (\rf+1)+\alpha_g
   \alpha_y^2 \left[\rc \left(\left(12 \zeta
   _{3}+\frac{89}{4}\right) \epsilon ^2+(132 \zeta_3+154) \epsilon
   +363 \zeta_3+\epsilon
   ^3+\frac{5445}{16}\right)\right.{}\nonumber\\ 
			& \hspace{0.5cm}\left.+\left(\frac{1659}{4}-12 \zeta_3\right)
   \epsilon ^2+(2475-132 \zeta_3) \epsilon -363 \zeta_3+23
   \epsilon ^3+\frac{78287}{16}\right]{}\nonumber\\ 
			& \hspace{0.5cm}+\alpha_y^3
   \left[\rc \left(\left(\frac{7}{3}-6 \zeta_3\right)
   \epsilon ^2+\left(\frac{77}{3}-66 \zeta_3\right) \epsilon
   -\frac{363 \zeta_3}{2}+\frac{847}{12}\right)-\frac{11 \epsilon
   ^4}{3}-100 \epsilon ^3-986 \epsilon ^2-\frac{25267 \epsilon
   }{6}-\frac{105875}{16}\right]
\end{align}

\begin{align}
    \beta_y^{(1)}\alpha_y^{-1} &= (13+2\epsilon)\alpha_y-6(1-\rc)\alpha_g,\\
    \beta_y^{(2)}\alpha_y^{-1} &= -\frac{1}{8}\left[(11+2\epsilon)(2\epsilon+35)-32\rc\right]\alpha_y^2 + (1-\rc)(8\epsilon+49)\alpha_g\alpha_y + \frac{1}{6}(1-\rc)\left((20\epsilon-93)+9\rc \right)\alpha_g^2{}\nonumber\\ 
			& \hspace{0.5cm}
   -4\left[(11+2\epsilon)(1+\rf)\right]\alpha_y\alpha_u +
   4(1+\rf)\alpha_u^2 +16\rf\alpha_u\alpha_v-8\rf\alpha_v\alpha_y(11+2\epsilon)+4\rf(1+\rf)\alpha_v^2,
\end{align}

\begin{align}
   \beta_y^{(3)}\alpha_y^{-1} &=\alpha_g \alpha_y^2
   \left[16 \rc^2+\rc \left(19 \epsilon ^2+\frac{445
   \epsilon }{2}+633\right)-19 \epsilon ^2-\frac{445 \epsilon
   }{2}-649\right],{}\nonumber\\ 
			& \hspace{0.5cm}+\alpha_g^2 \alpha_y
   \left[\rc^2 \left(\left(\frac{157}{8}-18 \zeta_3\right)
   \epsilon -81 \zeta_3+\frac{721}{16}\right)+\rc \left((54
   \zeta_3+92) \epsilon +279 \zeta_3+17 \epsilon
   ^2+31\right)\right.{}\nonumber\\ 
			& \hspace{0.5cm}\left.+\left(-36 \zeta_3-\frac{893}{8}\right) \epsilon
   -198 \zeta_3-17 \epsilon ^2-\frac{1217}{16}\right]{}\nonumber\\ 
			& \hspace{0.5cm}+\alpha_g^3 \left[\frac{129 \rc^3}{4}+\rc^2 ((23-24 \zeta
   _{3}) \epsilon -132 \zeta_3+62)+\rc \left(-\frac{70
   \epsilon ^2}{27}-\frac{856 \epsilon
   }{9}-\frac{2413}{12}\right)\right.{}\nonumber\\ 
			& \hspace{0.5cm}\left.+\left(24 \zeta
   _{3}+\frac{649}{9}\right) \epsilon +132 \zeta_3+\frac{70 \epsilon
   ^2}{27}+\frac{641}{6}\right]+4 \alpha_g \alpha_u
   \alpha_y (1-\rc) (\rf+1) \left(\epsilon
   +\frac{11}{2}\right){}\nonumber\\ 
			& \hspace{0.5cm}+8 \alpha_g \alpha_v
   \alpha_y \left(\epsilon +\frac{11}{2}\right)
   (1-\rc )\rf+\alpha_u \text{$\alpha_y
   $}^2 \left[60 \rc+30 \rf \left(\epsilon
   +\frac{11}{2}\right)+12 \epsilon ^2+162 \epsilon
   +528\right]{}\nonumber\\ 
			& \hspace{0.5cm}+\alpha_v \alpha_y^2 \left[24
   \rc \rf+48 \rc+60 \rf \left(\epsilon
   +\frac{11}{2}\right)\right]{}\nonumber\\ 
			& \hspace{0.5cm}+\alpha_y^3 \left[\rc
   ((6 \zeta_3-28) \epsilon +39 \zeta_3-162)-\frac{3 \epsilon
   ^3}{8}+\frac{59 \epsilon ^2}{16}+\frac{2595 \epsilon
   }{32}+\frac{17413}{64}\right]{}\nonumber\\ 
			& \hspace{0.5cm}-\alpha_u^2 \alpha_v \left(36 \rf+84 \right)\rf-\alpha_u
   \alpha_v^2 \left(96 \rf+24
   \right)\rf+\alpha_u \alpha_v \alpha_y \left[\rf^2 (80 \epsilon +440)+\rf (100 \epsilon
   +490)\right]{}\nonumber\\ 
			& \hspace{0.5cm}+\alpha_v^2 \alpha_y
   \left[\rf^2 \left(85 \epsilon
   +\frac{905}{2}\right)+\rf \left(5 \epsilon
   +\frac{25}{2}\right)\right]-\alpha_v^3 \left(16
   \rf^3+20 \rf^2+4 \rf\right)-\alpha_u^3
   (32 \rf+8){}\nonumber\\ 
			& \hspace{0.5cm}+\alpha_u^2 \alpha_y
   \left[\rf \left(85 \epsilon +\frac{905}{2}\right)+5 \epsilon
   +\frac{25}{2}\right]
\end{align}

\begin{align}
    \beta_u^{(1)} &=8\alpha_u^2+4\alpha_u\alpha_y-(11+2\epsilon)\alpha_y^2+24\alpha_u\alpha_v \rf,\\
    \beta_u^{(2)} &= -24(1+5\rf)\alpha_u^3-16\alpha_y\alpha_u^2-352\alpha_u^2\alpha_v \rf - 3(11+2\epsilon)\alpha_y^2\alpha_u-8\alpha_u\alpha_v^2 \rf(5+41\rf) {}\nonumber\\ 
			& \hspace{0.5cm}
   +10(1-\rc)\alpha_g\alpha_y\alpha_u-48\rf\alpha_y\alpha_u\alpha_v-2(11+2\epsilon)(1-\rc)\alpha_g\alpha_y^2+4(11+2\epsilon)\rf\alpha_y^2\alpha_v+(11+2\epsilon)^2\alpha_y^3,
\end{align}

\begin{align}
   \beta_u^{(3)} &=\alpha_g^2 \alpha_u \alpha_y
   \left[\rc^2 \left(36 \zeta
   _{3}-\frac{119}{4}\right)+\rc \left(-36 \zeta_3+8 \epsilon
   +\frac{53}{2}\right)-8 \epsilon +\frac{13}{4}\right]{}\nonumber\\ 
			& \hspace{0.5cm}+\alpha_g^2 \alpha_y^2 \left[\rc^2
   \left(\left(\frac{131}{4}-24 \zeta_3\right) \epsilon -132 \zeta
   _{3}+\frac{1441}{8}\right)+\rc \left((24 \zeta_3-66)
   \epsilon +132 \zeta_3-5 \epsilon ^2-\frac{847}{4}\right)\right.{}\nonumber\\ 
			& \hspace{0.5cm}\left.+5
   \epsilon ^2+\frac{133 \epsilon
   }{4}+\frac{253}{8}\right]+\alpha_g \alpha_v
   \alpha_y^2 \left[\rc \rf ((144 \zeta_3-112)
   \epsilon +792 \zeta_3-616)+\rf ((112-144 \zeta_3)
   \epsilon -792 \zeta_3+616)\right]{}\nonumber\\ 
			& \hspace{0.5cm}-\alpha_v \alpha_y^3 \left[\rc (96 \zeta_3+152)-64 \rf
   \left(\epsilon +\frac{11}{2}\right)\right]+\alpha_g
   \alpha_u^2 \alpha_y (1-\rc) (96 \zeta
   _{3}-102){}\nonumber\\ 
			& \hspace{0.5cm}+\alpha_g \alpha_u \alpha_y^2
   \left[\rc \left(\left(120 \zeta_3-\frac{149}{2}\right)
   \epsilon +660 \zeta
   _{3}-\frac{1639}{4}\right)+\left(\frac{149}{2}-120 \zeta_3\right)
   \epsilon -660 \zeta_3+\frac{1639}{4}\right]{}\nonumber\\ 
			& \hspace{0.5cm}+\alpha_g
   \alpha_y^3 \left[\rc \left((5-24 \zeta_3)
   \epsilon ^2+(55-264 \zeta_3) \epsilon -726 \zeta
   _{3}+\frac{605}{4}\right)\right.{}\nonumber\\ 
			& \hspace{0.5cm}\left.+(24 \zeta_3-5) \epsilon ^2+(264 \zeta
   _{3}-55) \epsilon +726 \zeta_3-\frac{605}{4}\right]{}\nonumber\\ 
			& \hspace{0.5cm}+\alpha_u \alpha_y^3 \left[\rc (12 \zeta
   _{3}-168)-\frac{315 \epsilon ^2}{4}-\frac{3209 \epsilon
   }{4}-\frac{32483}{16}\right]{}\nonumber\\ 
			& \hspace{0.5cm}+\alpha_y^4 \left[\rc
   ((20-24 \zeta_3) \epsilon -132 \zeta_3+110)+\frac{13 \epsilon
   ^3}{4}+\frac{265 \epsilon ^2}{8}+\frac{1111 \epsilon
   }{16}-\frac{2541}{32}\right]+\alpha_u \alpha_v^2
   \alpha_y \left[642 \rf^2+66
   \rf\right]{}\nonumber\\ 
			& \hspace{0.5cm}+\alpha_g \alpha_u \alpha_v
   \alpha_y \left[-\frac{9}{2}\rf(-4+\rf(11+2\epsilon)^2)(-17+16\zeta_3)\right]{}\nonumber\\ 
			& \hspace{0.5cm}+\alpha_u^3 \alpha_v
   \left[\rf^2 (6144 \zeta_3+6752)+\rf (1536 \zeta
   _{3}+2912)\right]-\alpha_u^2 \alpha_v^2 \left[280
   \rf-\rf^2 (9216 \zeta_3+12728)\right]{}\nonumber\\ 
			& \hspace{0.5cm}+\alpha_v^2 \alpha_y^2 \left[\rf^2 ((216 \zeta_3+136)
   \epsilon +1188 \zeta_3+748)+\rf (64 \epsilon
   +352)\right]{}\nonumber\\ 
			& \hspace{0.5cm}-\alpha_u \alpha_v^3
   \left[-\left(\rf^3 (5376 \zeta_3+6568)\right)-\rf^2
   (768 \zeta_3+1472)+104 \rf\right]{}\nonumber\\ 
			& \hspace{0.5cm}+\alpha_u^3
   \alpha_y (226 \rf+34)+648 \alpha_u^2
   \alpha_v \alpha_y \rf+\alpha_u^4
   [\rf (1152 \zeta_3+2360)+104]{}\nonumber\\ 
			& \hspace{0.5cm}+\alpha_u^2
   \alpha_y^2 [\rf ((216 \zeta_3+156) \epsilon +1188
   \zeta_3+858)+166 \epsilon +889]{}\nonumber\\ 
			& \hspace{0.5cm}+\alpha_u \alpha_v \alpha_y^2 \rf ((192 \zeta_3+734) \epsilon
   +1056 \zeta_3+3965)
\end{align}

\begin{align}
    \beta_v^{(1)} &= 12\alpha_u^2 +16\alpha_u\alpha_v+4\alpha_v\alpha_y+4(1+4\rf)\alpha_v^2,\\
    \beta_v^{(2)} &= -24\alpha_v^3\rf(3+7\rf)-8(1+4\rf)\alpha_y\alpha_v^2-352\alpha_u\alpha_v^2\rf-(11+2\epsilon)(3\alpha_v-4\alpha_u)\alpha_y^2-8(5+41\rf)\alpha_u^2\alpha_v{}\nonumber\\ 
			& \hspace{0.5cm}
   +10(1-\rc)\alpha_g\alpha_y\alpha_v-32\alpha_y\alpha_u\alpha_v-24\alpha_y\alpha_u^2 +(11+2\epsilon)^2\alpha_y^3-96\alpha_u^3,
\end{align}

\begin{align}
    \beta_v^{(3)} &=192 \alpha_u^3 \alpha_y+\alpha_y^4
   \left[-10 \epsilon ^3-183 \epsilon ^2-\frac{2211 \epsilon
   }{2}-\frac{8833}{4}\right]{}\nonumber\\ 
			& \hspace{0.5cm}+\alpha_g^2 \alpha_v
   \alpha_y \left[\rc^2 \left(36 \zeta
   _{3}-\frac{119}{4}\right)+\rc \left(-36 \zeta_3+8 \epsilon
   +\frac{53}{2}\right)-8 \epsilon +\frac{13}{4}\right]{}\nonumber\\ 
			& \hspace{0.5cm}+\alpha_g^2 \alpha_y^2 (1-\rc) \left(24 \epsilon ^2+264
   \epsilon +726\right)+\alpha_g \alpha_u^2
   \alpha_y (1-\rc) (144 \zeta_3-153)+\alpha_g \alpha_u \alpha_v \alpha_y
   (1-\rc) (192 \zeta_3-204){}\nonumber\\ 
			& \hspace{0.5cm}+\alpha_g \alpha_u \alpha_y^2 (1-\rc) [(112-144 \zeta_3)
   \epsilon -792 \zeta_3+616]{}\nonumber\\ 
			& \hspace{0.5cm}+\alpha_g \alpha_v
   \alpha_y^2 (1-\rc) \left[\left(\frac{149}{2}-120
   \zeta_3\right) \epsilon -660 \zeta
   _{3}+\frac{1639}{4}\right]{}\nonumber\\ 
			& \hspace{0.5cm}+\alpha_g \alpha_y^3
   \left[\rc \left((2-24 \zeta_3) \epsilon ^2+(22-264 \zeta
   _{3}) \epsilon -726 \zeta_3+\frac{121}{2}\right)\right.{}\nonumber\\ 
			& \hspace{0.5cm}\left.+(24 \zeta_3-2)
   \epsilon ^2+(264 \zeta_3-22) \epsilon +726 \zeta
   _{3}-\frac{121}{2}\right]{}\nonumber\\ 
			& \hspace{0.5cm}+\alpha_v \alpha_y^3
   \left[\rc (12 \zeta_3-136)-\frac{187 \epsilon
   ^2}{4}-\frac{1801 \epsilon
   }{4}-\frac{16995}{16}\right]+\alpha_v^3 \alpha_y
   \left(322 \rf^2+130 \rf\right){}\nonumber\\ 
			& \hspace{0.5cm}+\alpha_g
   \alpha_v^2 \alpha_y \left[\rf^2
   \left((204-192 \zeta_3) \epsilon ^2+(2244-2112 \zeta_3)
   \epsilon -5808 \zeta_3+6171\right)\right.{}\nonumber\\ 
			& \hspace{0.5cm}\left.+\rf \left((51-48 \zeta
   _{3}) \epsilon ^2+(561-528 \zeta_3) \epsilon -1260 \zeta
   _{3}+\frac{5355}{4}\right)+48 \zeta_3-51\right]{}\nonumber\\ 
			& \hspace{0.5cm}+\alpha_u^2 \alpha_v^2 \left[\rf^2 (8064 \zeta
   _{3}+10476)+\rf (1152 \zeta_3+6680)+12\right]+\alpha_u \alpha_v^3 \left[\rf^2 (6144 \zeta
   _{3}+10544)+1264 \rf\right]{}\nonumber\\ 
			& \hspace{0.5cm}+\alpha_v^4
   \left[\rf^3 (2112 \zeta_3+2960)+\rf^2 (960 \zeta
   _{3}+1844)+132 \rf\right]+\alpha_u^2 \alpha_v \alpha_y (642 \rf+66)+648 \alpha_u
   \alpha_v^2 \alpha_y \rf{}\nonumber\\ 
			& \hspace{0.5cm}+\alpha_u^4 [\rf (1536 \zeta_3+1700)+384 \zeta
   _{3}+772]+\alpha_u^3 \alpha_v [\rf (4608
   \zeta_3+9600)+480]{}\nonumber\\ 
			& \hspace{0.5cm}+\alpha_u \alpha_v
   \alpha_y^2 [\rf ((528 \zeta_3+132) \epsilon +2904
   \zeta_3+726)+(96 \zeta_3+152) \epsilon +528 \zeta
   _{3}+788]{}\nonumber\\ 
			& \hspace{0.5cm}+\alpha_v^2 \alpha_y^2 \left[\rf
   ((192 \zeta_3+268) \epsilon +1056 \zeta_3+1426)+41 \epsilon
   +\frac{427}{2}\right]{}\nonumber\\ 
			& \hspace{0.5cm}+\alpha_u^2 \alpha_y^2
   \left[(192 \zeta_3+187) \epsilon +1056 \zeta
   _{3}+\frac{1985}{2}\right]{}\nonumber\\ 
			& \hspace{0.5cm}+\alpha_u \alpha_y^3
   \left[(-96 \zeta_3-88) \epsilon ^2+(-1056 \zeta_3-904)
   \epsilon -2904 \zeta_3-2310\right]
\end{align}

\subsection{Anomalous dimension of fields and scalar mass squared}\label{sec:ADM_fields_mass}
The gauge-independent scalar field anomalous dimension reads
\begin{align}
    \gamma_H^{(1)}&=\alpha_y,\\
    \gamma_H^{(2)}&=2 \alpha _u^2(\rf+1)-\frac{5}{2} \alpha _y\alpha _g (\rc-1) -\frac{3}{4} \alpha _y^2  (2 \epsilon +11) +8  \alpha _u \alpha _v\rf+2\alpha _v^2 \rf
   (\rf+1) ,\\
   \gamma_H^{(3)}&=-4 \alpha _u^3(4 \rf+1)  +\alpha _y \left(-\frac{15}{2} (\rf+1) \alpha _u^2-30 \rf \alpha _u \alpha _v-\frac{15}{2} \rf (\rf+1) \alpha _v^2\right) \nonumber\\&
   \alpha _y^2 \left(\frac{5}{2} (\rf+1) (2 \epsilon +11) \alpha _u+5 \rf (2 \epsilon +11) \alpha _v\right)
   +\alpha _y^3 \left(3 \zeta _3 \rc-2 \rc+\frac{1}{64} (2 \epsilon +11) (10 \epsilon +183)\right)\nonumber\\&
  + \alpha _g \alpha _y^2 \frac{1}{16} \left(48 \zeta _3-5\right) (\rc-1) (2 \epsilon +11) 
  +\frac{1}{16}  \alpha _g^2 \alpha _y (\rc-1)\left(144 \zeta _3 \rc-119 \rc+32 \epsilon -13\right)
  \nonumber\\&
  -12  \alpha _u \alpha _v^2\rf (4 \rf+1)-6 \alpha _u^2 \alpha _v\rf (3 \rf+7) -2 \alpha _v^3\rf (\rf+1) (4 \rf+1).
\end{align}
The scalar mass term respects $U_L(N_f)\times U_R(N_f)$ global symmetry and its anomalous dimension is given by
\begin{align}
    \gamma_{m^2}^{(1)} &= 8\alpha_u +4\alpha_v +2\alpha_y,\\
    \gamma_{m^2}^{(2)} &= -20 (\alpha_u^2 -  \rf ) - 8 \alpha_v \alpha_y( 1+ \rf) - 16 \alpha_u\alpha_y \nonumber \\
		       & +5 \alpha_g \alpha_y (1-\rc) -\frac{3}{2}\alpha_y^2( 11+2\epsilon) - 20\alpha_v^2\rf(1+\rf) - 80\alpha_u \alpha_v \rf,\\
    \gamma_{m^2}^{(3)} &= \alpha_u^3(888 \rf+240) +
    \alpha _u^2 \alpha _v \left(984 \rf^2+2388 \rf+12\right) +\alpha _y \left(\left(33 \rf^2+33 \rf\right) \alpha _v^2+(33 \rf+33) \alpha _u^2+132 \rf \alpha _u \alpha _v\right) \nonumber\\ &
    + \alpha _y^2 \left[\alpha _u \left(264 \zeta _3+264 \zeta _3 \rf+48 \zeta _3 \rf \epsilon -6 \rf \epsilon -33 \rf+48 \zeta _3 \epsilon +76 \epsilon
   +394\right)\right. \nonumber\\ & \left.+\alpha _v \left(528 \zeta _3 \rf+96 \zeta _3 \rf \epsilon +29 \rf \epsilon +\frac{295 \rf}{2}+41 \epsilon
   +\frac{427}{2}\right)\right] \nonumber\\ &
   +\alpha _g \alpha _y \left[\alpha _v \left(48 \zeta _3-48 \zeta _3 \rc-48 \zeta _3 \rc \rf+51 \rc \rf+51 \rc+48 \zeta _3
   \rf-51 \rf-51\right)\right. \nonumber\\& \left.
   +\alpha _u\left(96 \zeta _3-96 \zeta _3 \rc+102 \rc-102\right) \right] \nonumber\\&
   +\alpha _y^3 \left(30 \zeta _3 \rc-24 \rc-\frac{187 \epsilon ^2}{8}-\frac{1801 \epsilon }{8}-\frac{16995}{32}\right) \nonumber\\&+ \alpha_g\alpha_y^2\left(-330 \zeta _3+330 \zeta _3
   \rc+60 \zeta _3 \rc \epsilon -\frac{149 \rc \epsilon }{4}-\frac{1639 \rc}{8}-60 \zeta _3 \epsilon +\frac{149 \epsilon
   }{4}+\frac{1639}{8}\right) \nonumber\\&
   +\alpha _g^2 \alpha _y \left(18 \zeta _3 \rc^2-\frac{119 \rc^2}{8}-18 \zeta _3 \rc+4 \rc \epsilon +\frac{53 \rc}{4}-4 \epsilon
   +\frac{13}{8}\right) \nonumber\\&
   +\alpha _u \alpha _v^2 \left(2664 \rf^2+720 \rf\right)  + \alpha _v^3\left(444 \rf^3+564 \rf^2+120 \rf\right)
\end{align}

The anomalous dimensions of fermion ($\gamma_\psi$), gluon ($\gamma_G$), and ghost ($\gamma_c$) fields depend on the gauge-fixing parameter $\xi$. The results up to three-loop level are given by   

\begin{align}
    \gamma_{\psi}^{(1)}&=\frac{1}{4}\alpha _y (2\epsilon +11) - \frac{1}{2}  \xi  \alpha _g (\rc-1),\\
    \gamma_{\psi}^{(2)}&=-\frac{1}{8}\alpha _g^2 (\rc-1)  (\xi (8 + \xi)+3 \rc-4 \epsilon )+
    \frac{1}{2} \alpha _g \alpha _y(\rc-1) (2 \epsilon +11) +
    -\frac{1}{32} \alpha _y^2(2 \epsilon +11) (2 \epsilon +23),\\
    \gamma_{\psi}^{(3)}&=-\frac{11}{8} \alpha _u^2 \alpha _y(\rf+1) (2 \epsilon +11) +\alpha _y^2 \left(\left(2 \rc+\frac{1}{2} (2 \epsilon +11)^2\right) \alpha _u+4 \rc \alpha _v\right)\nonumber\\&
    \frac{1}{256}\alpha _y^3 (2 \epsilon +11)  \left(192 \zeta _3 \rc-128 \rc+4 (41-3 \epsilon ) \epsilon +1217\right)\nonumber\\&
    -\frac{1}{32} \alpha _g \alpha _y^2(\rc-1) (2 \epsilon +11)  \left(48 \zeta _3+24 \epsilon +137\right)+\frac{1}{64} \alpha _g^2 \alpha _y (\rc-1) (2 \epsilon +11) \left(48 \zeta _3 (\rc+4)-51 \rc-12 \epsilon -77\right)\nonumber\\&
    +\frac{1}{576}\alpha _g^3   \left[
    -160 \epsilon^2 ( \rc-1) + 24 \epsilon (  \rc-1) (109 + 9 \rc) - 90  \xi^3 ( \rc-1) - 27\xi^2 ( \rc-1)  (13 + 4 \zeta_3) \right. \nonumber\\ & \left.- 18 ( \rc-1) (-331 + 208 \rc + 6 \rc^2 - 6 (7 + 16 \rc) \zeta_3) + 
 \xi (612 \epsilon (\rc-1) -27 (\rc-1) (-37 + 8 \zeta_3))\right] \nonumber\\&
  -\frac{11}{2} \alpha _u \alpha _v \alpha _y\rf (2 \epsilon +11) -\frac{11}{8}\alpha _v^2 \alpha _y \rf (\rf+1) (2 \epsilon +11) 
\end{align}

\begin{align}
    \gamma_G^{(1)}&=\frac{1}{6} \alpha _g (3 \xi +4 \epsilon +9),\\
    \gamma_G^{(2)}&=-\frac{1}{8} \alpha _g^2 (-11 \xi -2\xi^2 +\rc (8 \epsilon +44)-28 \epsilon -95) -\alpha _g \alpha _y\frac{1}{4} (2 \epsilon +11)^2, \\
    \gamma_G^{(3)}&=\frac{1}{8} \alpha _g \alpha _y^2(2 \epsilon +11)^2 (3 \epsilon +20) +\frac{1}{32} \alpha _g^2 \alpha _y(6 \rc-31) (2 \epsilon +11)^2 \nonumber\\&
    +\frac{1}{288} \alpha _g^3 \left[-81 \xi - 27\xi^2(11+2\zeta_3)-63\xi^3-36 \rc^2 (2 \epsilon +11)-54 \zeta _3 (-4 \xi +16 \rc (2 \epsilon +11)+16 \epsilon +85)\right.\nonumber\\&\left.+2 \rc (2 \epsilon +11)
   (44 \epsilon +273)-4 \epsilon  (72 \xi +196 \epsilon +347)+6117\right]
\end{align}

\begin{align}
    \gamma_c^{(1)} &= \frac{1}{4}\alpha _g (\xi -3),\\
    \gamma_c^{(2)} &= \frac{1}{48} \alpha _g^2 (-3 \xi +20 \epsilon +15),\\
    \gamma_c^{(3)} &=-\alpha _g^2 \alpha _y\frac{23}{64} (2 \epsilon +11)^2 \nonumber\\&
    +\frac{1}{1728}\alpha _g^3 \left[
     81 \xi^3 - 162 \xi^2 (-1 + \zeta_3) - 
 108 \xi (7 \epsilon + 6 (5 + \zeta_3)) \right. \nonumber \\ & \left.+ 
 12 \epsilon (983 + 216 \zeta_3 + 27 \rc (-15 + 16 \zeta_3)) + 560 \epsilon^2 +
 9 (3569 + 1530 \zeta_3 + 198 \rc (-15 + 16\zeta_3))
    \right]
\end{align}

\subsection{Anomalous dimensions for dimension-three operators}\label{sec:ADM_3}
Here we collect the matrix elements of the $4\times 4$ anomalous dimension for a set of dimension-three operators. Note that we assume that the operators are rescaled according to Eq.~\eqref{eq:dim3_ops_rescaled}: 
{\allowdisplaybreaks
\begin{align}
	(\gamma_{O}^{(1)})_{11} & = 
		-3 \alpha_g (1 - \rc) 
		+\frac{\alpha_y}{2}\left(11 + 2 \epsilon\right) 
		,\label{eq:op3_gamma_11_1l}\\
	(\gamma_{O}^{(2)})_{11} & = 
		2 \alpha_g \alpha_y(11 + 2 \epsilon) (1 - \rc) 
		+\frac{\alpha_g^2 }{12} (1 - \rc) \left(9 \rc+20 \epsilon -93\right)
		+\alpha_y^2 \left[2 \rc-\frac{1}{4} \epsilon  (\epsilon +17)-\frac{253}{16}\right]
		,\label{eq:op3_gamma_11_2l}\\
	(\gamma_{O}^{(3)})_{11} & = 
		\alpha_g^2 \alpha_y \left[\frac{9}{2} (11 + 2 \epsilon) (1 - \rc) \left(\rc-2\right)\zeta_{3} -\frac{1 - \rc}{32} \left(2 \epsilon  \left(157 \rc+136 \epsilon +861\right)+959 \rc+1243\right)\right] \nonumber\\
				     & 
		+\alpha_g^3 \left[6 (11 + 2 \epsilon) (1-\rc^2) \zeta_{3}
		-\frac{1 - \rc}{216}  \left(27 (92 \epsilon +377) \rc+3483 \rc^2-4 \epsilon  (70 \epsilon +1947)-11538\right)\right] \nonumber\\
				     & 
		+\alpha_g \alpha_y^2 \left[3 (11 + 2 \epsilon) (1 - \rc) \zeta_{3}-\frac{1 - \rc}{16}  \left(128 \rc+2 \epsilon  (76 \epsilon +895)+5247\right)\right] 
		-11 \alpha_u \alpha_v \alpha_y (11 + 2 \epsilon) \rf \nonumber\\
				     & 
		+\alpha_u \alpha_y^2 \left((11 + 2 \epsilon)^2+12 \rc\right)
		-\frac{11\alpha_v^2 \alpha_y  }{4} (11 + 2 \epsilon) (1 + \rf) \rf
		+\alpha_v \alpha_y^2 \left(4 \rc \left(\rf+3\right)\right) \nonumber\\
				     & 
		+\alpha_y^3 \left[\frac{3}{2} (11 + 2 \epsilon) \rc \zeta_{3}  -\frac{11 + 2 \epsilon}{128}  (4 \epsilon  (3 \epsilon -41)-1217)-(14 \epsilon +79) \rc\right]
		-\frac{11 (11 + 2 \epsilon) (1 + \rf)}{4} \alpha_u^2 \alpha_y
		,\label{eq:op3_gamma_11_3l}
\end{align}
% g12
\begin{align}
	(\gamma_{O}^{(1)})_{12} \cdot \alpha_y^{-3/2} & = 
		-4 (11 + 2 \epsilon) 
		,\label{eq:op3_gamma_12_1l}
\\
	(\gamma_{O}^{(2)})_{12} \cdot \alpha_y^{-3/2} & = 
		-8 (11 + 2 \epsilon) (1 - \rc) \alpha_g 
		+ 8 \alpha_v (11 + 2 \epsilon) \rf
		+ 4 (11 + 2 \epsilon)^2 \alpha_y
		,\label{eq:op3_gamma_12_2l}
\\
	(\gamma_{O}^{(3)})_{12} \cdot \alpha_y^{-3/2} & = 
		\alpha_g^2 (11 + 2 \epsilon) (1 - \rc)\left[48  \zeta_{3} \rc+\frac{1}{2}  \left(-131 \rc+20 \epsilon +23\right)\right] 
		+ \alpha_g \alpha_u  (11 + 2 \epsilon) (1 - \rc)\left[72  -96  \zeta_{3}\right] 
		\nonumber\\
				     &
		+ \alpha_y^2 (11 + 2 \epsilon) \left[\frac{1}{8}  \left(320 \rc+4 \epsilon  (13 \epsilon +61)-231\right)-48  \zeta_{3} \rc\right] 
		+ \alpha_g \alpha_y (11 + 2 \epsilon)^2 (1 - \rc) \left[24 \zeta_{3}-5 \right] 		\nonumber \\
				     &
		 + \alpha_g \alpha_v  (11 + 2 \epsilon) (1 - \rc) \rf \left[112 -144 \zeta_{3} \right] 
+ \alpha_v^2 (11 + 2 \epsilon) \rf \left[96 \zeta_{3} \rf+32  \left(5 \rf+2\right) \right] \nonumber\\
				     &
		+ \alpha_u^2  (11 + 2 \epsilon) \left[96 \zeta_{3} \rf+32 \left(5 \rf+2\right)\right]
		+ \alpha_u \alpha_v  (11 + 2 \epsilon) \rf \left[96  \zeta_{3} +480  \right] \nonumber \\
				     &
		- \alpha_u \alpha_y \left[96 \zeta_{3} \rc+2 \left(180 \rc+4 \epsilon  (25 \epsilon +263)+2761\right)\right] 
		+ \alpha_v \alpha_y \left[64 \left((11 + 2 \epsilon) \rf-6 \rc\right)-288 \zeta_{3} \rc\right] 
		,\label{eq:op3_gamma_12_3l}
\end{align}
% g12
\begin{align}
	(\gamma_{O}^{(2)})_{13} \cdot \alpha_y^{-3/2} & = 
		%8 ( 11+ 2 \epsilon) \alpha_u + 
		4 (11 + 2 \epsilon)^2 ( \alpha_y + 2 \alpha_u)
		,\label{eq:op3_gamma_13_2l}
\\
	(\gamma_{O}^{(3)})_{13} \cdot \alpha_y^{-3/2} & = 
		96 (11 + 2 \epsilon) \alpha_u^2 (1 +  \zeta_{3})  		
		+ \alpha_g \alpha_u  (11 + 2 \epsilon) (1 - \rc) (112 -144  \zeta_{3})
		\nonumber \\
							   &
		+ \alpha_g \alpha_v (11 + 2 \epsilon) (1 - \rc)(72 -96 \zeta_{3})
		+ \alpha_g \alpha_y (11 + 2 \epsilon)^2 (1 - \rc) \left(24  \zeta_{3}-2 \right)
		\nonumber \\
							   &
		+ \alpha_u \alpha_v (11 + 2 \epsilon) \left[48  \zeta_{3} \left(5 \rf+1\right)+16 \left(11 \rf-1\right)\right] 
		+ 24 \alpha_g^2 (11 + 2 \epsilon)^2 (1 - \rc) 
		\nonumber\\
							   &
		- \alpha_u \alpha_y(11 + 2 \epsilon) \left[72 (11 + 2 \epsilon) \zeta_{3}+4 (22 \epsilon +105)\right]
		+ 96 \alpha_v^2 (11 + 2 \epsilon) \rf \left(1 + \zeta_{3}\right) 
		\nonumber \\ 
							   &
		- \alpha_v \alpha_y \left[96 \zeta_{3} \rc+8 \left(32 \rc+12 \epsilon  (\epsilon +10)+297\right)\right]
		-(11 + 2 \epsilon)^2 (10 \epsilon +73) \alpha_y^2
		,\label{eq:op3_gamma_13_3l}
\end{align}
% g21
\begin{align}
	(\gamma_{O}^{(2)})_{21} \cdot \alpha_y^{-3/2} &=
		-\frac{(11 + 2 \epsilon) (1 + \rf)}{2} 
		,\label{eq:op3_gamma_21_2l}
\\
	(\gamma_{O}^{(3)})_{21} \cdot \alpha_y^{-3/2} & = 
		\frac{\alpha_g}{4}(11 + 2 \epsilon) (1 - \rc) (1 + \rf)
		+ \alpha_u (11 + 2 \epsilon) \left(1 + 6 \rf\right)
		+ \frac{\alpha_v }{2} (11 + 2 \epsilon) \rf \left(9 + 5 \rf\right) \nonumber \\
							   &
		+ \frac{\alpha_y }{8} \left[36 \rc+4 \epsilon  \left(5 \rf+\epsilon +16\right)+110 \rf+231\right]
		,\label{eq:op3_gamma_21_3l}
\end{align}
%\hrule
% g22
\begin{align}
	(\gamma_{O}^{(1)})_{22} & = 
		3 \alpha_y
		+ 8 \alpha_u
		+ 12 \rf \alpha_v
		,\label{eq:op3_gamma_22_1l}
\\
	(\gamma_{O}^{(2)})_{22} & = 
		-16 \alpha_u \alpha_y
		+ \frac{15 \alpha_g \alpha_y }{2} (1 - \rc)
		-248 \rf \alpha_u \alpha_v
		-24 \rf \alpha_v \alpha_y \nonumber\\
				     &
		 -2 \alpha_u^2\left(61 \rf+13\right) 
		 -6 \alpha_v^2\left(19 \rf+3\right) \rf 
		-\frac{9 \alpha_y^2}{4} (11 + 2 \epsilon)
		,\label{eq:op3_gamma_22_2l}
\\
	(\gamma_{O}^{(3)})_{22} & = 
		\alpha_g \alpha_u \alpha_y (1 - \rc)(96  \zeta_{3}-102 ) 
		- \alpha_g^2 \alpha_y  (1 - \rc)\left[27 \zeta_{3} \rc-\frac{3}{16}  \left(119 \rc-32 \epsilon +13\right)\right]
		\nonumber\\
				     &
		+ \alpha_g \alpha_v \alpha_y(1 - \rc) \rf \left(144  \zeta_{3} - 153 \right)
		+ \alpha_g \alpha_y^2 (11 + 2 \epsilon) (1 - \rc)\left[\frac{303 }{16}-33 \zeta_{3}\right]
		+ 454 \rf \alpha_u \alpha_v \alpha_y
		\nonumber\\
				     &
		+ \alpha_u^2 \alpha_v \left[1152 \zeta_{3} \rf \left(4 \rf+1\right)+2 \rf \left(2511 \rf+1195\right)\right]
		+ \alpha_u^3 \left[1152 \zeta_{3} \rf+108 \left(22 \rf+1\right)\right]
		\nonumber\\
				     &
		+ \alpha_u \alpha_v^2 \rf \left[4608 \zeta_{3} \rf+140 \left(50 \rf-1\right) \right]
		+ \alpha_u \alpha_y^2 \left[84 (11 + 2 \epsilon) \zeta_{3} \rf+\left(71 \epsilon  +\frac{781}{2} \right) \rf +129 \epsilon +\frac{1371}{2}\right]
		\nonumber \\
				     &
		+ \alpha_v^3 \rf \left[192 \zeta_{3} \left(7 \rf+1\right) \rf+6 \left(7 \rf \left(46 \rf+15\right)-9\right)\right]
		+ \alpha_v \alpha_y^2 \rf \left[48 (11 + 2 \epsilon) \zeta_{3} +\frac{3}{2} (166 \epsilon +889) \right]
		\nonumber\\
				     &
		- \alpha_y^3 \left[\frac{3}{64} (11 + 2 \epsilon) (310 \epsilon +1321) + (76 - 33 \zeta_3 ) \rc\right]
		+ \frac{\alpha_u^2 \alpha_y}{2} \left(83 + 467 \rf\right) 
		+ \frac{3 \alpha_v^2 \alpha_y}{2} \rf \left(17 + 145 \rf\right)
		,\label{eq:op3_gamma_22_3l}
\end{align}
% g23
\begin{align}
	(\gamma_{O}^{(1)})_{23} & = 
		12 \alpha_u
		+ 8 \alpha_v
		,\label{eq:op3_gamma_23_1l}
\\
	(\gamma_{O}^{(2)})_{23} & = 
		-16 \alpha_v \alpha_y
		-24 \alpha_u \alpha_y
		+ 2( 11 + 2 \epsilon ) \alpha_y^2
		- 96 \alpha_u^2
		-112 \rf \alpha_v^2
		-24 \left(1 + 9 \rf\right) \alpha_u \alpha_v
		,\label{eq:op3_gamma_23_2l}
\\
	(\gamma_{O}^{(3)})_{23} & = 
		\alpha_g \alpha_u \alpha_y(1 - \rc)  [144 \zeta_{3}-153 ] 
		+ 192 \alpha_u^2 \alpha_y
		+ \alpha_g \alpha_v \alpha_y (1 - \rc) [96  \zeta_{3}-102 ] 
\nonumber\\
				     &
		+ \alpha_g \alpha_y^2 (11 + 2 \epsilon) (1 - \rc)(28 -36 \zeta_{3})
		+ \alpha_u^2 \alpha_v \left[3456 \zeta_{3} \rf+8 \left(901 \rf+31\right)\right]
		\nonumber\\
				     &
		+ \alpha_u^3 \left[384 \zeta_{3} \left(4 \rf+1\right)+1700 \rf+772\right]
		+ \alpha_u \alpha_v^2 \rf  \left[576 \zeta_{3} \left(7 \rf+1\right)+4 \left(1233 \rf+769\right)\right] 
		\nonumber \\
				     &
		+ \alpha_u \alpha_y^2 \left[72 (11 + 2 \epsilon) \zeta_{3}+139 \epsilon +\frac{1457}{2}\right]
		+ \alpha_v^3 \rf \left[1536 \zeta_{3} \rf+8 \left(259 \rf+47\right) \right]
		\nonumber \\
				     &
		+ \alpha_v \alpha_y^2 \left[12 (11 + 2 \epsilon) \zeta_{3} \left(8 \rf+1\right)+2 \epsilon  \left(8 \rf+37\right)+88 \rf+383\right]
		\nonumber\\
				     &
		- \alpha_y^3 (11 + 2 \epsilon)\left[6 (11 + 2 \epsilon) \zeta_{3}+ (22 \epsilon +105)\right]
		+ 224 \rf \alpha_v^2 \alpha_y
		+ 48 \alpha_u \alpha_v \alpha_y \left(1 + 9 \rf\right) 
		,\label{eq:op3_gamma_23_3l}
\end{align}
%\hrule
% g31
\begin{align}
	(\gamma_{O}^{(2)})_{31} \cdot \alpha_y^{-3/2}& = 
		-(11 + 2 \epsilon) \rf
		,\label{eq:op3_gamma_31_2l}
\\
	(\gamma_{O}^{(3)})_{31} \cdot \alpha_y^{-3/2} & = 
		\frac{\alpha_g}{2} (11 + 2 \epsilon) (1 - \rc) \rf
		+ \frac{\alpha_u}{2} (11 + 2 \epsilon)  \left(5 \rf+9\right) \rf
		\nonumber \\
							   &
		+ \alpha_v (11 + 2 \epsilon)  \left(6 \rf+1\right) \rf 
		 % + \alpha_y \rf \left(2 \rc +3 \frac{\rc}{\rf} +5 \epsilon +\frac{55}{2}\right)
							   + \alpha_y \rf \left[2 \rc  
							   %+ \frac{3}{4} (11 + 2\epsilon)^2 +\frac{5}{2} (11 + 2 \epsilon)\right]
							    + \frac{473}{4} + 38 \epsilon + 3 \epsilon^2 \right]
		,\label{eq:op3_gamma_31_3l}
\end{align}
%\hrule
% g32
\begin{align}
	(\gamma_{O}^{(1)})_{32} & = 
		12 \rf \alpha_u
		,\label{eq:op3_gamma_32_1l}
\\
	(\gamma_{O}^{(2)})_{32} & = 
		2 \alpha_y^2 (11 + 2 \epsilon) \rf
		-112 \alpha_u^2 \rf  
		-24 \left(1 + 9 \rf\right) \alpha_u \alpha_v \rf  
		-24 \alpha_u \alpha_y \rf  
		,\label{eq:op3_gamma_32_2l}
\\
	(\gamma_{O}^{(3)})_{32} & = 
		\alpha_g \alpha_u \alpha_y (1 - \rc) \rf \left[144  \zeta_{3} -153 \right]
		+ \alpha_g \alpha_y^2 (11 + 2 \epsilon) (1 - \rc) \rf \left[28 -36  \zeta_{3} \right]
		+ 48 \alpha_u \alpha_v \alpha_y\left[1 + 9 \rf\right]  \rf  
		\nonumber\\
				     &
		+ \alpha_u^2 \alpha_v \rf \left[4608 \zeta_{3} \rf+16 \left(361 \rf-8\right) \right]
		+ \alpha_u^3 \rf \left[384 \zeta_{3}  \left(1 + 4 \rf\right)+4 \left(437 \rf+141\right)\right]
		\nonumber\\
				     &
		+ \alpha_u \alpha_v^2 \rf \left[576 \zeta_{3} \left(1 + 7 \rf\right) \rf+12 \left(\rf \left(387 \rf+71\right)-4\right)\right]
		+ \alpha_u \alpha_y^2 \rf \left[24 (11 + 2 \epsilon) \zeta_{3} +\frac{1}{2} (470 \epsilon +2513) \right]
		\nonumber\\
				     &
		+ \alpha_v \alpha_y^2 (11 + 2 \epsilon) \rf \left[84  \zeta_{3} \rf+4 \left(4 + 7 \rf\right) \right]
		%+ \alpha_y^3 \rf \left[8 \left(-7 \rc/\rf +4 \epsilon +22 \right)-24 \zeta_{3} \rc/\rf\right]
		- \alpha_y^3 \rf (11 + 2 \epsilon) \left[2 ( 69 + 14 \epsilon) +6 (11 + 2 \epsilon) \zeta_3\right]
		+ 224  \alpha_u^2 \alpha_y \rf 
		,\label{eq:op3_gamma_32_3l}
\end{align}
% g33
\begin{align}
	(\gamma_{O}^{(1)})_{33} & = 
		3 \alpha_y
		+ 8 \alpha_u
		+ 4 \left(1 + 4 \rf\right) \alpha_v
		,\label{eq:op3_gamma_33_1l}
\\
	(\gamma_{O}^{(2)})_{33} & = 
		-16 \alpha_u \alpha_y
		+ \frac{15\alpha_g \alpha_y }{2} (1 - \rc) 
		-248 \rf \alpha_u \alpha_v
		-2 \alpha_v^2\left(85 \rf+37\right) \rf  
		\nonumber \\
				     &
		-6 \alpha_u^2 \left(3 + 19 \rf\right) 
		-8 \alpha_v \alpha_y\left(1 + 4 \rf\right) 
		-\frac{9 \alpha_y^2}{4} (11 + 2 \epsilon) 
		,\label{eq:op3_gamma_33_2l}
\\
	(\gamma_{O}^{(3)})_{33} & = 
		\alpha_g \alpha_u \alpha_y (1 - \rc)(96  \zeta_{3}-102) 
		- \alpha_g^2 \alpha_y (1 - \rc) \left[27 \zeta_{3} \rc+\frac{3}{16}  \left(-119 \rc+32 \epsilon -13\right)\right] 
		\nonumber \\
	&
		+ \alpha_g \alpha_v \alpha_y (1 - \rc) \left(1 + 4 \rf\right)\left[48  \zeta_{3} -51 \right]
		+ \alpha_g \alpha_y^2 (11 + 2 \epsilon) (1 - \rc) \left[\frac{303 }{16}-33  \zeta_{3}\right]
	\nonumber\\	
	&
		+ \alpha_u^2 \alpha_v \left[576 \zeta_{3} \left(1 + 7 \rf\right) \rf+5562 \rf^2+3646 \rf+12\right]
		+ \alpha_u^3 \left[1152 \zeta_{3} \rf+4 \left(602 \rf+59\right)\right]
		\nonumber\\
	&
		+ \alpha_u \alpha_v^2 \rf \left[4608 \zeta_{3} \rf+60 \left(142 \rf+15\right) \right]
		+ \alpha_u \alpha_y^2 \left[12 (11 + 2 \epsilon) \zeta_{3} \left(2 + 9 \rf\right)+23 \epsilon  \rf+\frac{253 \rf}{2}+81 \epsilon +\frac{843}{2}\right]
		\nonumber \\
	&
		+ \alpha_v^3 \rf \left[192 \zeta_{3} \left(11 \rf+5\right) \rf+2 \left(\rf \left(1484 \rf+927\right)+67\right) \right]
		+ \frac{\alpha_v^2 \alpha_y}{2}  \left(659 \rf+275\right) \rf 
		\nonumber\\
	&
		+ \alpha_v \alpha_y^2 \left[72 (11 + 2 \epsilon) \zeta_{3} \rf+\epsilon  \left(210 \rf+41\right)+1107 \rf+\frac{427}{2}\right]
		+ 454 \rf \alpha_u \alpha_v \alpha_y
		\nonumber \\
	&
		- \alpha_y^3 \left[\frac{9}{64} (11 + 2 \epsilon) (82 \epsilon +323) +(70-33 \zeta_{3}) \rc-70 \rc\right]
		+ \frac{3}{2} \alpha_u^2 \alpha_y\left(145 \rf+17\right) 
		\label{eq:op3_gamma_33_3l}
\end{align}
\begin{align}
	(\gamma_{O}^{(1)})_{44} & = 
		\alpha_y
		,\label{eq:op3_gamma_44_1l}
\\
	(\gamma_{O}^{(2)})_{44} & = 
		2 \alpha_u^2 \left(1 + \rf\right) 
		+ 2 \alpha_v^2 \rf \left(1 + \rf\right) 
		+ 8 \alpha_u \alpha_v\rf 
		 -\frac{3}{4} (11 + 2 \epsilon) \alpha_y^2
		 +\frac{5}{2} \alpha_g \alpha_y \left(1 - \rc\right) 
		,\label{eq:op3_gamma_44_2l}
\\
	(\gamma_{O}^{(3)})_{44} & = 
		\alpha_g^2 \alpha_y (1 - \rc)\left[\frac{1}{16}  \left(119 \rc-32 \epsilon +13\right)-9 \zeta_{3} \rc\right]
		- 2 \alpha_v^3 (1 + \rf) \rf \left(1 + 4 \rf\right)
		\nonumber \\
				&
		+ \alpha_g \alpha_y^2 (11 + 2 \epsilon) (1 - \rc)\left[\frac{5 }{16}-3 \zeta_{3}\right]
		-\frac{15\alpha_v^2 \alpha_y  }{2}(1 + \rf) \rf 
		+ 5 \alpha_v \alpha_y^2 (11 + 2 \epsilon) \rf 
		\nonumber\\
				&
		+ \alpha_y^3 \left[\frac{1}{64} (11 + 2 \epsilon) (10 \epsilon +183)+(3 \zeta_{3}-2 )\rc\right]
		-30 \alpha_u \alpha_v \alpha_y \rf  
		-4  \alpha_u^3\left(1 + 4 \rf\right) 
	\nonumber  \\
				&
		+ \frac{5 \alpha_u \alpha_y^2}{2}  (11 + 2 \epsilon) (1 + \rf) 
		- 12 \alpha_u \alpha_v^2  \rf \left(1 + 4 \rf\right) 
		-6 \alpha_u^2 \alpha_v \rf \left(3 \rf+7\right) 
		-\frac{15 \alpha_u^2 \alpha_y}{2} (1 + \rf)
		,\label{eq:op3_gamma_44_3l}
		\\
% g14
	(\gamma_{O}^{(1)})_{14}\cdot \alpha_y^{-1/2}  & = 
		-4
		,\label{eq:op3_gamma_14_1l}
\\
	(\gamma_{O}^{(2)})_{14}\cdot \alpha_y^{-1/2} & = 
		-10 (1 - \rc) \alpha_g
		+ 3 (11 + 2 \epsilon) \alpha_y
		,\label{eq:op3_gamma_14_2l}
\\
	(\gamma_{O}^{(3)})_{14}\cdot \alpha_y^{-1/2} & = 
		-5 \alpha_u \alpha_y (11 + 2 \epsilon) (1 + \rf) 
		+ \alpha_g^2 (1 - \rc) \left[36  \zeta_{3} \rc-\frac{1}{4}  \left(119 \rc-32 \epsilon +13\right)\right]
		-10 \alpha_v \alpha_y  (11 + 2 \epsilon) \rf
		\nonumber\\
				&
		+ \alpha_g \alpha_y(11 + 2 \epsilon) (1 - \rc) \left(12  \zeta_{3}-\frac{5 }{4}\right)
		- \alpha_y^2 \left[\frac{1}{16} (11 + 2 \epsilon) (10 \epsilon +183)+( 12 \zeta_{3} +8 ) \rc\right]
		,\label{eq:op3_gamma_14_3l}
		\\
%%
% g24
	(\gamma_{O}^{(2)})_{24} & = 
		-2 \alpha_u(1 + \rf) 
		-4 \alpha_v \rf 
		,\label{eq:op3_gamma_24_2l}
\\
	(\gamma_{O}^{(3)})_{24} & = 
		\frac{15 \alpha_u \alpha_y}{2}  (1 + \rf) 
		+ 15 \alpha_v \alpha_y \rf  
		+ 4 \alpha_u^2 \left(1 + 4 \rf\right) 
		\nonumber\\
				     &
		+ 4 \alpha_u \alpha_v \rf \left(3 \rf+7\right) 
		+ 4 \alpha_v^2 \rf \left(1 + 4 \rf\right) 
		-\frac{5 \alpha_y^2}{4} (11 + 2 \epsilon) (1 + \rf)
		,\label{eq:op3_gamma_24_3l}
		\\
%%
% g34
	(\gamma_{O}^{(2)})_{34} & = 
		-2 \alpha_v (1 + \rf) \rf
		-4  \alpha_u \rf
		,\label{eq:op3_gamma_34_2l}
\\
	(\gamma_{O}^{(3)})_{34} & = 
		2 \alpha_v^2 (1 + \rf) \rf \left(1 + 4 \rf\right)
		+ \frac{15 \alpha_v \alpha_y }{2} (1 + \rf) \rf
		-\frac{5 \alpha_y^2 }{2} (11 + 2 \epsilon) \rf
		\nonumber \\
				     &
		+ 15  \alpha_u \alpha_y \rf
		+ 2 \alpha_u^2 \rf \left(3 \rf+7\right) 
		+ 8 \alpha_u \alpha_v \rf \left(1 + 4 \rf\right) 
		.\label{eq:op3_gamma_34_3l}
\end{align}
}
\section{Other approximations for finite-$N_c$ factors}\label{sec:other_fits}
In this appendix we provide the approximations for finite-$N_c$ factors obtained by fitting numerical data to the ratio of two second-order polynomials in $N_c$. The fixed point factors are given as
\begin{equation}
\begin{aligned}
    f_{g}^{(1)}&=\frac{N_c^2}{N_c^2-\frac{110}{19}}, \qquad f_{g}^{(2)}=\frac{N_c^2+6.750}{N_c^2-7.669}, \qquad f_{g}^{(3)}=\frac{N_c^2+26.937}{N_c^2-8.402},\\
    f_{y}^{(1)}&=\frac{N_c^2-1}{N_c^2-\frac{110}{19}}, \qquad f_{y}^{(2)}=\frac{N_c^2+5.578}{N_c^2-7.518}, \qquad f_{y}^{(3)}=\frac{N_c^2+23.618}{N_c^2-8.379},\\
    f_{u}^{(1)}&=\frac{N_c^2-0.9724}{N_c^2-\frac{110}{19}}, \qquad f_{u}^{(2)}=\frac{N_c^2+5.612}{N_c^2-7.545}, \qquad f_{u}^{(3)}=\frac{N_c^2+25.240}{N_c^2-8.395},\\
    f_{v}^{(1)}&=\frac{N_c^2-0.9393}{N_c^2-\frac{110}{19}}, \qquad f_{v}^{(2)}=\frac{N_c^2+4.233}{N_c^2-7.355}, \qquad f_{v}^{(3)}=\frac{N_c^2+17.972}{N_c^2-8.278}.
\end{aligned}
    \label{eq:fit_2nd_order}
\end{equation}
The factors for the field anomalous dimensions and that of $m^2_{\phi}$ are 
\begin{equation}
\begin{aligned}
    f_{H}^{(1)}&=\frac{N_c^2-1}{N_c^2-\frac{110}{19}}, \qquad f_{H}^{(2)}=\frac{N_c^2+5.855}{N_c^2-7.543}, \qquad f_{H}^{(3)}=\frac{N_c^2+27.656}{N_c^2-8.415},\\
    f_{\psi}^{(1)}&=\frac{N_c^2-1}{N_c^2-\frac{110}{19}}, \qquad f_{\psi}^{(2)}=\frac{N_c^2+8.285}{N_c^2-7.747}, \qquad f_{\psi}^{(3)}=\frac{N_c^2+31.882}{N_c^2-8.448},\\
    f_{m^2}^{(1)}&=\frac{N_c^2-0.999}{N_c^2-\frac{110}{19}}, \qquad f_{m^2}^{(2)}=\frac{N_c^2+21.855}{N_c^2-7.977}, \qquad f_{m^2}^{(3)}=\frac{N_c^2+114.062}{N_c^2-8.562},\\
    f_{G(c)}^{(1)}&=\frac{N_c^2}{N_c^2-\frac{110}{19}}, \qquad f_{G(c)}^{(2)}=\frac{N_c^2+6.959}{N_c^2-7.707}, \qquad f_{G(c)}^{(3)}=\frac{N_c^2+29.766}{N_c^2-8.437}.
\end{aligned}
    \label{eq:fit_2nd_order_fields_and_mm}
\end{equation}
The critical exponents corresponding to dimension-4 operators are corrected by 
\begin{equation}
\begin{aligned}
    f_{\theta_1}^{(2)} & =\frac{N_c^2 }{N_c^2 - \frac{110}{19}}, & 
    f_{\theta_1}^{(3)} & = \frac{N_c^2 + 4.15}{N_c^2 - 7.085} &
f_{\theta_1}^{(4)} & = \frac{N_c^2 + 30.8}{N_c^2 - 8.403} ,\\
	f_{\theta_2}^{(1)} & =\frac{N_c^2 - 1}{N_c^2 - \frac{110}{19}}, & 
f_{\theta_2}^{(2)} & = \frac{N_c^2 + 5.71}{N_c^2 - 7.507}, &
f_{\theta_2}^{(3)} & = \frac{N_c^2 + 29.7}{N_c^2 - 8.473}, \\
f_{\theta_3}^{(1)} & = \frac{N_c^2 - 0.9103}{N_c^2 - 5.78323}, & 
f_{\theta_3}^{(2)} & = \frac{N_c^2 + 5.01}{N_c^2 - 7.5524}, &
f_{\theta_3}^{(3)} & = \frac{N_c^2 + 32.9}{N_c^2-8.4346}, \\
f_{\theta_4}^{(1)} & = \frac{N_c^2 - 1.175}{N_c^2-5.8011}, & 
f_{\theta_4}^{(2)} & = \frac{N_c^2 + 29.5}{N_c^2 - 7.891}, &
f_{\theta_4}^{(3)} & = \frac{N_c^2 + 144}{N_c^2 - 8.543}.
\end{aligned}
	\label{eq:theta_2nd_order}
\end{equation}
The approximate expressions for the two nontrivial eigenvalues for $\gamma^*_O$ are given as
\begin{equation}
\begin{aligned}
f_{\gamma_3}^{(1)} & = \frac{N_c^2 - 1.153}{N_c^2 - 5.7996}, & 
f_{\gamma_3}^{(2)} & = \frac{N_c^2 + 15.14}{N_c^2 - 7.856}, &
f_{\gamma_3}^{(3)} & = \frac{N_c^2 + 71.03}{N_c^2 -8.5345}, \\
f_{\gamma_4}^{(1)} & = \frac{N_c^2 - 0.919}{N_c^2-5.7839}, & 
f_{\gamma_4}^{(2)} & = \frac{N_c^2 + 6.04}{N_c^2 - 7.6215}, &
f_{\gamma_4}^{(3)} & = \frac{N_c^2 + 32}{N_c^2 - 8.4419}.
\end{aligned}
	\label{eq:gamma_2nd_order}
\end{equation}

%\bibliography{ls433.bib}

\end{document}